\RequirePackage{pdf14}
\documentclass[%
aip,amsmath,amssymb,
preprint
]{revtex4-1}
\usepackage[utf8]{inputenc}
\usepackage[T1]{fontenc}
\usepackage{mathptmx}
\usepackage{lipsum}
\usepackage{esint}
\usepackage{graphicx,subfigure}
\usepackage{siunitx}
\usepackage[]{algorithm2e}
\usepackage{csquotes}
\usepackage{dcolumn}
\usepackage{bm}
\usepackage[colorlinks=true,urlcolor=cyan,citecolor=cyan,linkcolor=cyan,bookmarks=true]{hyperref}

\usepackage{rotating}
\usepackage{float}
\usepackage{multirow}

\usepackage{color, colortbl}
\definecolor{LightCyan}{rgb}{0.88,1,1}
\definecolor{Gray}{gray}{0.9}

\usepackage{tikz,pgfplots}
\usetikzlibrary{patterns}
\usetikzlibrary{calc}
\usetikzlibrary{arrows,shapes}
\usetikzlibrary{positioning}

\DeclareMathAlphabet{\mathsfit}{T1}{\sfdefault}{\mddefault}{\updefault}
\SetMathAlphabet{\mathsfit}{bold}{T1}{\sfdefault}{\bfdefault}{\updefault}
\renewcommand{\vec}[1]{\mathbf{{#1}}}
\renewcommand{\cite}[1]{\citep{#1}}

\newlength{\FigureHeight}
\newlength{\FigureHeightHalf}
\newcommand{\FigureXYLabel}[5]{%
\settoheight{\FigureHeight}{#1}%
\setlength{\FigureHeightHalf}{0.5\FigureHeight}%
\begin{center}%
\raisebox{\FigureHeightHalf}{\makebox{#4\makebox[#5]{}}}%
#1\\%
\vspace{#3}%
#2\\%
\end{center}%
}

\makeatother
\begin{document}

\title{Couette flow turbulence reduction by the flow spanwise reflection symmetry breaking: On the universality of the control strategy}

\author{George Khujadze}
\affiliation{Schie\ss bergstr 60a, Siegen, Germany}
\email{georgekh@yahoo.com}
\author{David Gogichaishvili}
\affiliation{Institute of Geophysics, Tbilisi State University, Georgia}
\affiliation{Kutaisi International University, Georgia}
\author{George Chagelishvili}
\affiliation{Institute of Geophysics, Tbilisi State University, Georgia}
\affiliation{Georgian National astrophysical Observatory, Georgia}

\begin{abstract}
A novel turbulence control strategy  for wall-bounded shear flow  is proposed by Chagelishvili et al, 2014. The essence of this strategy  involves  continuously imposition of specially designed seed velocity perturbations with spanwise asymmetry near the flow wall. The configuration of this imposed velocity field, enhanced due to the shear flow non-normality,  breaks the flow spanwise reflection symmetry, specifically,  resulting in  the generation of a secondary nonuniform spanwise mean flow. Consequently, this secondary flow  significantly reduces  flow turbulence. In Chagelishvili et al, 2014, the first steps were taken towards developing this new turbulence control strategy and demonstrating its efficiency. The plane Couette flow  was considered, as a representative example, and a theoretical and hypothetical weak near-wall volume forcing was designed, which   though theoretical, provided valuable insights into the  characteristics of the seed velocity field. Obviously, the practical significance of this control strategy should be confirmed by evaluating its effectiveness  in flows at  various Reynolds numbers, and  with different  parameters of the imposed seed velocity field. In this paper, we investigate the effectiveness and universality of the turbulence control strategy for Couette flow at various Reynolds numbers and different locations of the volume forcing. Through  direct numerical simulations, we show  universality of the discussed turbulence control method.  The application of a specially designed, weak near-wall volume forcing with a fixed configuration and amplitude  results in   the same effective turbulence control, reducing turbulence kinetic energy production  by 30-40\% across a wider range Reynolds numbers, $Re_{\tau} = 52,92,128,270$ and various localizations. 
\end{abstract}

\maketitle

\section{Introduction}
Investigations of problems related to the reduction of energy costs of the moving various bodies, such as airplanes, ships, wind turbines in fluids, as well as the transportation of fuels through pipelines, have a century-long history and are mostly focused on drag reduction studies. The significance of successful drag reduction applications in economic and environmental aspects can not be to overestimated. A wide variety of active and passive flow control and drag reduction strategies have been suggested over the years \citep{GadelHak,Bewley2001,Kim2003,Dean,Marusic2021,Ricco2021}. Despite extensive research, practical drag reduction and turbulent flow control methods are still far from the required developments. In this paper, we briefly overview of active and passive control strategies. For a comprehensive review of turbulence control methods and their practical realizations, readers are referred to the work by \citet{Ricco2021}.

A wide variety of active and passive flow control strategies for the drag reduction have been developed over the years \citep{Choi1993,GadelHak,Bewley2001,Kim2003,Marusic2012,Dean,Ricco2021}. Surface riblets are one of the few passive drag-reduction techniques that have been successfully demonstrated in theory and applied in practice. However, a maximum friction-drag reduction of about $8\%$ was achieved \citep{Dean}. Other passive approaches are skewed wavy surfaces and circular or tear-shaped dimples, but they yielded either no drag reduction at all or very modest levels of order of a few percents \citep{Ricco2021}. As for active control strategy, it can be applied via different wall-based forcing methods, introducing in a flow \textit{finite-amplitude}, unsteady or steady perturbations to create a mean spanwise flow: among others are methods of blowing and suction, \citep{Marusic2012,Kametani2011}, wall oscillations  \citep{Baron1995,Choi1998,Quadrio2012,Jovanovic2012,Touber2012,Blesbois2013,Agostini2014,Skote2011,Skote2013,Skote2014,Zaki2014,Marusic2021}, streamwise and spanwise traveling waves \citep{Karniadakis2003,Quadrio2009,Duque2012,Gallorini2022,Fukagata2024} and theoretically interesting ``opposition control'' \citep{Choi1994,Stroh2015,Xia2015,Cheng2021,Marusic2023a,Marusic2023b}.
The spatial oscillations of a segment of the wall under a turbulent boundary layer was used to study the drag reduction\cite{Skote2011,Skote2013,Skote2014}. The results of uniform  and intermittent blowing or suction was presented by \citet{Kametani2011,Kametani2015,Kametani2016}. They performed the dynamical decomposition of the local skin friction coefficient, the FIK-identity\cite{FIK}, and found that the control efficiency of the uniform blowing is higher than the efficiency of other advanced
active control methods proposed for internal flows. Intensive study of global effects of local skin friction drag reduction in boundary layer flow was reported in the paper by \citet{Stroh2016}, where two locally applied drag-reduction control methods were investigated, damping near-wall turbulence and wall-normal constant mass flux. It was shown that both methods give the same results within the control region, but different behavior was observed downstream, after the control region,  where the drag was increased in the former case, while persistent drag reduction happened in the latter case.
Here, it is worth to note that all these methods described above introduce in a flow finite-amplitude perturbations, that, of course, reduces the efficiency of a control due to the large input power. It is well understood that balance between the latter and drag reduction is very subtle and depends on many control parameters. Therefore, it is very important to reduce power input to get a positive net energy balance.

Besides the way of generation of a mean spanwise flow used in the above described control strategies,
which \emph{directly alter} or destroy the near-wall structures responsible for high friction drag,
there exists another, \emph{indirect way} of generation of a mean flow proposed by \citet{Chagelishvili2014}. This strategy is based on a subtle, weak near-wall forcing that initiates the imposition of specially designed secondary velocity field in the flow and, finally, causing the breaking of the spanwise reflection symmetry of turbulence, leads to a reduction of turbulent kinetic energy (TKE) production. This \emph{new control strategy} is not a wall-based method in a sense that it doesn't directly influence, modify near-wall structures, it changes the statistical characteristics of whole turbulent flow leading to the indirect control by the following scheme:\\
$\bullet$ A specially designed spanwise non-symmetric weak near-wall forcing, when implemented in shear flow, generates specially designed (optimal) seed velocity perturbations. These perturbations extract energy from the shear flow and undergo substantial transient growth in the characteristic dynamical time of the flow system \citep{Farr931,Farrell2000}; \\
$\bullet$ these amplified non-symmetric velocity perturbations trigger the breaking of the spanwise reflection symmetry of the flow, causing the generation of a mean spanwise flow, which, \\ 
$\bullet$ changes the TKE and stress balances (statistics) of the turbulence, leading to a substantial reduction of its level.

Overall, the specially designed imposed/seed perturbations, due to the fast growth, become active participant of nonlinear dynamics and drastically change the course of events of uncontrolled turbulent flow. As a result, the final dynamical balance between linear and nonlinear processes is achieved at a substantially lower level of turbulence compared to the uncontrolled flow.

In this paper, we investigate the effectiveness and universality of the turbulence control strategy proposed in \citet{Chagelishvili2014} for Couette flow at various Reynolds numbers and different locations of the volume forcing. The outline of the paper is as follows. Section \ref{Performed} provides the details about the numerical requirements of the numerical simulations performed;  Section \ref{forcing_model} gives the description of the weak, helical near-wall forcing; In section \ref{results} the analysis of the results of the flow control are presented. Finally the conclusions are given.

\section{Model of turbulent plane Couette flow control}
First of all, we provide an overview of the numerical details of the performed DNS of the turbulent plane Couette flow at various Reynolds numbers and in different simulation boxes. 
\subsection{Details of numerical study}
\label{Performed}
The incompressible Navier-Stokes equations are discretized in an orthogonal coordinate system $(x, y, z)$ representing the streamwise, wall-normal and spanwise directions, using Fourier and Chebyshev decompositions in horizontal and wall-normal directions. The pseudo-spectral code (SIMSON) developed at the Royal Institute of Technology, Stockholm\citep{Chevalier2007} was used in our study. The Reynolds number is defined as $Re = U_wh/\nu$, where $U_w$ is the half of the velocity difference of the walls, $h$ is the channel's half-height, and $\nu$ is the kinematic viscosity. The simulations, parameters of which are given are given in Table \ref{tab:control1}, were performed at four different Reynolds numbers, considering two simulation boxes. A comprehensive study of grid convergence was done (see Table \ref{tab:control1} and the figure \ref{evol750s} for resolution details.) 
The simulation parameters in our DNS are in agreement with the standards accepted within the turbulence research community  \citep{Tsukahara2006,Pirozzoli2014,LeeMoser2018}.

\begin{table}
\caption{Computational parameters for turbulent Couette flow cases. The bulk and friction Reynolds numbers are defined as $Re = hU_w/\nu$ and  $Re_{\tau} = hu_{\tau}/\nu$, respectively.  $\Delta x^+, ~\Delta z^+$ represent the grid spacing in wall (viscous) units in parallel directions, while $\Delta y^+_w$ and $\Delta y^+_c$ denote the grid spacing in wall-normal direction at the wall and in the center of the channel, respectively.}
%
\begin{tabular}{l cccccccccccccc}
{case } ~~&~~$Re$~~ & ~~$Re_{\tau}$ ~~&~~  $L_x$~~& ~~$L_z$ ~~& ~~$N_x$ ~~& ~~$N_y$ ~~& ~~$N_z$ ~~& ~~$\Delta x^+$  ~~&~~ $\Delta y^+_w$ ~~ &~~ $\Delta y^+_c$ ~~&~~ $\Delta z^+$ \\
\hline \hline
$1^a$ &$750$  & $52$  & $8\pi$&$4\pi$ & $384$ & $129$ & $256$ & $ 3.4 $ & $ 0.016$  & $ 1.3 $ & $2.6$    \\
$1^b$ &$750$  & $52$  & $8\pi$&$4\pi$ & $512$ & $513$ & $256$ & $ 2.6 $ & $ 0.001 $ & $ 0.3 $ & $2.6$        \\
$1^c$ &$750$  & $52$  & $16\pi$&$8\pi$ & $1024$ & $257$ & $256$ & $ 2.6 $ & $ 0.001 $ & $ 0.3 $ & $2.6$  \\ 
\hline
$2^a$ &$1500$ & $92$  & $8\pi$ & $4\pi$ & $768$ & $257$  & $512$ & $3.0$ & $0.007$ & $1.1 $ & $2.3$                 \\
$2^b$ &$1500$ & $92$  & $8\pi$ & $4\pi$ & $768$ & $385$  & $512$ & $3.0$ & $ 0.003 $  & $ 0.75 $ & $2.3$  \\
$2^c$ &$1500$ & $92$  & $16\pi$ & $8\pi$ & $1024$ & $513$ & $512$ & $4.5$ & $0.002$  & $0.57$ & $4.5$   \\ 
\hline
$3^a$ &$2150$ & $128$ & $8\pi$ &$4\pi$ & $512$  & $385$ & $384$ & $6.28$ & $0.0043$ & $1.05$ & $4.19$ \\
$3^b$ &$2150$ & $128$ & $8\pi$ &$4\pi$ & $1024$ & $513$ & $512$ & $3.14$ & $0.0024$ & $0.79$ & $3.14$  \\
$3^c$ &$2150$ & $128$ & $16\pi$ & $8\pi$ & $1024$ & $385$  & $512$ & $6.23$ & $0.004$  & $1.04$ &  $6.23$     \\ 
\hline
$4^a $ &$5000$ & $270$ &$8\pi$  & $4\pi$ & $1536$ & $769$  & $512$ & $4.4$  & $0.0023$ & $1.1$  &  $6.6 $  \\
$4^b$ &$5000$ & $270$ & $8\pi$  & $4\pi$ & $2048$ & $769$  & $512$ & $3.3$  & $0.0023$ & $1.1$  &  $6.6 $  \\
\end{tabular}
\label{tab:control1}
\end{table}

\subsection{Near-wall weak volume forcing}
\label{forcing_model}
The model of weak volume forcing presented is based on the transient growth of perturbations in shear flows, due to the non-normality of the linearized dynamical operators in such flow systems. This energy growth of perturbations forms the basis of the dynamical activity of smooth shear flows. It is well-known that such flows support a set of
optimal perturbations that undergo large transient growth during
the dynamical timescale of the turbulence for sufficiently high Reynolds numbers. This timescale can be defined by the characteristic time of nonlinear processes, which is order of $\mathcal{O} (1/A)$.
Generally, in smooth shear flows a robust growth appears for $3D$ perturbations satisfying the following conditions\cite{Craik1986,Farr931}:\\
-- the length scales in streamwise and spanwise directions are of
the same order, but larger than the viscous dissipative length scale,
${\ell}_x \simeq {\ell}_z \gg {\ell}_{\nu}$, or, in terms of
wavenumbers, $~k_x,k_z \ll k_{\nu}$ (here, $k_{\nu} \equiv
\sqrt{Re} \approx 1/{\ell}_{\nu}$);\\
-- the perturbations are tilted with the background shear, or, in
terms of wavenumbers, $k_y/k_x < 0$. \\
Based on these conditions the following near-wall helical forcing for the flow under consideration was first presented in \citet{Chagelishvili2014}:
\begin{multline}
F_i (\vec{x}) = A_i f(y) \sum_{n,m=0}^{N,M} \left [ Z_m \delta_{ix} -\delta_{iy} - X_n\delta_{iz}  \right ] 
e^{- X_n^2 - Z_m^2}e^{-\frac{(X_n -\cos \phi)^2}{{l_{x}}^2}-\frac{(Z_m+\sin \phi)^2}{{l_{z}}^2} },~ i = (x,y,z)
\label{VF}
\end{multline}
\begin{eqnarray}
\text{where,} && X_n(a,\phi)=\frac{x}{a}-2n\cos \phi, 
~~~ Z_m(a,\phi)=\frac{z}{a}-1-(4m+1)(1-\sin \phi),  \nonumber \\
&& f(y) = \sin(\pi y) e^{ - {{(|y|-y_{peak})^2}/{l_y^2}} }, ~~~ y \in [-1,1]. \nonumber
\end{eqnarray}
The streamwise and spanwise lengths of simulation box are related to the forcing parameters as: 
\begin{equation}
L_x = 2a (N-1) \cos \phi, ~~L_z = 4a M (1-\sin \phi). \nonumber     
\end{equation}
The numbers of the forcing centers in the streamwise and spanwise directions are $(N+1)$ and $(M+1)$, respectively; $a$ and $\phi$ define the size of the forcing ``cells'' and the orientation in $xz$-plane, respectively. For example, at $\phi=\pi/4$ quasi-equipartition of the forcing between the streamwise and spanwise directions takes place; $l_x,$ $l_y$ and $l_z$ are the length scales of the  forcing localization in the streamwise, wall-normal and spanwise directions; $A$ and $B$ are the forcing amplitudes in parallel and wall-normal directions, correspondingly; ${X}_n(a,\phi)$ and ${Z}_m(a,\phi)$ define the forcing localization centers in parallel directions, while the forcing localization center in the wall-normal direction is defined by $y_{peak}$. 

\begin{figure}[ht!]
\centering
\includegraphics[width = .59\linewidth]{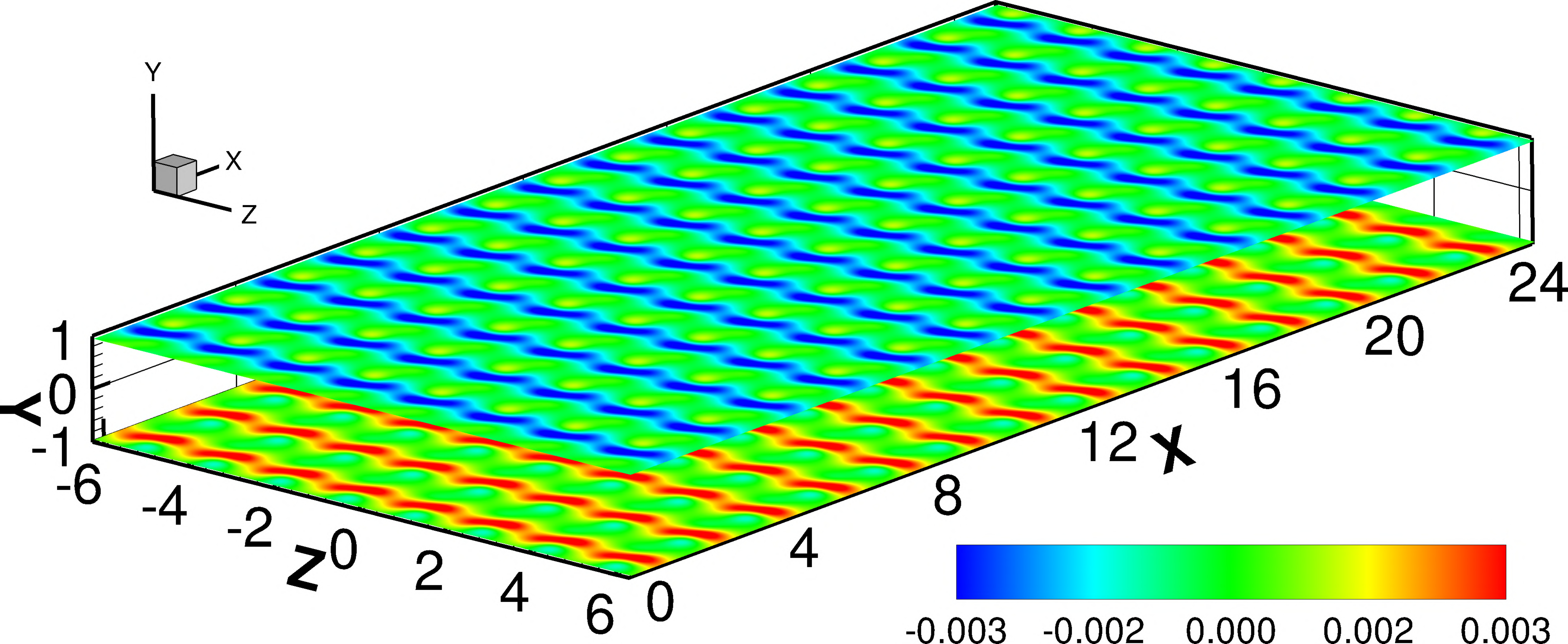}
\hfill
\includegraphics[width = .4\linewidth]{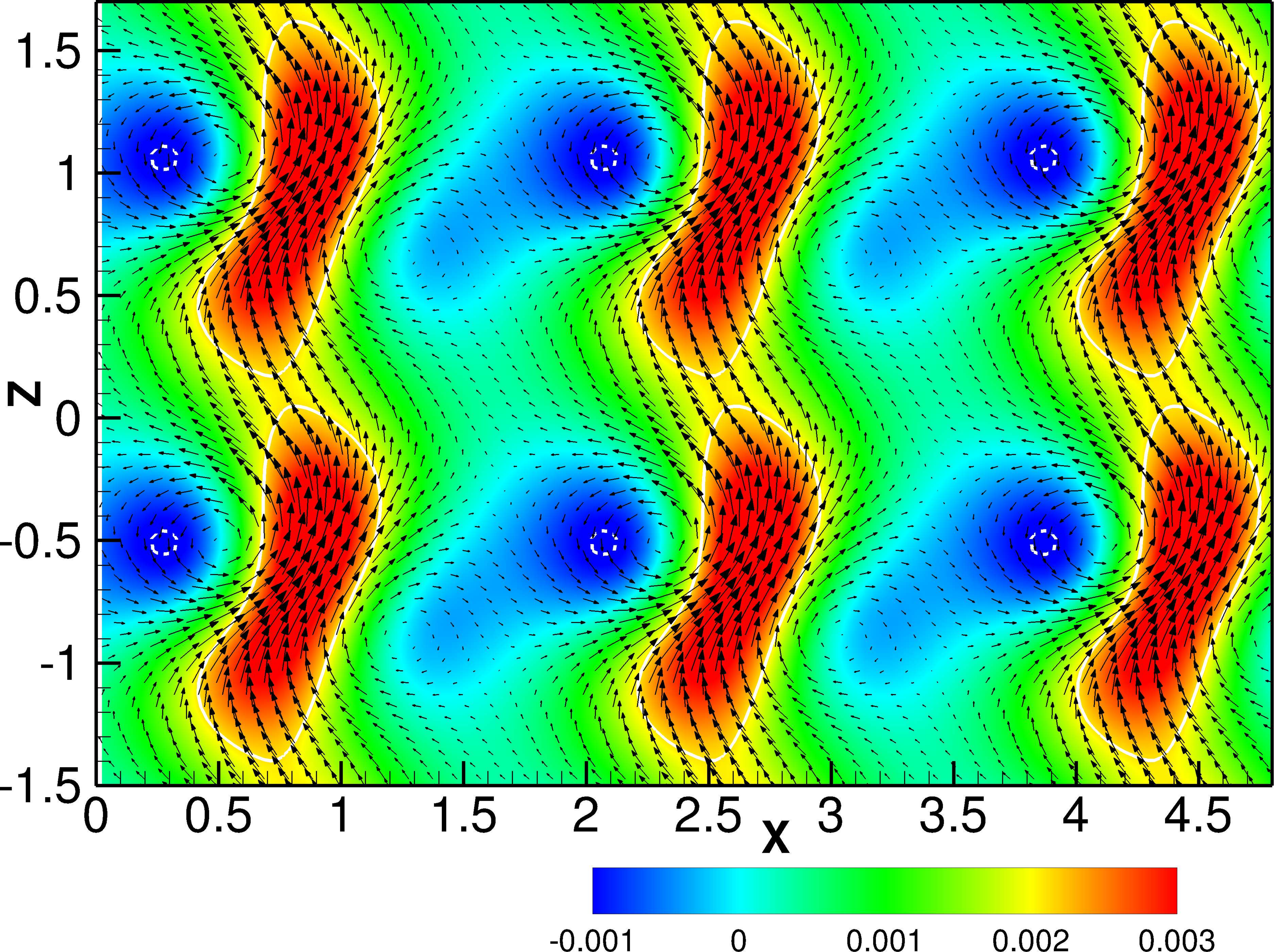}
\caption{Left plot: Spanwise velocity field generated by the volume force. Right plot: Zoom of the velocity field. $u_xu_z$-vector field of seed perturbations is presented with (with barely noticeable) solid and dashed white curves showing the positive and negative values of spanwise velocity component, $u_z^{p} = 0.002$ and $u_z^{n} = -0.001$.}
\label{SketchVolF}
\end{figure}
Plots in Figure \ref{SketchVolF} show the velocity field generated by the forcing. The spanwise velocity field near the walls at $y_{peak}$ is shown in the left plot. The amplitudes of the velocity perturbations generated by the forcing are very small, infinitesimal, order of $10^{-3}$. The zoom of the  field of velocity perturbations $(u_xu_z)$ is shown in the second figure, alongside spanwise velocity contours at negative ($u_z = -0.001$) and positive ($u_z = 0.002$) values, denoted by the solid and dashed white curves, correspondingly.  

\section{Results and discussion}
\label{results}
\begin{figure}[ht!]
\centering
\begin{minipage}[c]{.5\linewidth}
\begin{tabular}{l cc c}
~~$Re$ ~~& ~~$y_{peak}$ ~~& ~~$y_{peak}^+$~~&  ~~$R\%$~~\\
\hline \hline
$750$ & $0.003$ & $0.16$   &$45$     \\
$1500$ & $0.003$ & $0.28$  & $40$    \\
$1500$ & $0.03$  & $2.76$  &$40$    \\
$2150$ & $0.003$ & $0.38$  &$30$     \\
$2150$ & $0.02$  & $2.56$  &$30$     \\
$5000$ & $0.003$ & $0.81$  &$30$    \\ 
\end{tabular}
\end{minipage}
\hspace{.3cm}
\begin{minipage}[c]{.4\linewidth}
\hspace{-.3cm}
\FigureXYLabel{\includegraphics[width = 1.\textwidth]{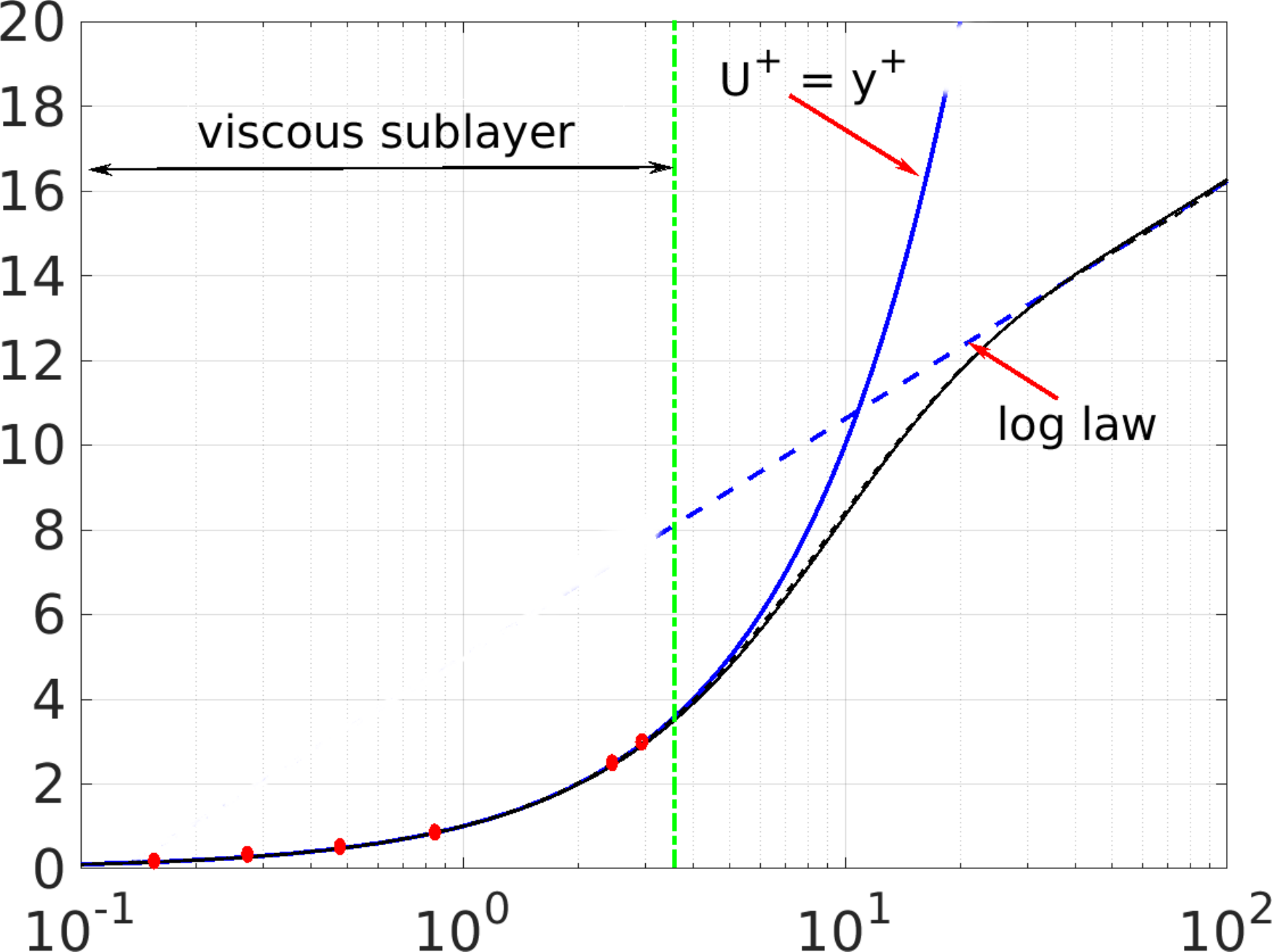}}
{${y^+}$}{-2mm}{\begin{rotate}{90} ${U^+}$ \end{rotate}}{1mm}
\end{minipage}
\caption{Table 2: Forcing localisation and reduction of TKE production for different Reynolds numbers; \emph{Right plot}: Mean velocity profile. Red dots show the forcing localization $y^+_{peak}$ in wall-normal direction (see the table  for the values). All points are located in the viscous sub-layer, $U^+=y^+$ (blue solid line). The log law, $U^+={1}/{0.41}\log(y^+)+5.0$, are presented by the dashed blue line. Black (collapsed dashed and solid) lines represent the results of DNS of the flow at $Re = 1500, 2150$. }
\label{Table_Umean}
\end{figure}

In this section we provide the results of numerical study of the flow turbulence control. The control caused reduction in TKE production is summarized in the table 2 in the figure \ref{Table_Umean}. 

Mean velocity profile is presented in the figure \ref{Table_Umean} on the left plot with the forcing localization points that are shown with red thick dots. All positions are located in the viscous sub-layer, $U^+ = y^+$ (blue solid line) of the turbulent flow. For comparison, log-law with dashed blue line is also shown. The black solid and dashed lines  correspond to the turbulent flow at $Re = 1500, 2150$ Reynolds numbers. 

Once the forcing parameters are determined, the procedure of the control investigation is the following: (i) Initially, the fully developed turbulence is achieved then (ii) the control is activated and the simulation is resumed.
The results are presented in the following subsections.
\subsection{Iso-surfaces of spanwise velocity field: comparison of natural/uncontrolled and controlled turbulent flows}
\begin{figure}[ht!]
\centering
\begin{tikzpicture}[overlay]
\fill[fill=white]
(4.0,1.35) node[fill=white] {$Re=750$}
(2.0,1.35) node[fill=white] {turbulent}
(-.1,.7) node[fill=white] {$\bf y$};
\end{tikzpicture}
\includegraphics[width = .7\linewidth]{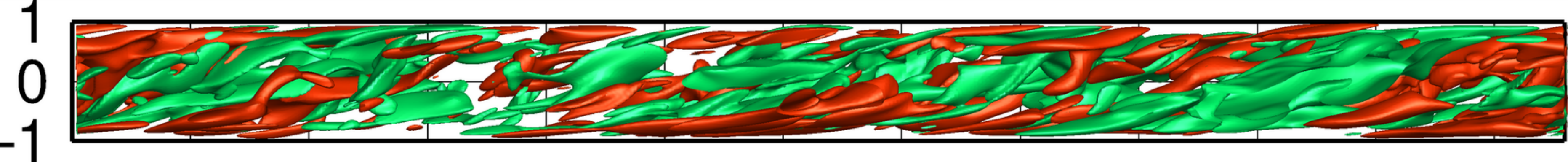} \\
\vspace{.3cm}
\begin{tikzpicture}[overlay]
\fill[fill=white]
(4.0,1.35) node[fill=white] {$Re=750$}
(2.0,1.35) node[fill=white] {controlled}
(-.1,.7) node[fill=white] {$\bf y$};
\end{tikzpicture}
\includegraphics[width = .7\linewidth]{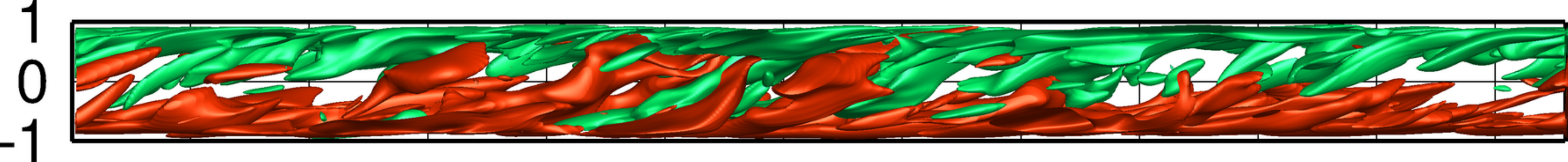}\\
\vspace{0.6cm}
\begin{tikzpicture}[overlay]
\fill[fill=white]
(4.0,1.4) node[fill=white] {$Re=1500$}
(2.0,1.4) node[fill=white] {turbulent}
(-.1,.7) node[fill=white] {$\bf y$};
\end{tikzpicture}
\includegraphics[width = .7\linewidth]{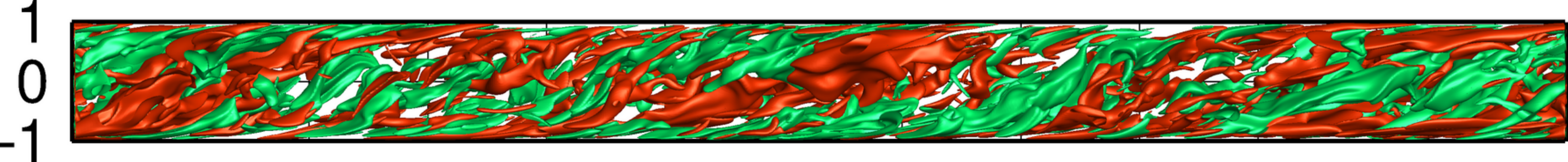}\\
\vspace{0.3cm}
\begin{tikzpicture}[overlay]
\fill[fill=white]
(4.0,1.35) node[fill=white] {$Re=1500$}
(2.0,1.35) node[fill=white] {controlled}
(-.1,.7) node[fill=white] {$\bf y$};
\end{tikzpicture}
\includegraphics[width = .7\linewidth]{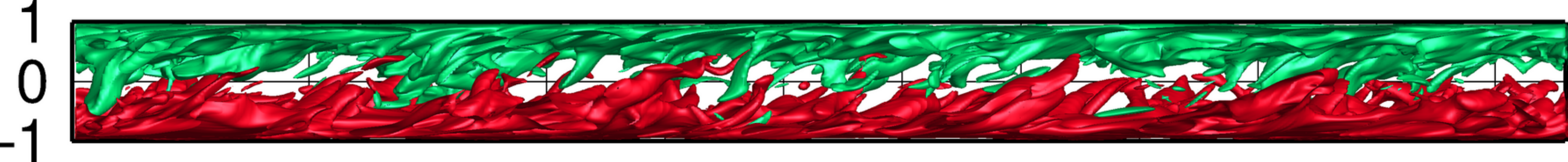}\\
\vspace{0.6cm}
\begin{tikzpicture}[overlay]
\fill[fill=white]
(4.0,1.4) node[fill=white] {$Re=2150$}
(2.0,1.4) node[fill=white] {turbulent}
(-.1,.7) node[fill=white] {$\bf y$};
\end{tikzpicture}
\includegraphics[width = .7\linewidth]{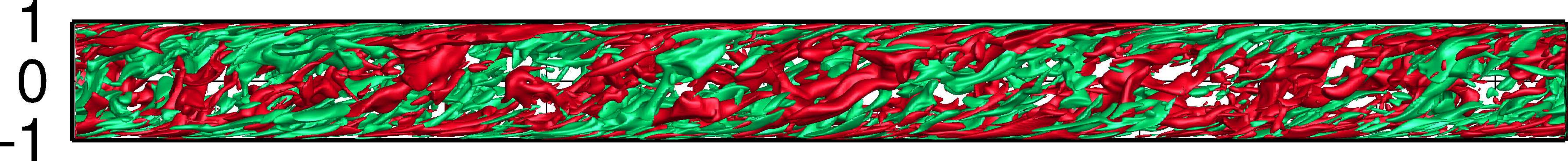} \\
\vspace{.3cm}
\begin{tikzpicture}[overlay]
\fill[fill=white]
(4.0,1.35) node[fill=white] {$Re=2150$}
(2.0,1.35) node[fill=white] {controlled}
(-.1,.7) node[fill=white] {$\bf y$};
\end{tikzpicture}
\includegraphics[width = .7\linewidth]{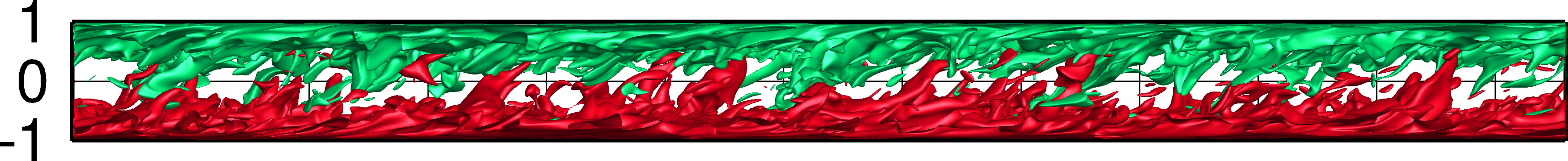}\\
\vspace{0.5cm}
\begin{tikzpicture}[overlay]
\fill[fill=white]
(4.0,1.3) node[fill=white] {$Re=5000$}
(2.0,1.3) node[fill=white] {turbulent}
(-.1,.7) node[fill=white] {$\bf y$};
\end{tikzpicture}
\includegraphics[width = .7\linewidth]{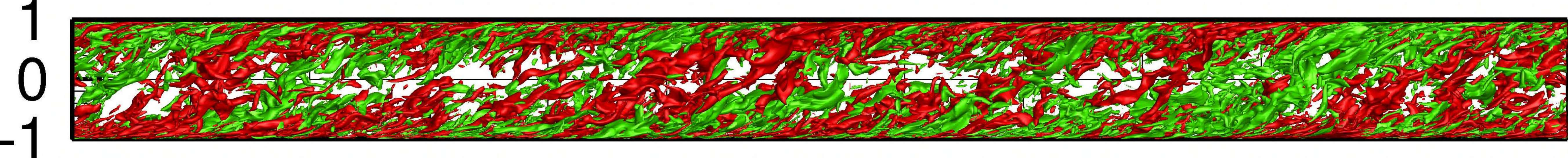} \\
\vspace{.3cm}
\begin{tikzpicture}[overlay]
\fill[fill=white]
(4.0,1.7) node[fill=white] {$Re=5000$}
(2.0,1.7) node[fill=white] {controlled}
(-.1,.8) node[fill=white] {$\bf y$}
(7.,-.3) node[fill=white] {$\bf x$};
\end{tikzpicture}
\includegraphics[width = .7\linewidth]{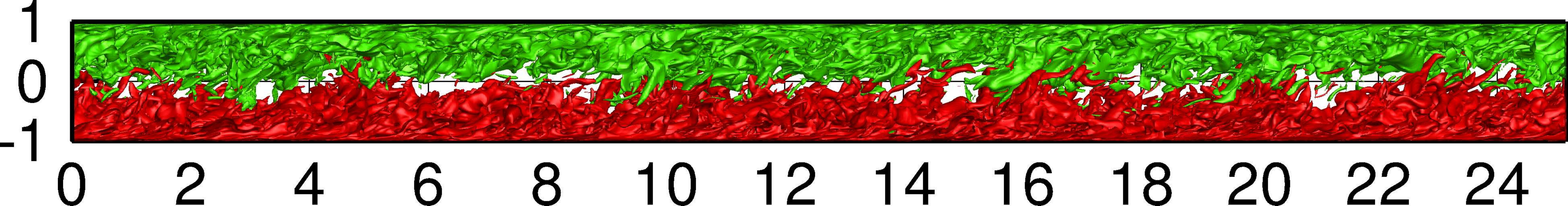}
\caption{$XY$-slices of spanwise velocity iso-surfaces $u_z = -0.15, 0.15$ for turbulent and controlled cases at Reynolds numbers, $Re=750, 1500, 2150, 5000$, from top to bottom. }
\label{Couette2D}
\end{figure}
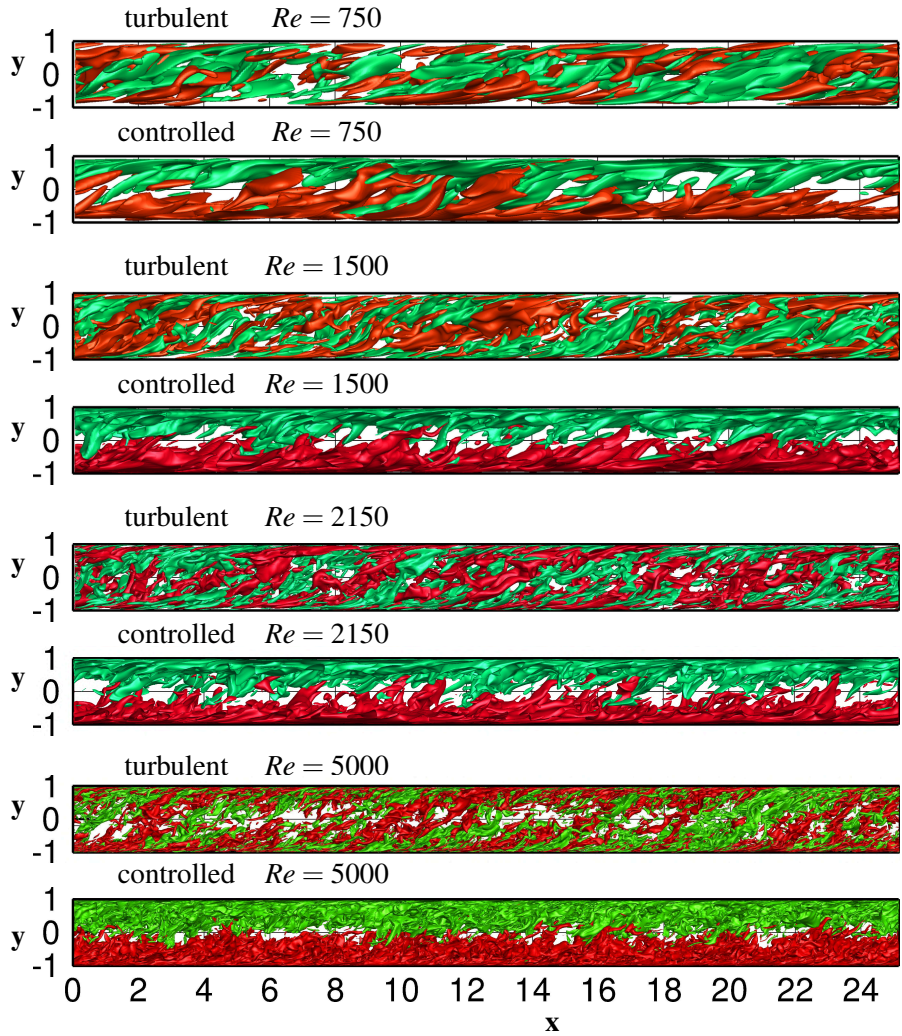

\begin{figure}[ht!]
\begin{minipage}[c]{.45\linewidth}
\FigureXYLabel{\includegraphics[width = .9\linewidth]{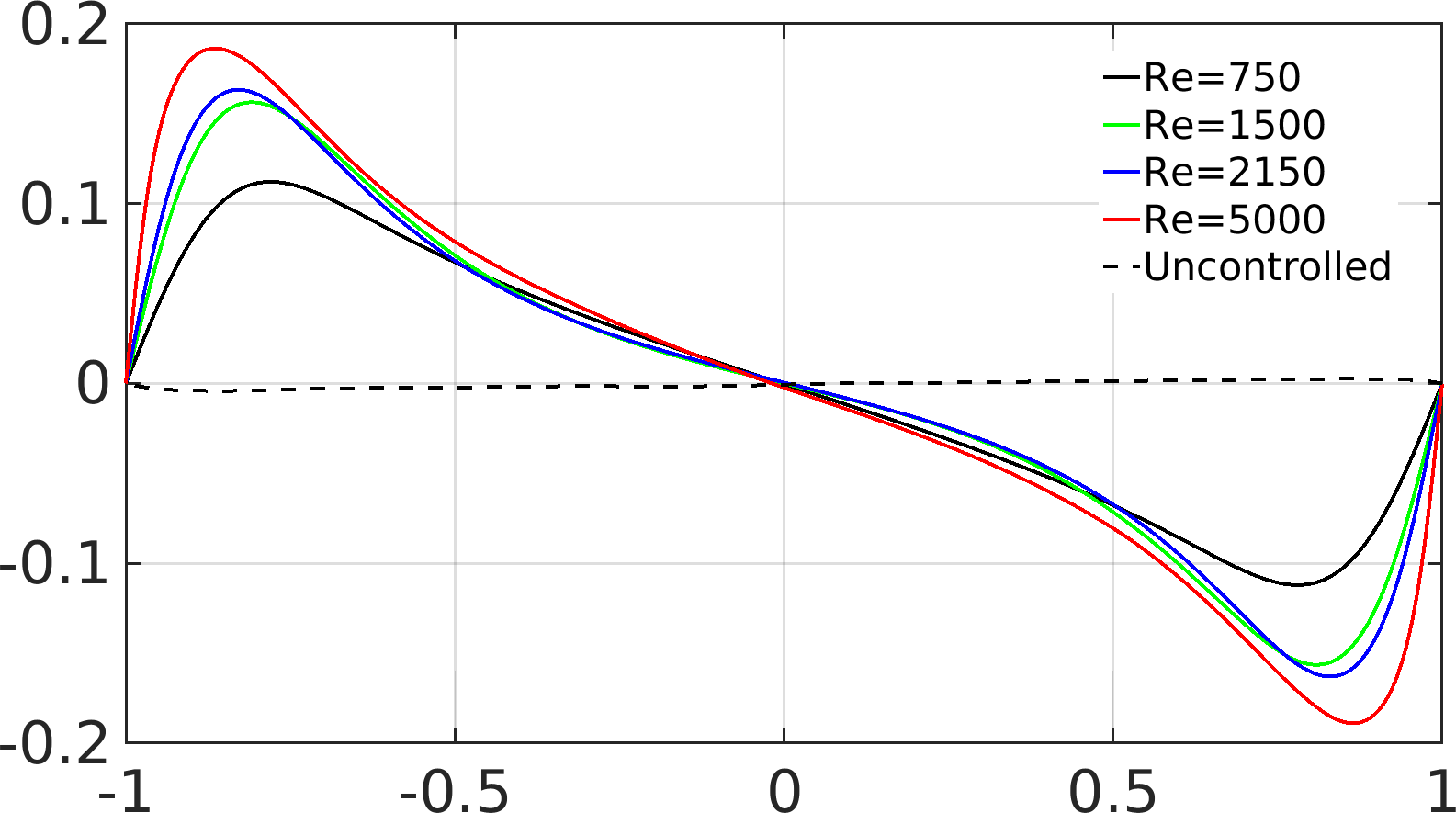}}
{$y$}{-2mm}{\begin{rotate}{90} $\overline{U_z}$\end{rotate}}{1mm}
\end{minipage}
\caption{Mean spanwise velocity profiles of turbulent and controlled flows, solid and dashed lines. $Re= 750,~ 1500, ~2150,~5000$, black, green, blue and red lines, respectively.}
\label{Stat_Uz}
\end{figure}

Figure \ref{Couette2D} the iso-surfaces of the spanwise component of velocity of uncontrolled and controlled turbulent flows are presented for Reynolds numbers $Re=750, 1500, 2150, 5000$. 
As observed, the iso-surfaces with the positive and negative values in the case of uncontrolled turbulent flows are mixed in a manner that results in a zero mean spanwise velocity profile, as expected. In the case of controlled turbulent flows, those iso-surfaces are ``decoupled'' generating a non-zero mean spanwise velocity profile presented on the plot \ref{Stat_Uz}. Plots in Figure \ref{Couette2D} demonstrate that the control with the same parameters (forcing amplitudes, localization in wall-normal direction, and number of forcing centers) yields nearly the same results for the flow control. The detailed analysis of TKE reduction for various Reynolds numbers is provided in the next section. 

The similar results, shown in Figure \ref{Couette2D}, were obtained for the longer simulation box $16\pi \times 2\times 8\pi$ as well, although they are not presented here for the sake of brevity.

\subsection{Production of turbulent kinetic energy: Time evolution}
\begin{figure}[ht!]
\centering
\begin{minipage}[c]{.45\linewidth}
\FigureXYLabel{\includegraphics[width = 1.\textwidth]{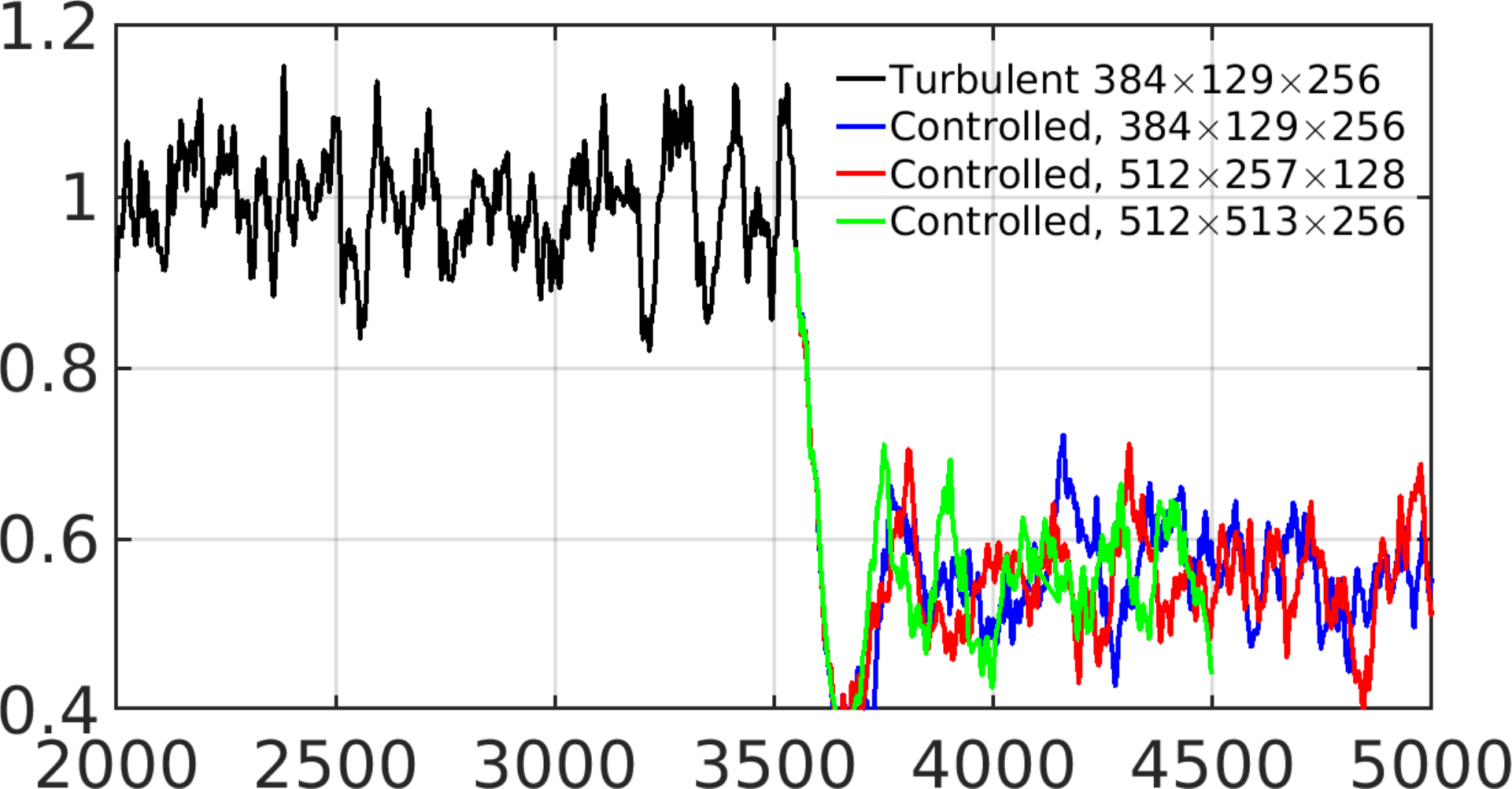}}
{${time}$}{-1mm}{\begin{rotate}{90} ${Pr_x}$ \end{rotate}}{1mm}
\end{minipage}
\begin{tikzpicture}[overlay]
\fill[fill=white]
(-6,.1) node[fill=white] {(a)};
\end{tikzpicture}
\hfill
\begin{minipage}[c]{.48\linewidth}
\FigureXYLabel{\includegraphics[width = 1.\textwidth]{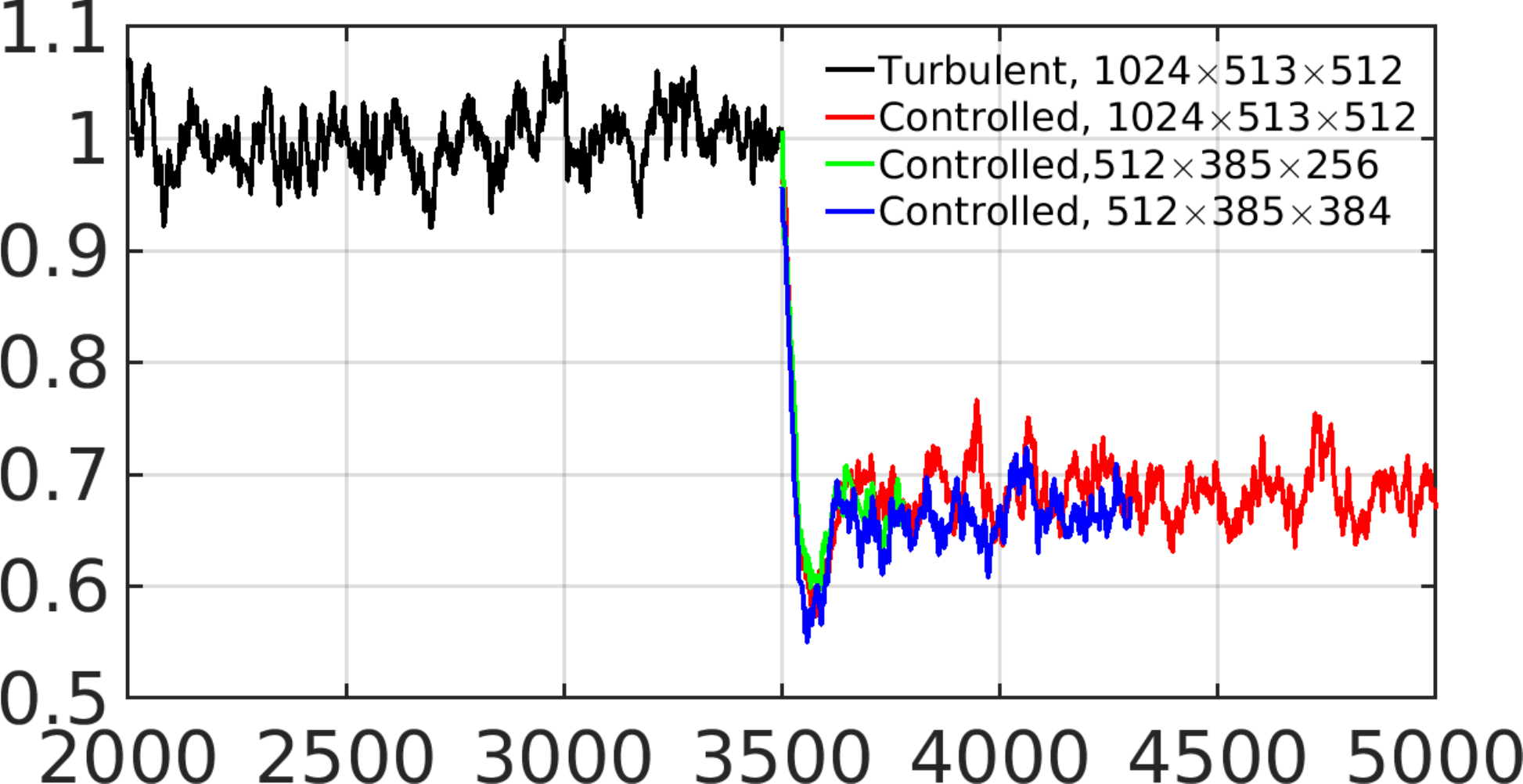}}
{${\small time}$}{-1mm}{\begin{rotate}{90} ${\small Pr_x}$ \end{rotate}}{1mm}
\end{minipage}
\begin{tikzpicture}[overlay]
\fill[fill=white]
(-6,.1) node[fill=white] {(b)};
\end{tikzpicture}
\caption{Time evolution of TKE production for the turbulent and controlled flows at the forcing location $y_{loc} = 0.003$ from the wall. Plot (a): The case of $Re_{\tau}=52$ is presented with different resolutions: $12$, $16$, $67$ million grid points; Plot (b): The case of $Re_{\tau}=128$ is presented with different resolutions: $63$, $77$ and $270$ million grid points. The figures show the robustness of the control when changing the grid.  The only notable change is that the amplitude of TKE production fluctuation varies slightly with grid change.}
\label{evol750s}
\end{figure}
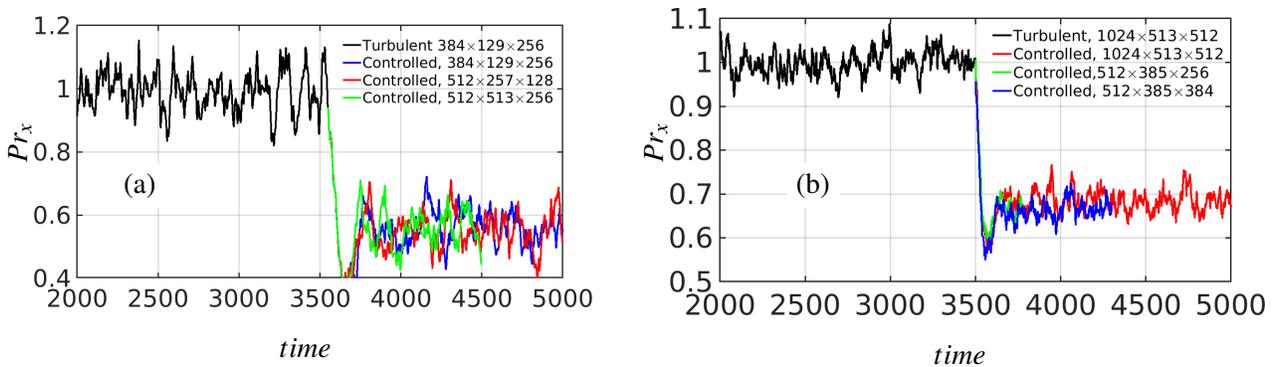

In this subsection, we present the reduction of the production of TKE due to the flow turbulence control. As mentioned earlier, we keep the same forcing parameters while varying the Reynolds number of the flow. First of all, we show the results grid convergence study of the numerical simulations. 
The cases of two Reynolds numbers, $Re_{\tau}=52, 128$ are illustrated in Figure \ref{evol750s}, plots (a) and (b), showing the reduction of TKE production at different number of grid points. It is evident that the control mechanism exhibits robustness while changing the grid. We performed studies of grid convergence in all/three directions of the simulation box. 

\begin{figure}[ht!]
\centering
\begin{minipage}[c]{.9\linewidth}
\FigureXYLabel{\includegraphics[width = .9\linewidth]{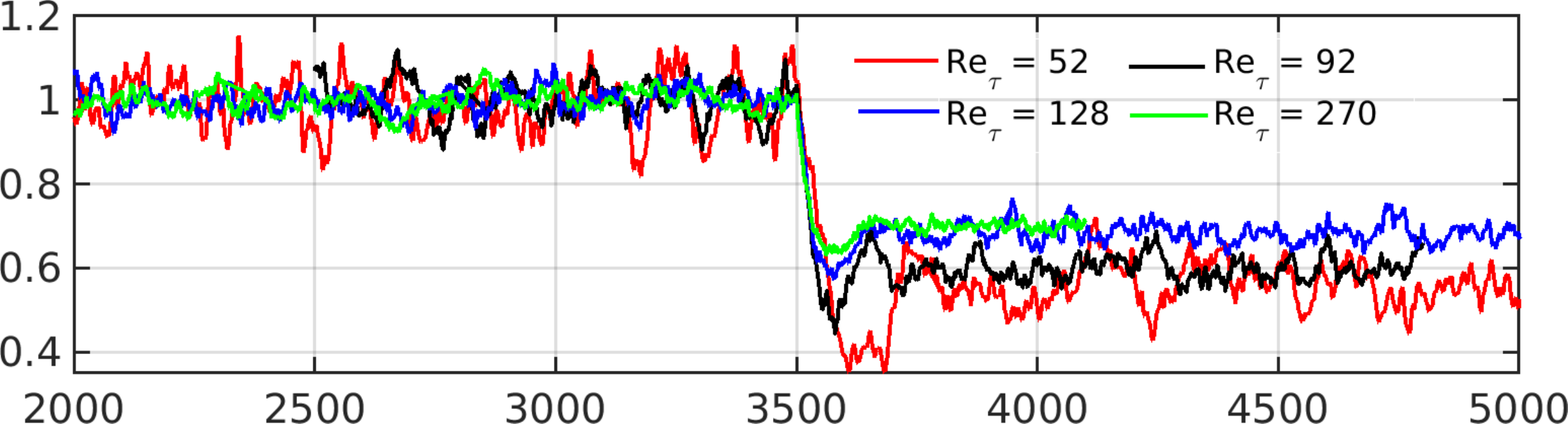}}
{${\small time}$}{-1mm}{\begin{rotate}{90} ${\small Pr_x}$ \end{rotate}}{1mm}
\end{minipage}
\caption{Time evolution of TKE production for the uncontrolled and controlled turbulent flows at forcing localisation $y_{loc} = 0.003$ from the wall. The cases at $Re_{\tau}=52,~ 94,~ 128, ~270$ are presented with red, black, blue and green lines, correspondingly.}
\label{evolAll}
\end{figure}

The comparison of the reduction of TKE at different Reynolds numbers are presented in Figure \ref{evolAll} represents(see Table 2 for achieved reductions). Once again, irrespective of the Reynolds numbers, the control strategy developed and presented in this paper, consistently achieves the reduction of TKE production in the range of $30-40\%$, showing the universality of the control mechanism.  Although the reduction at high Reynolds numbers is slightly less than that at low Reynolds numbers, it still remains sufficiently high to substantiate the earlier statement. It is obvious that a small modification of forcing parameters, such as the amplitude of the forcing, will yield a higher reduction, although this was not the focus of the current paper.

\begin{figure}[ht!]
\centering
\begin{minipage}[c]{.48\linewidth}
\FigureXYLabel{\includegraphics[width = 1.\textwidth]{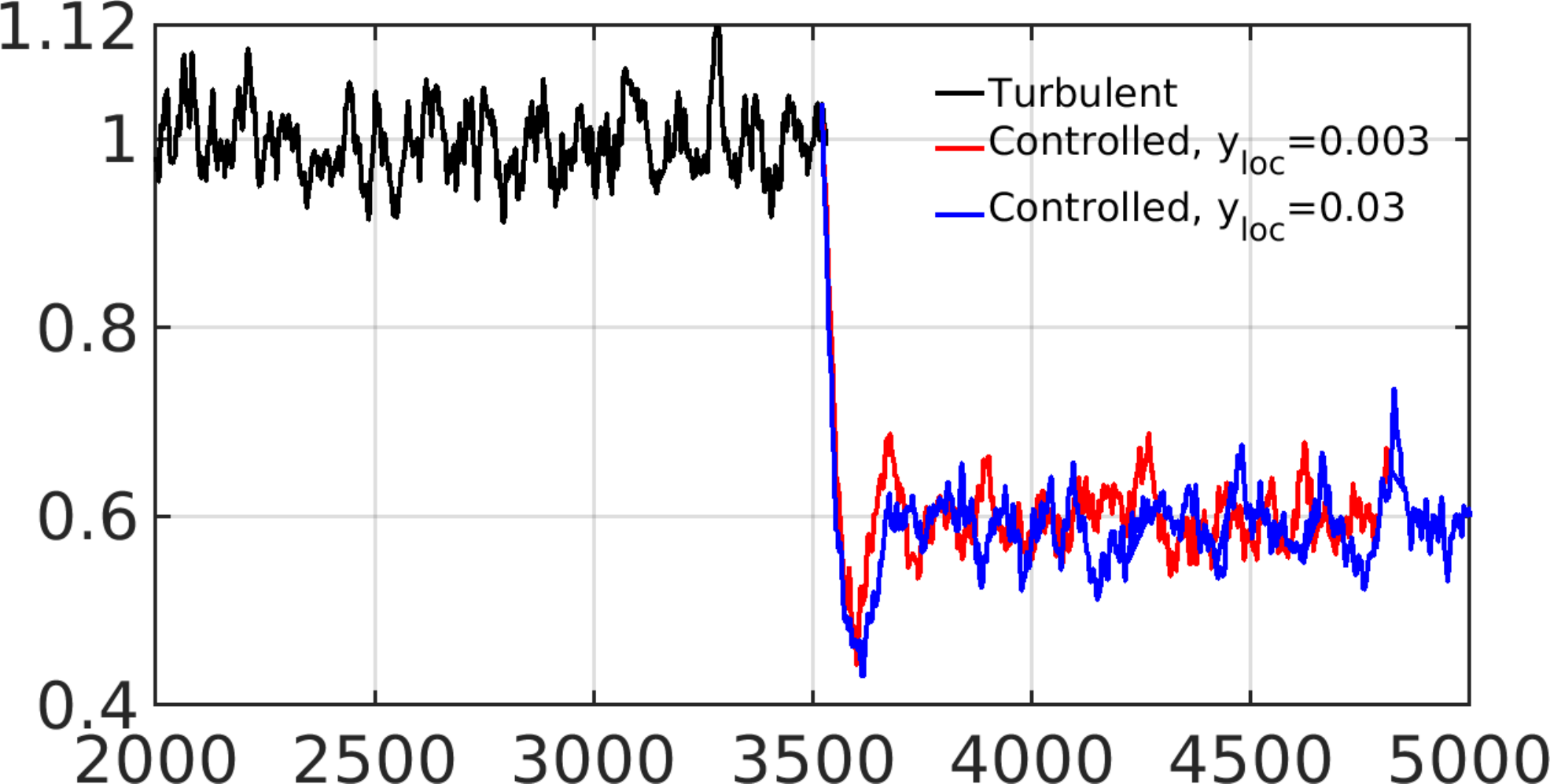}}
{${time}$}{-1mm}{\begin{rotate}{90} ${Pr_x}$ \end{rotate}}{1mm}
\end{minipage}
\begin{tikzpicture}[overlay]
\fill[fill=white]
(-6.,.3) node[fill=white] {(a)};
\end{tikzpicture}
\hfill
\begin{minipage}[c]{.48\linewidth}
\FigureXYLabel{\includegraphics[width = 1.\textwidth]{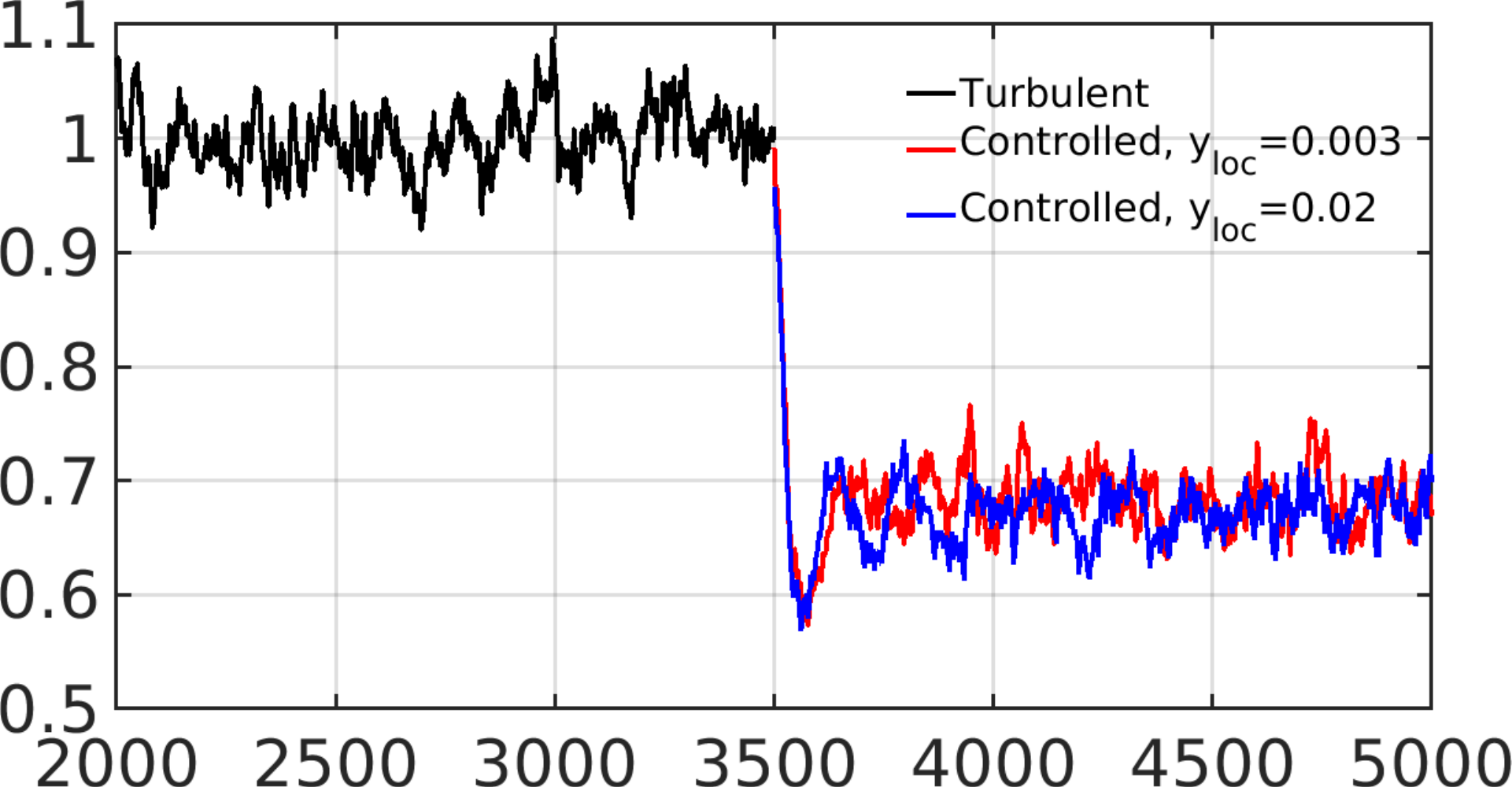}}
{${\small time}$}{-1mm}{\begin{rotate}{90} $Pr_x$ \end{rotate}}{1mm}
\end{minipage}
\begin{tikzpicture}[overlay]
\fill[fill=white]
(-6.,.3) node[fill=white] {(b)};
\end{tikzpicture}
\caption{Time evolution of TKE production for the uncontrolled and controlled turbulent flows at the different forcing localisation from the wall.
Plots show the flows with $Re_{\tau}=92$ (a) and $Re_{\tau}=128$ (b), correspondingly. Black line shows the uncontrolled turbulent flow. The red line shows $y_{loc}=0.003$ for both plots, while the blue line  corresponds to the left (a) $y_{loc}=0.03$ and right $y_{loc}=0.02$  plots (b), correspondingly.}
\label{evol2}
\end{figure}
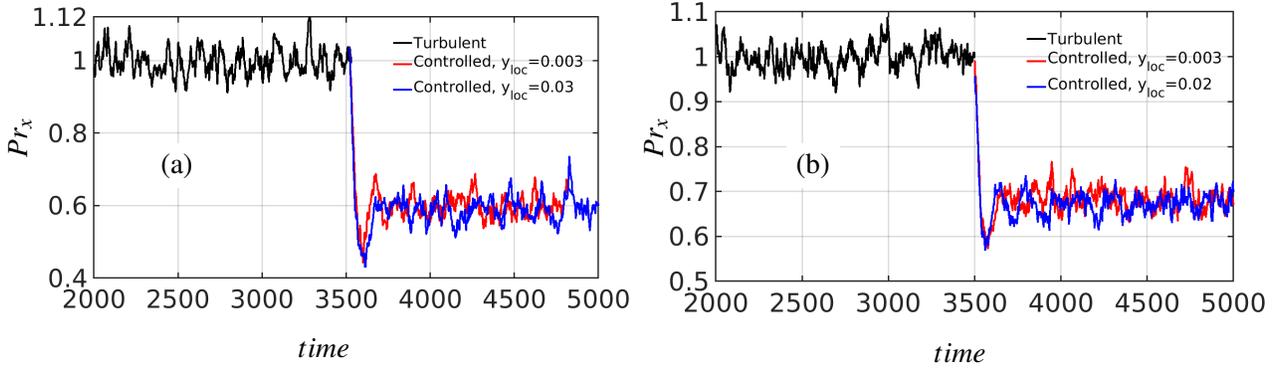

The time evolution of TKE production for the uncontrolled and controlled turbulent flows at two different forcing localisations from the wall are presented in Figure \ref{evol2}. The results for the two different Reynolds numbers, $Re_{\tau}=94, 128$ are shown. For the Reynolds number $Re_{\tau}=94$, the forcing localizations at $y_{loc}=0.003$ and $y_{loc}=0.03$ (in wall units, $y_{loc}^+=0.28$ and $y_{loc}^+=2.76$) were considered. At the higher Reynolds number $Re_{\tau}=128$, the localisation distances were $y_{loc}=0.003, ~0.02$ ($y_{loc}^+=0.38,~2.56$). Although the difference between localization distances is nearly an order of one, the results of the control are the same for both Reynolds numbers (see the red and blue curves representing the controlled cases). This observation highlights the robustness of the control strategy concerning the distance from the wall in the viscous sub-layer confirming  its universality. 

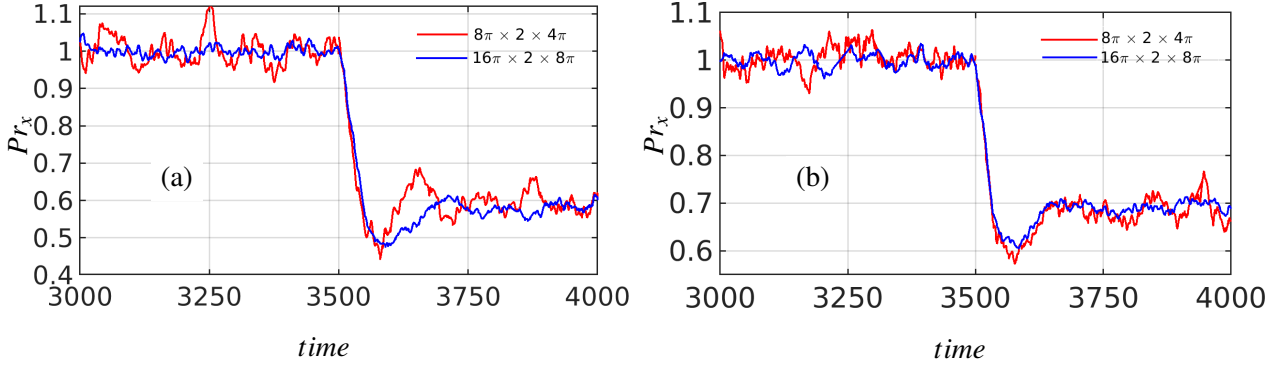
\begin{figure}[ht!]
\centering
\begin{minipage}[c]{.48\linewidth}
\FigureXYLabel{\includegraphics[width = 1.\textwidth]{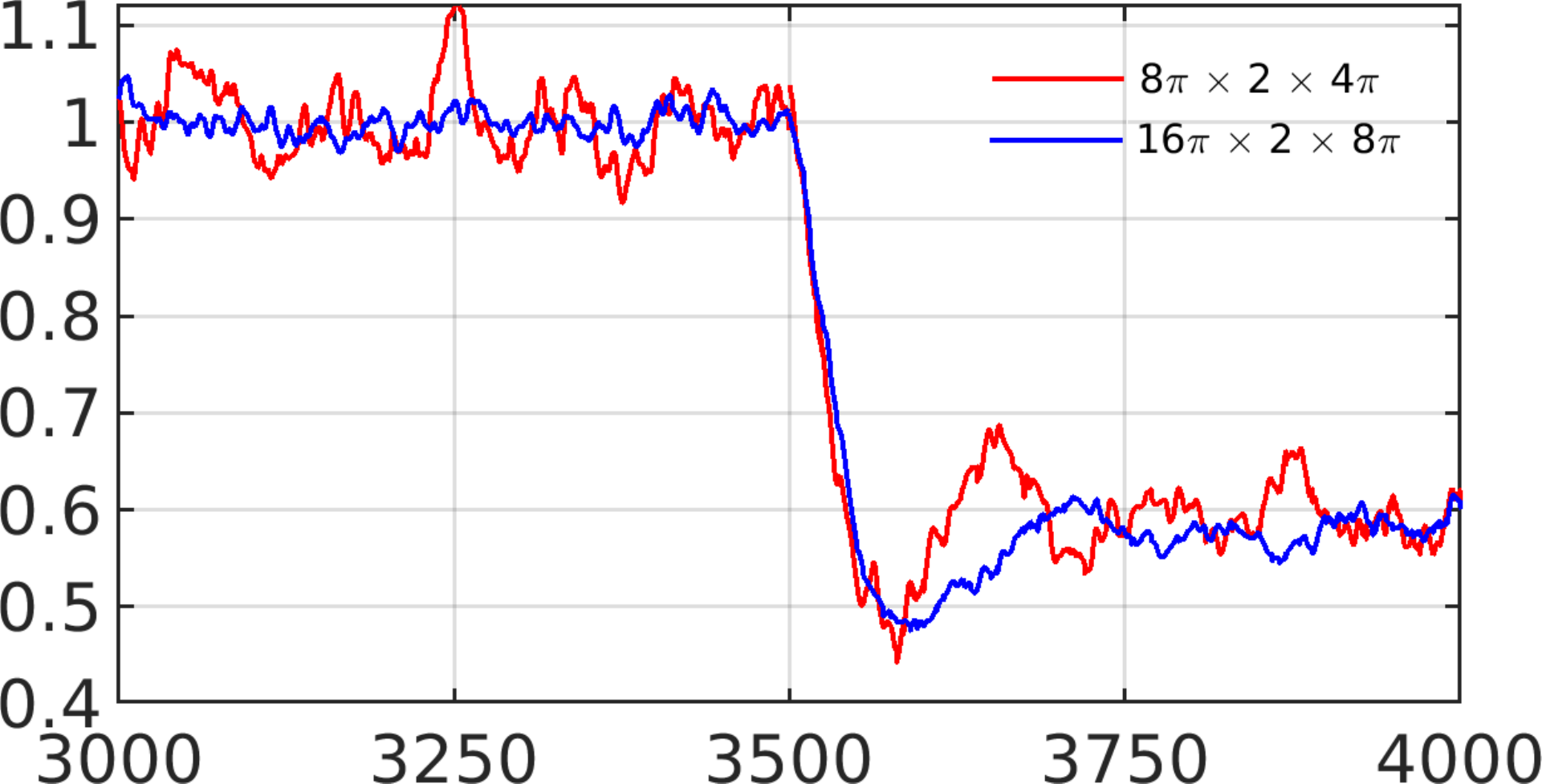}}
{${time}$}{-1mm}{\begin{rotate}{90} ${Pr_x}$ \end{rotate}}{1mm}
\end{minipage}
\begin{tikzpicture}[overlay]
\fill[fill=white]
(-6.,.1) node[fill=white] {(a)};
\end{tikzpicture} 
\hfill
\begin{minipage}[c]{.48\linewidth}
\FigureXYLabel{\includegraphics[width = 1.\textwidth]{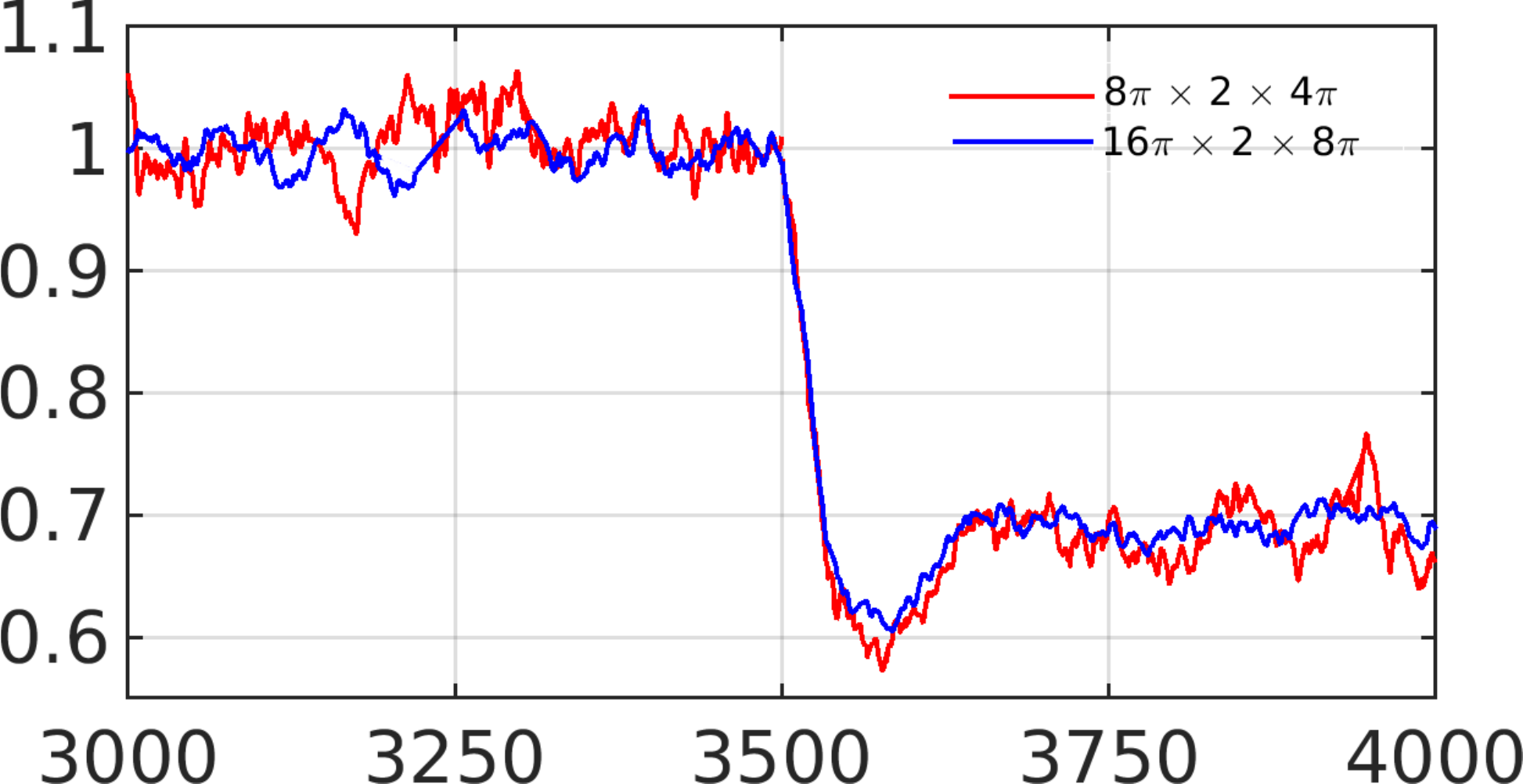}}
{${\small time}$}{-1mm}{\begin{rotate}{90} $Pr_x$ \end{rotate}}{1mm}
\end{minipage}
\begin{tikzpicture}[overlay]
\fill[fill=white]
(-6.,.1) node[fill=white] {(b)};
\end{tikzpicture}
\caption{Time evolution of TKE production for the uncontrolled and controlled turbulent flows at the same forcing localisation from the wall but for the different simulation box sizes.
The flows with $Re_{\tau}=92$ and $Re_{\tau}=128$ are presented on the top and bottom plots, correspondingly. The red and blue lines show the results for $8\pi \times 2 \times 4\pi$ and $16\pi \times 2 \times 8\pi$ boxes, respectively.}
\label{evol3}
\end{figure}

The time evolution of TKE production for the different simulation boxes, but with the same control parameters ($y_{loc}$, forcing amplitudes, centers, etc.) are presented in Figure \ref{evol3}. The cases of two Reynolds numbers, $Re_{\tau}=92, 128$ are presented. As one can see the control efficiency is the same for the both simulation boxes. 

\subsection{Production of turbulent kinetic energy: Space evolution}
\begin{figure}[ht!]
\centering
\begin{minipage}[c]{.32\linewidth}
\FigureXYLabel{\includegraphics[width = .9\textwidth]{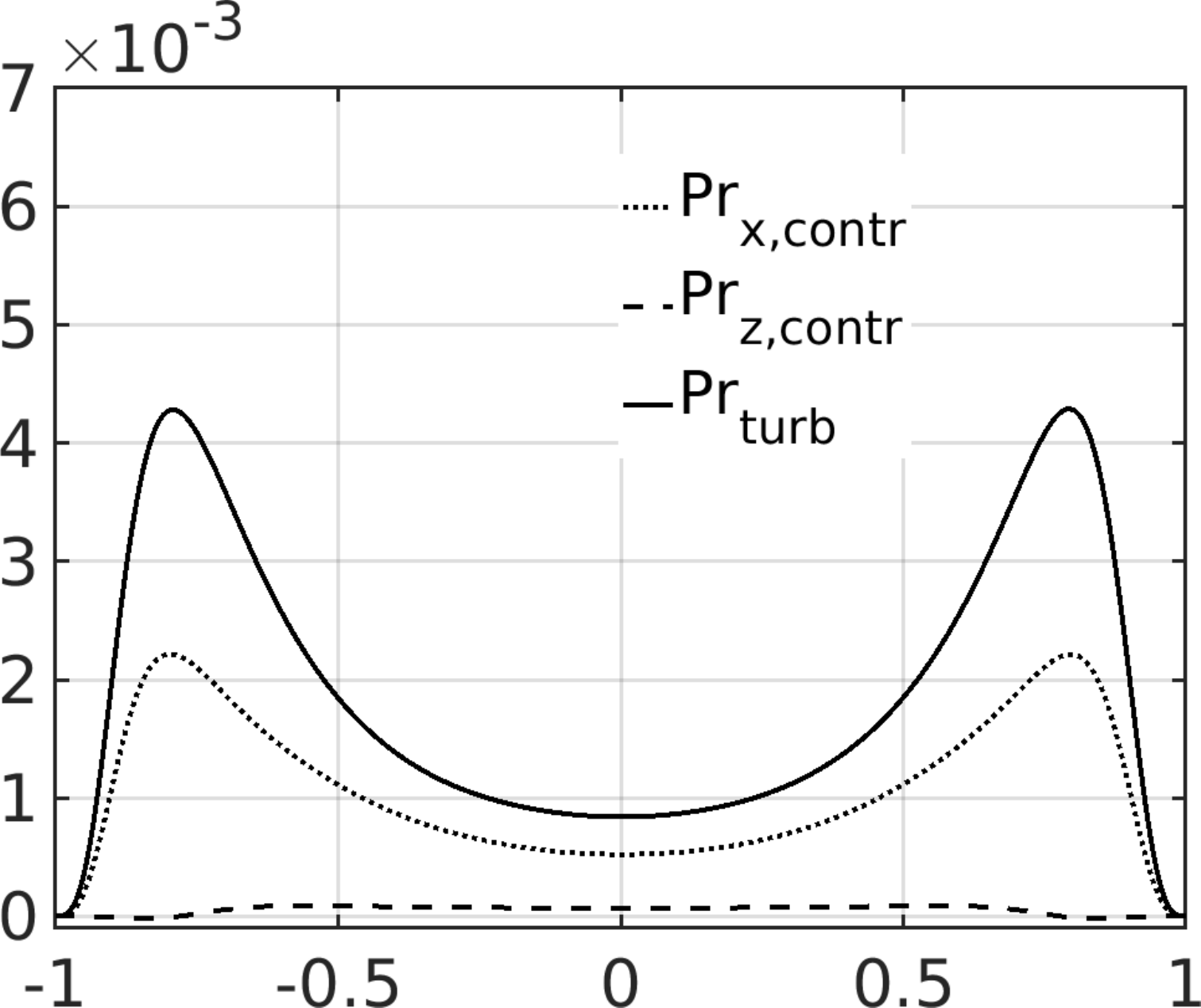}}
{${\small y}$}{-1mm}{\begin{rotate}{90} $Pr$ \end{rotate}}{1mm}
\end{minipage}
\begin{tikzpicture}[overlay]
\fill[fill=white]
(-4.2,1.8) node[fill=white] {(a)};
\end{tikzpicture} 
\begin{minipage}[c]{.32\linewidth}
\FigureXYLabel{\includegraphics[width = .9\textwidth]{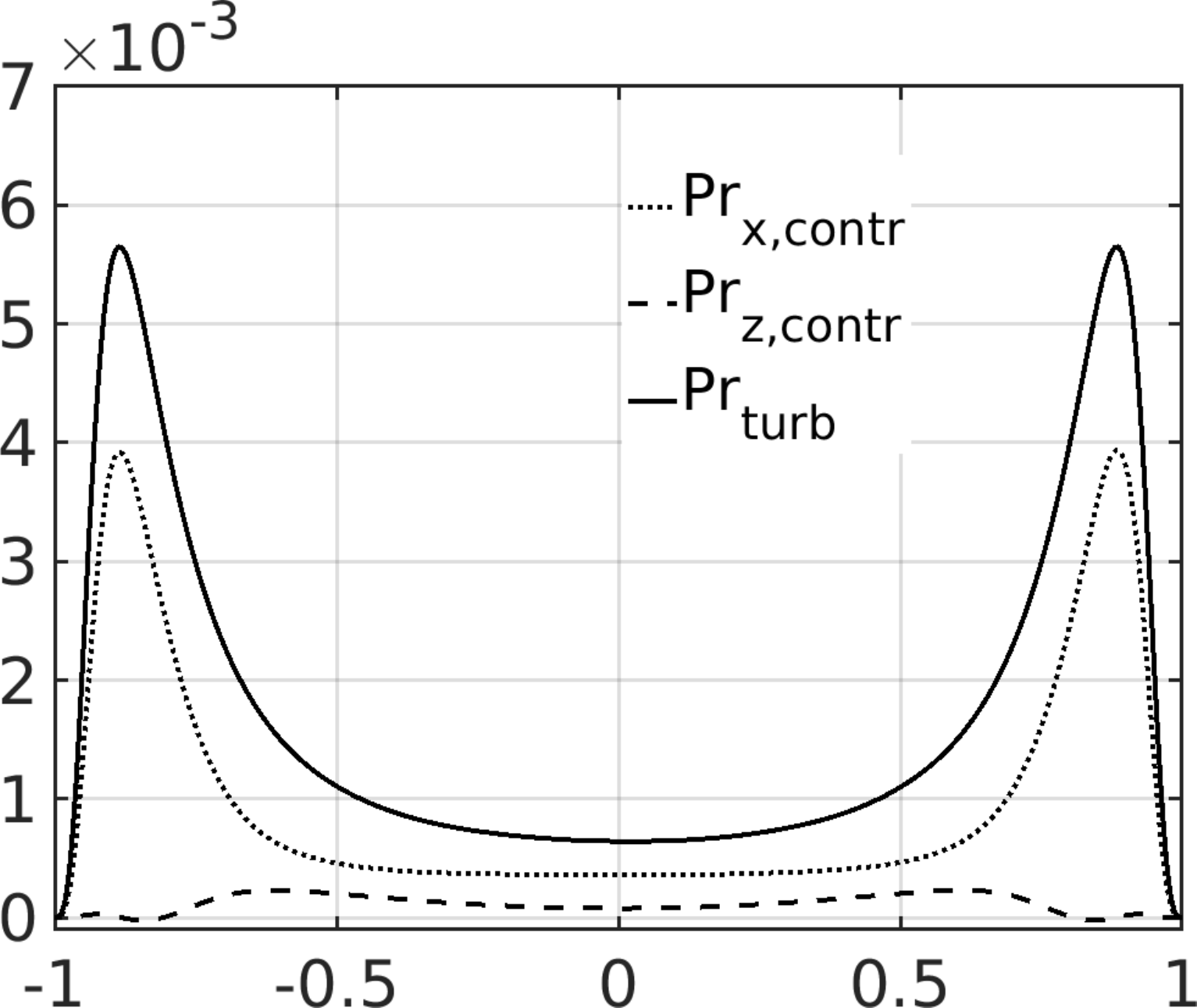}}
{${\small y}$}{-1mm}{\begin{rotate}{90} $Pr$ \end{rotate}}{1mm}
\end{minipage}
\begin{tikzpicture}[overlay]
\fill[fill=white]
(-4.2,1.8) node[fill=white] {(b)};
\end{tikzpicture} \\
\begin{minipage}[c]{.32\linewidth}
\FigureXYLabel{\includegraphics[width = .9\textwidth]{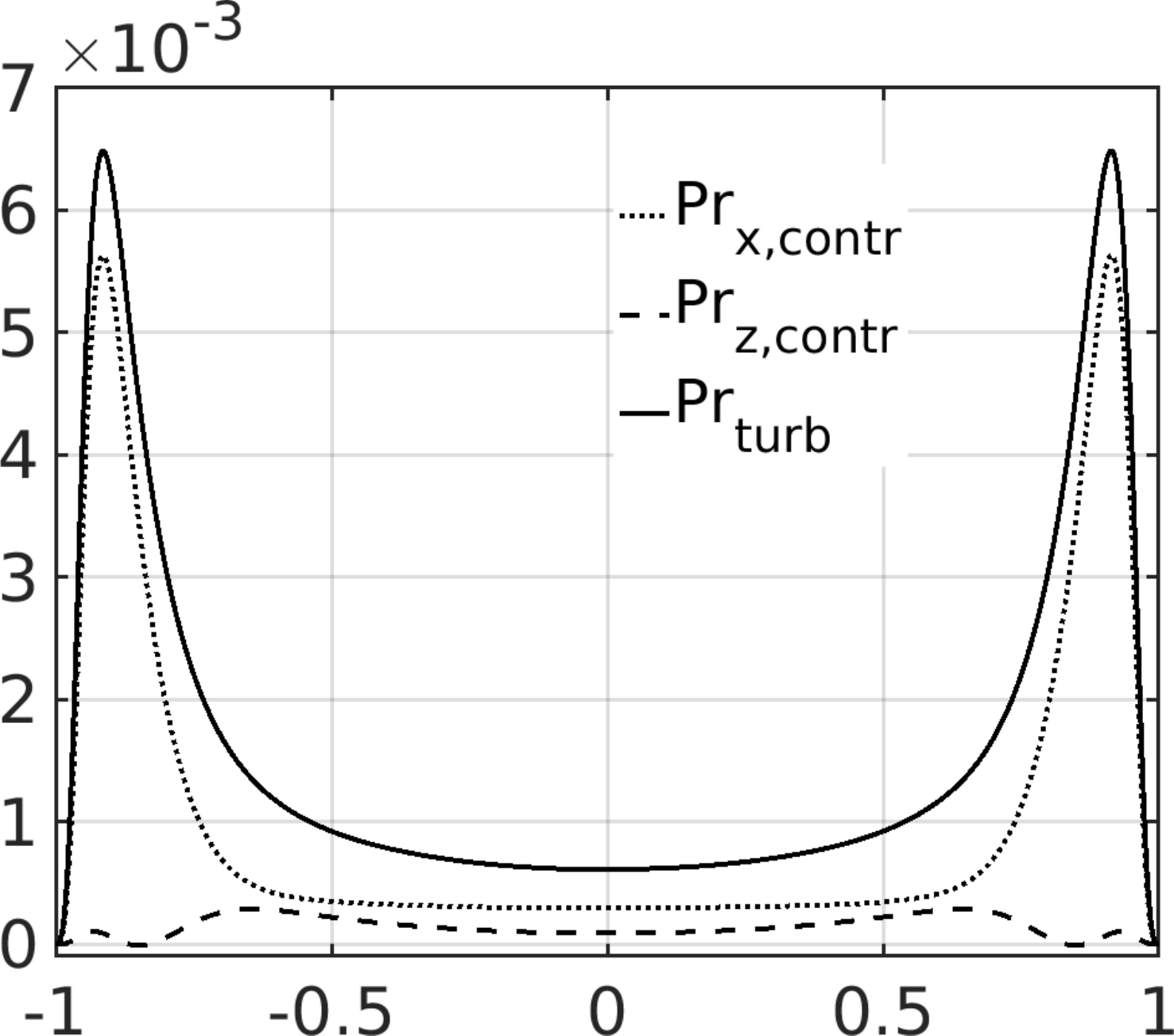}}
{${\small y}$}{-1mm}{\begin{rotate}{90} $Pr$ \end{rotate}}{1mm}
\end{minipage}
\begin{tikzpicture}[overlay]
\fill[fill=white]
(-4.2,1.8) node[fill=white] {(c)};
\end{tikzpicture}
\begin{minipage}[c]{.32\linewidth}
\FigureXYLabel{\includegraphics[width = .9\textwidth]{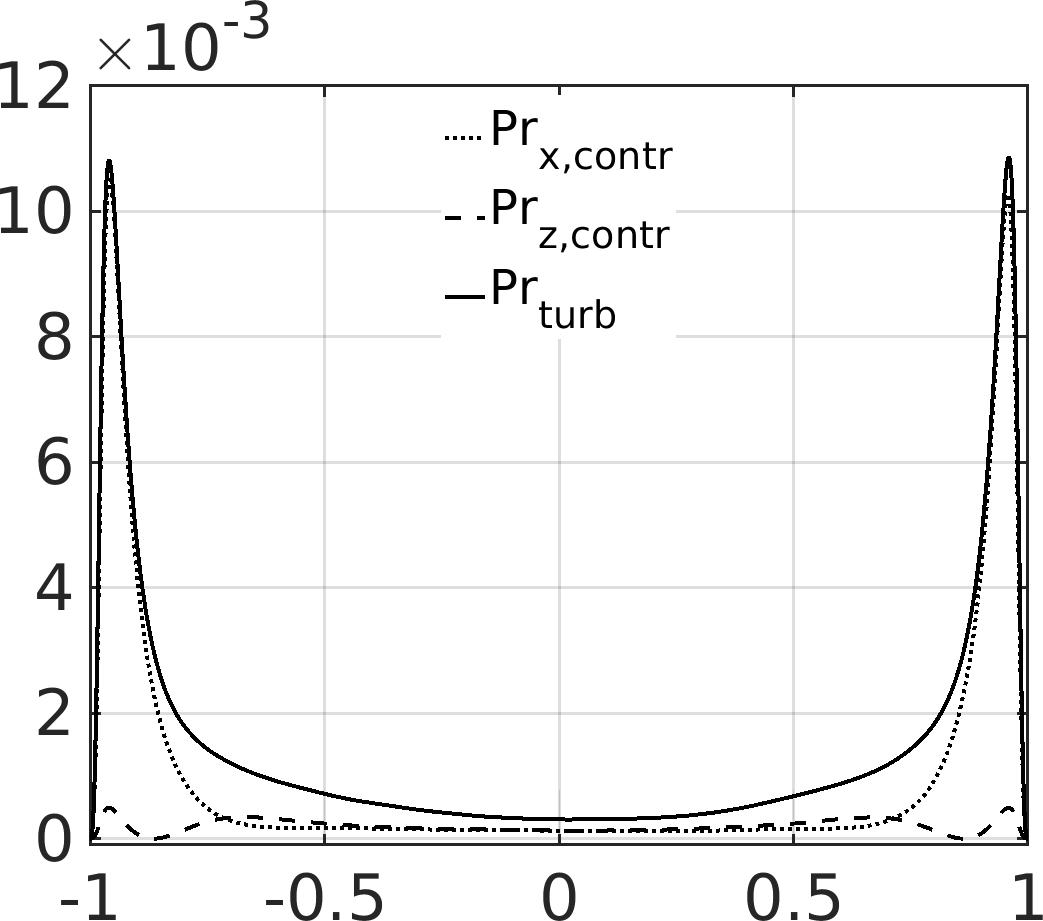}}
{${\small y}$}{-1mm}{\begin{rotate}{90} $Pr$ \end{rotate}}{1mm}
\end{minipage}
\begin{tikzpicture}[overlay]
\fill[fill=white]
(-4.,1.8) node[fill=white] {(d)};
\end{tikzpicture}
\caption{TKE production of the  uncontrolled and controlled turbulent flows. Plots (a), (b) and (c) correspond to Reynolds numbers, $Re=750, 1500, 2150$. Streamwise ($Pr_{x,contr}$) and spanwise ($Pr_{z,contr}$) components of productions of the controlled flow are presented with dotted and dashed lines. Solid lines show the production of the uncontrolled case.}
\label{Stat_Prods}
\end{figure}
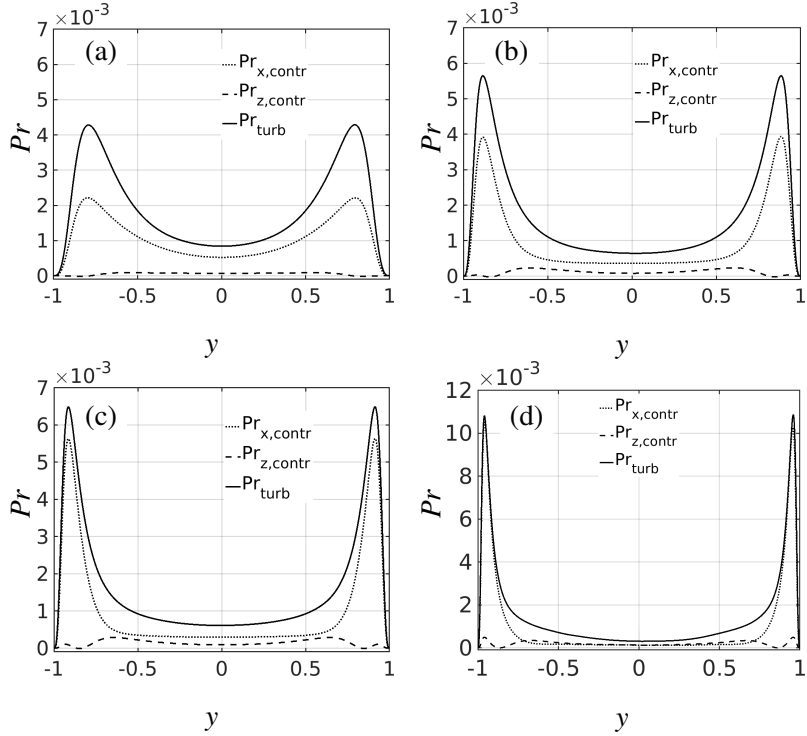
In Figure \ref{Stat_Prods} the productions of uncontrolled/natural and controlled turbulent flows (averaged in time and in space in parallel directions) are shown for Reynolds numbers $Re=750, 1500, 2150$ from left to right. Streamwise and spanwise components of productions of TKE are defined as follows:
\begin{equation}
    Pr_x = -\overline{u^\prime_xu^\prime_y}\frac{d\overline U_x}{dy}, ~~~
 Pr_z = -\overline{u^\prime_yu^\prime_z}\frac{d\overline U_z}{dy}.
\end{equation}
$Pr_z$ is non-zero only for the controlled turbulent flow. 
The solid lines correspond to the uncontrolled turbulent flows, while black dotted and dashed ones to the streamwise and spanwise productions, $Pr_{x,contr}$ and $Pr_{z,contr}$, of the controlled cases. The $Pr_z$ for uncontrolled turbulent flow is zero and not shown here. Streamwise components TKE production are reduced in the case of controlled turbulent flows for all Reynolds number in the whole channel. Spanwise components of production for the control turbulent flow are non-zero (dashed lines) but much smaller than streamwise ones. 

\subsection{Two-point correlations in streamwise and spanwise directions}
\begin{figure}[ht!]
\centering
\begin{minipage}[c]{.36\linewidth}
\FigureXYLabel{\includegraphics[width = 1.\textwidth]{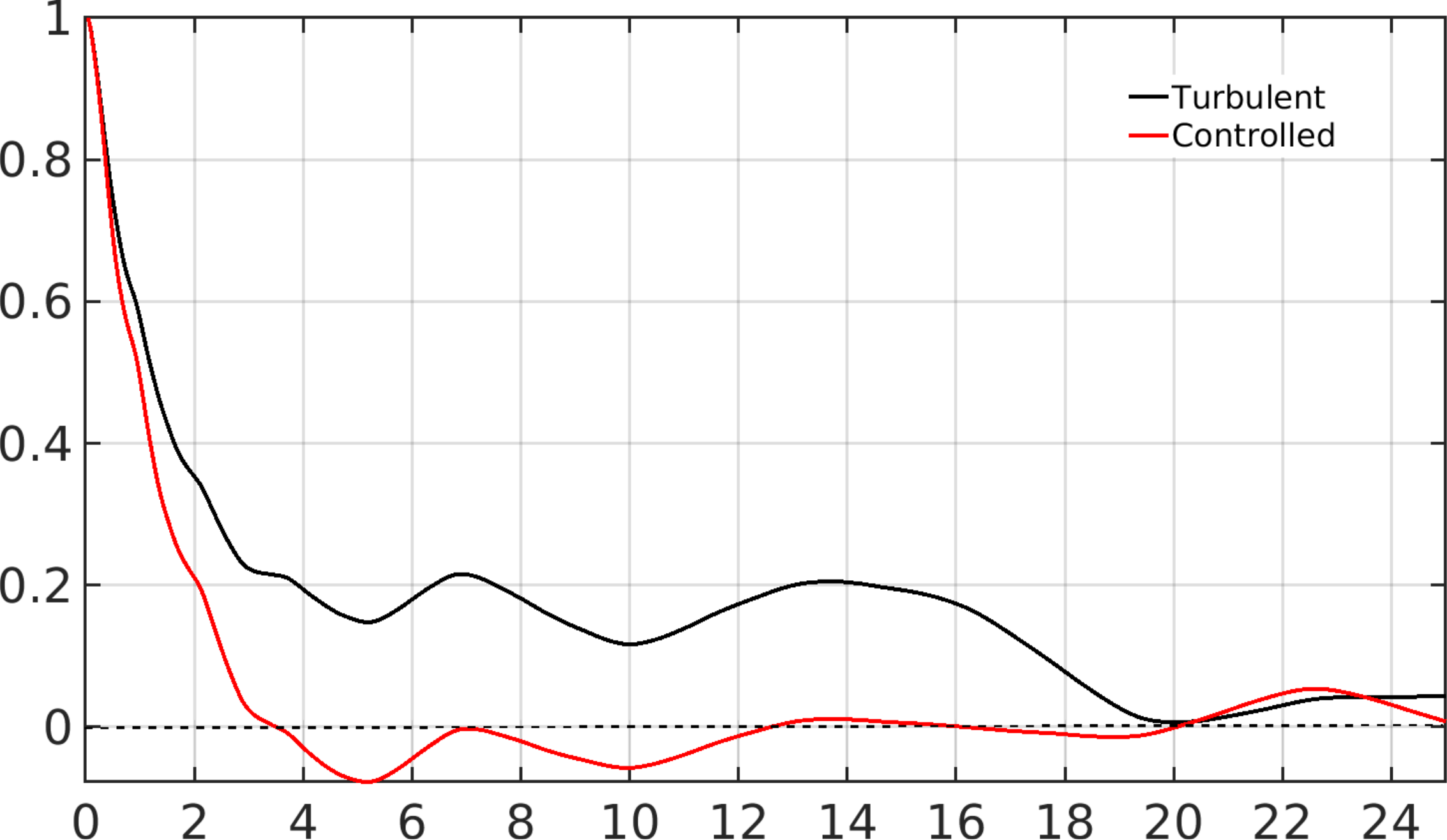}}
{${\small \Delta x}$}{-2mm}{\begin{rotate}{90} ${\small R_{uu}(\Delta x)}$ \end{rotate}}{1mm}
\end{minipage}
\begin{tikzpicture}[overlay]
\fill[fill=white]
(-4.7,1.5) node[fill=white] {(a)};
\end{tikzpicture}
\hspace{0.3cm}
\begin{minipage}[c]{.37\linewidth}
\FigureXYLabel{\includegraphics[width = 1.\textwidth]{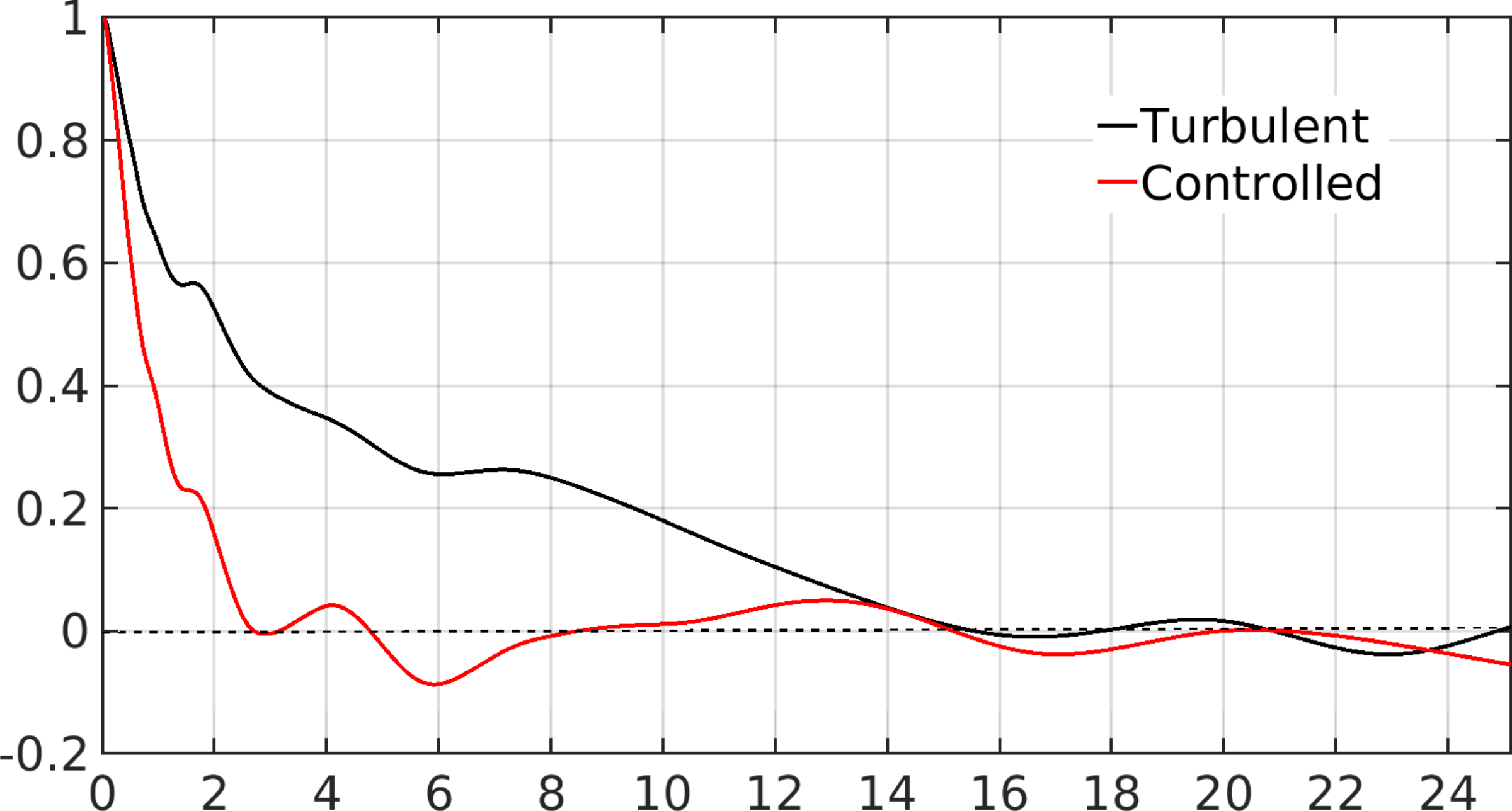}}
{${\small \Delta x}$}{-2mm}{\begin{rotate}{90} ${\small R_{uu}(\Delta x)}$ \end{rotate}}{1mm}
\end{minipage}
\begin{tikzpicture}[overlay]
\fill[fill=white]
(-4.7,1.45) node[fill=white] {(b)};
\end{tikzpicture}\\
\begin{minipage}[c]{.37\linewidth}
\FigureXYLabel{\includegraphics[width = 1.\textwidth]{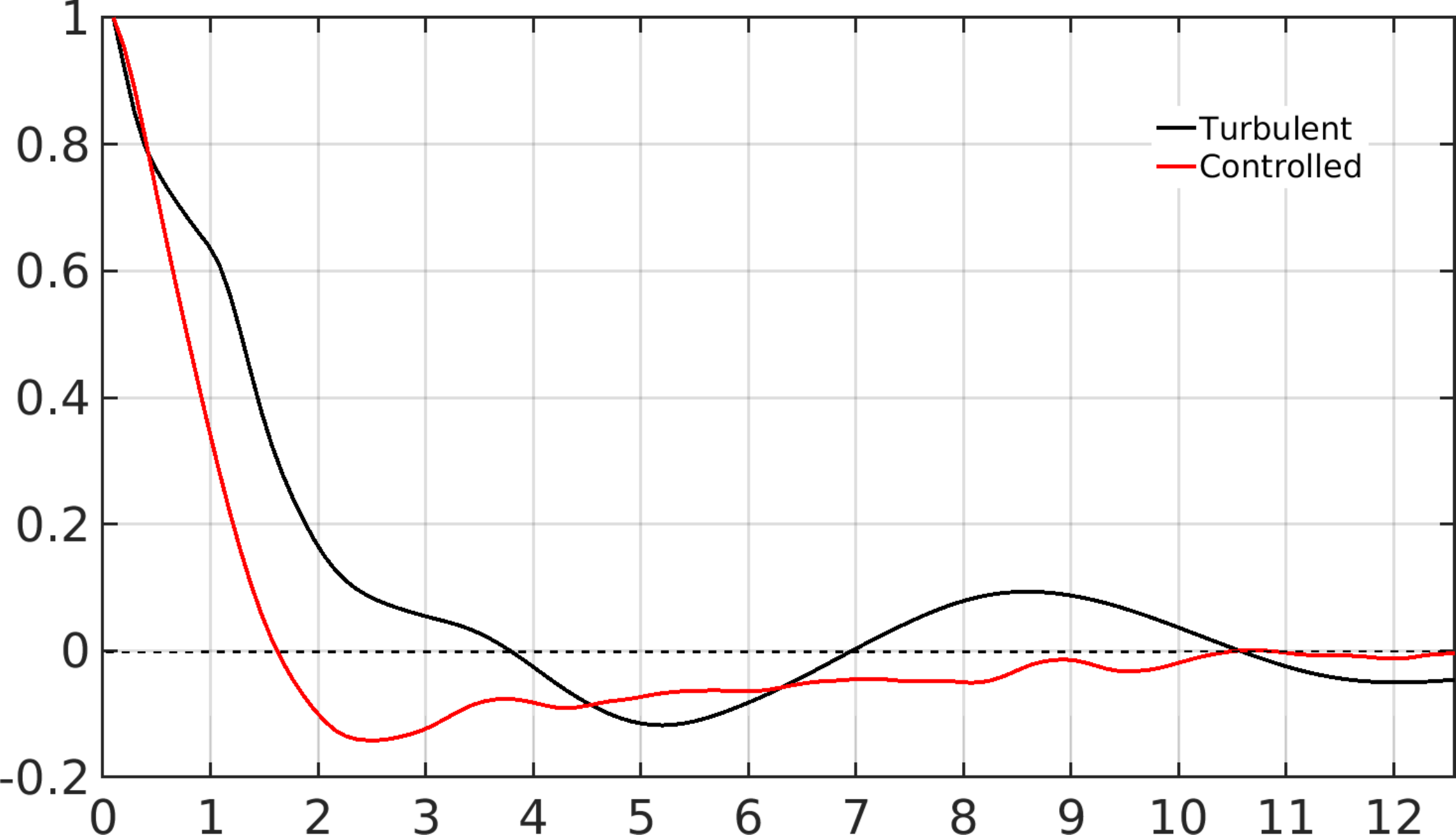}}
{${\small \Delta z}$}{-2mm}{\begin{rotate}{90} ${\small R_{uu}(\Delta z)}$ \end{rotate}}{1mm}
\end{minipage}
\begin{tikzpicture}[overlay]
\fill[fill=white]
(-4.7,1.45) node[fill=white] {(c)};
\end{tikzpicture}
\hspace{0.3cm}
\begin{minipage}[c]{.37\linewidth}
\FigureXYLabel{\includegraphics[width = 1.\textwidth]{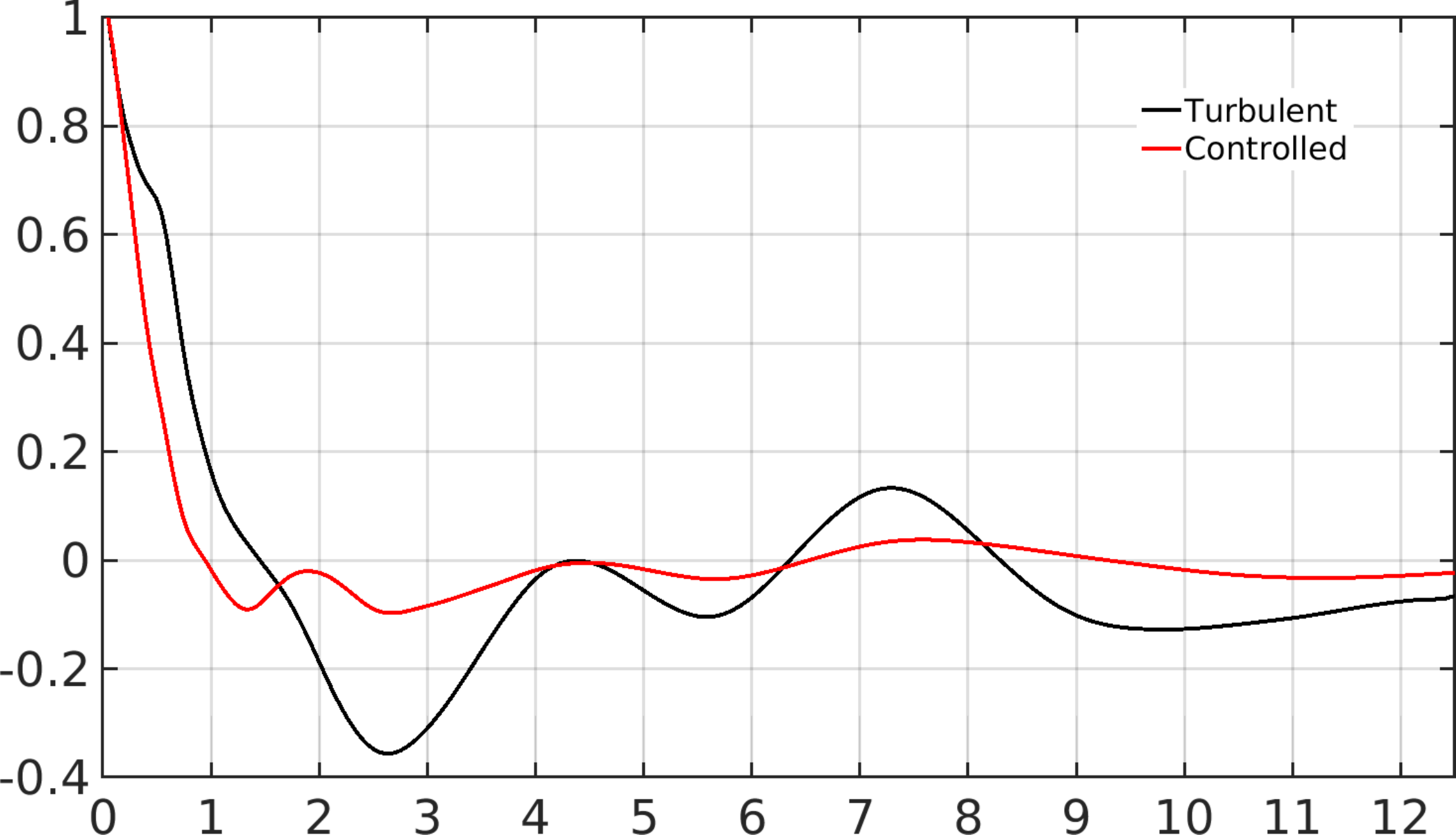}}
{${\small \Delta z}$}{-2mm}{\begin{rotate}{90} ${\small R_{uu}(\Delta z)}$ \end{rotate}}{1mm}
\end{minipage}
\begin{tikzpicture}[overlay]
\fill[fill=white]
(-4.7,1.45) node[fill=white] {(d)};
\end{tikzpicture}
\caption{The streamwise TPC coefficients ($R_{uu}( \Delta x)$) for the uncontrolled (black line) and controlled (red line) turbulent flows at $Re = 1500$ (a) and $Re = 2150$ (b) at the channel centerline; The spanwise TPC coefficients ($R_{uu}( \Delta z)$) for the uncontrolled (black line) and controlled (red line) turbulent flows at $Re = 1500$ (c) and $Re = 2150$ (d) at the channel centerline.  TPCs are shown for the half-domain of the simulation box $16\pi \times 2 \times 8\pi$; }
\label{R11_Re1500_Re2150}
\end{figure}
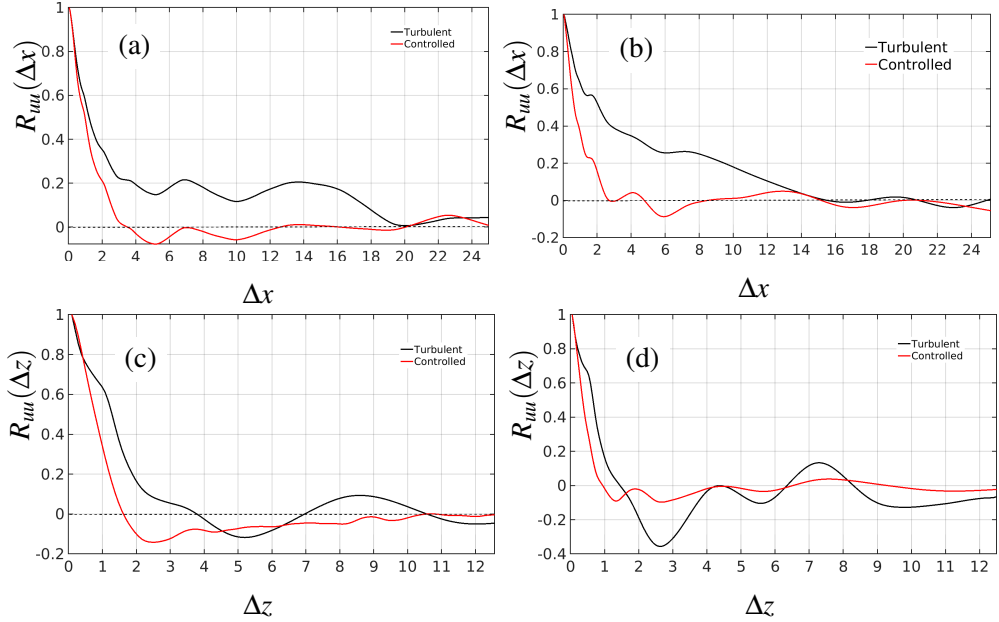

In the figure \ref{R11_Re1500_Re2150} the streamwise and spanwise two-point correlation (TPC) coefficients are presented for the uncontrolled (black line) and controlled turbulent (red line) cases for $Re=1500, 2150$ on the plots (a) and (b). As usual, TPC are plotted in the half of the simulation box at the centerline of the channel. The correlation coefficient is defined as 
\begin{equation} 
R_{uu}(\Delta x_i) = \frac{\overline{u^\prime (x_i)u^\prime (x_i+\Delta x_i)}}{\overline{u^\prime (x_i)u^\prime (x_i)}}, ~~~ i = (x,y,z)
\end{equation}
and normalized on rms of velocity. It is well-known that to avoid the unnatural confinement effects on the flow due to the streamwise and spanwise periodicity of the simulation box, it is necessary to have correlation coefficients decaying to zero in the half of the domain \citep{Tsukahara2006,Pirozzoli2014}. The spanwise TPC coefficients, $R_{uu}( \Delta z)$ are shown on the plots (c) and (d). As one can see, TPC coefficient for the uncontrolled case slowly decreases to zero in the middle of the box that means that its length is long enough to capture the random evolution of large scale structures within the computational box. As for the controlled case, TPC coefficient (red lines) decreases much faster to zero than in the controlled one. 

Streamwise and spanwise integral length scales are defined as 
\begin{equation}
\label{length_scales}
\Lambda_{uux} = \int_0^\infty R_{uu}(\Delta x)d\Delta x; ~~~ \Lambda_{uuz} = \int_0^\infty R_{uu}(\Delta z)d\Delta z.
\end{equation}
The following values of the streamwise length scales for the uncontrolled and controlled turbulent flows at $Re=1500$ can be obtained $\Lambda_{uux} =4.0,~ 0.6$, respectively. 
For the higher Reynolds number $Re=2150$, the values of the length scales are $ \Lambda_{uux} =3.5,~ 0.67$.
As we see, in the controlled cases the streamwise length scales are approximately by order of one smaller than in the controlled ones. 
As for the spanwise length scales, we obtained the following values for uncontrolled and controlled turbulent flows: At $Re=1500$ and $Re=2150$ cases we have $\Lambda_{uuz} =0.28,~ 0.14$ and $\Lambda_{uuz} =0.21,~ 0.08$, correspondingly. The length scales in spanwise direction decreased twice because of the flow control.

\begin{figure}[ht!]
\centering
\begin{minipage}[c]{.45\linewidth}
\includegraphics[width = .9\textwidth]{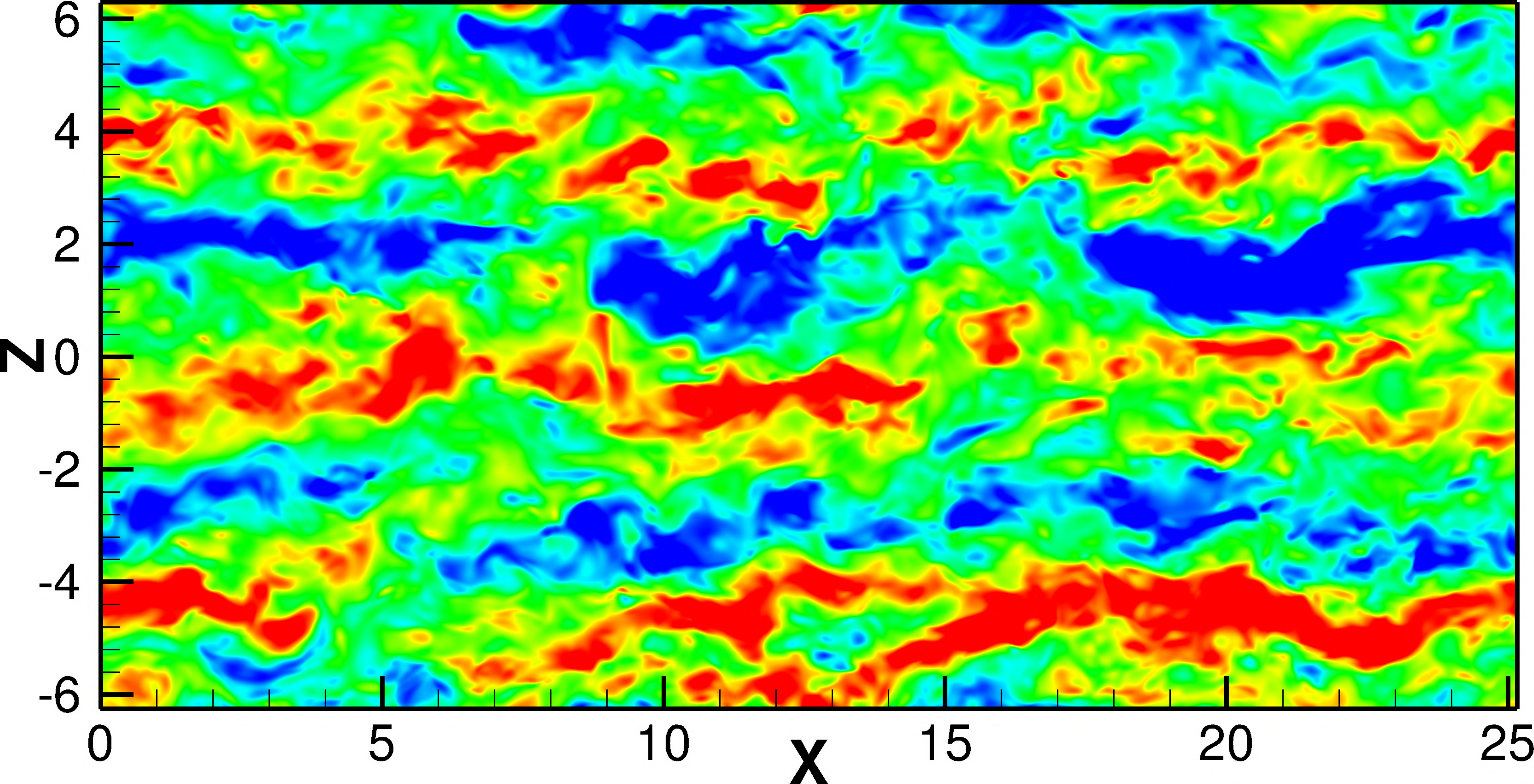}
\end{minipage}
\hspace{0.1cm}
\begin{minipage}[c]{.45\linewidth}
\includegraphics[width = .9\textwidth]{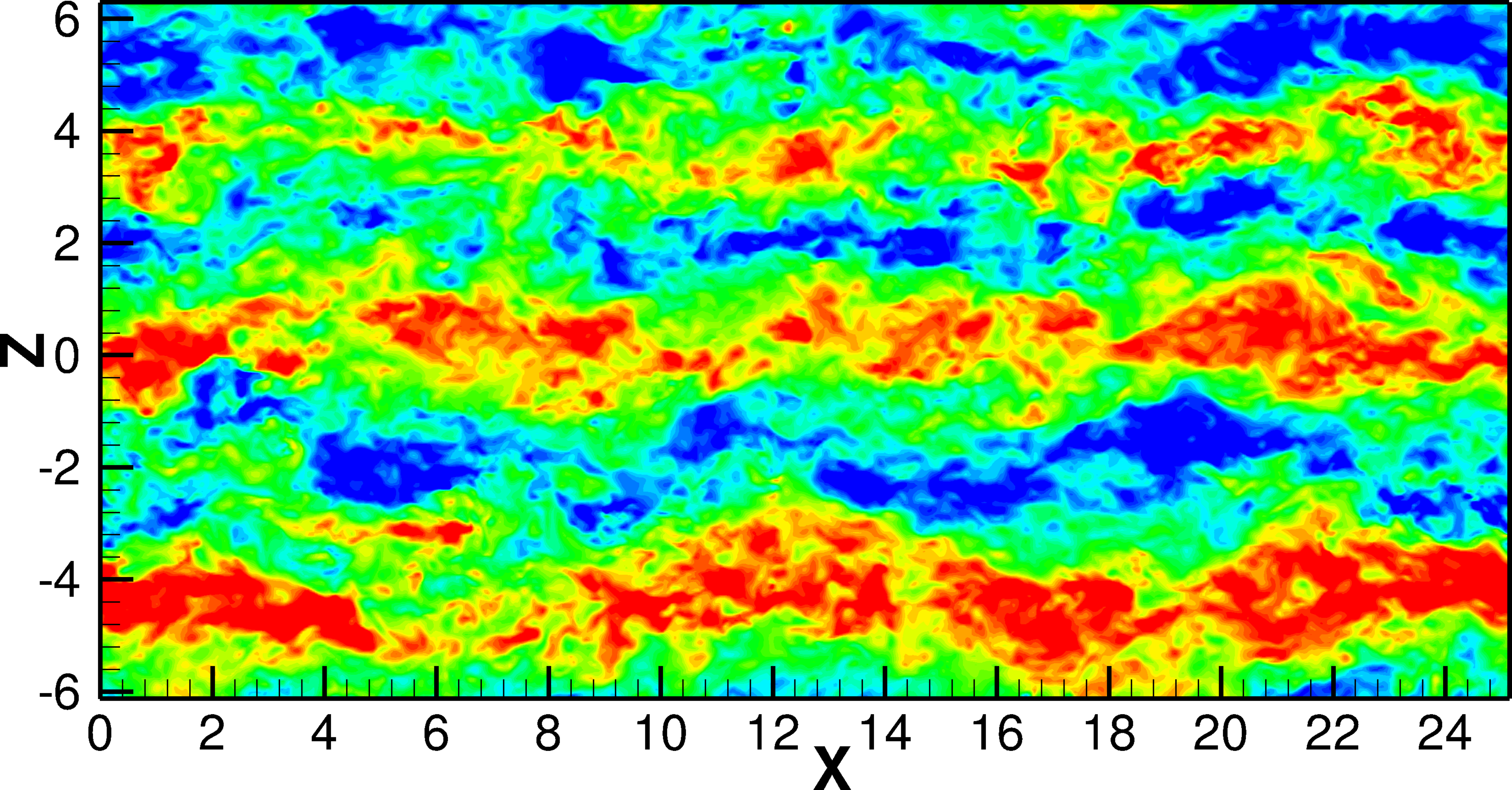}
\end{minipage}\\
\vspace{0.2cm}
\begin{minipage}[c]{.45\linewidth}
\includegraphics[width = .9\textwidth]{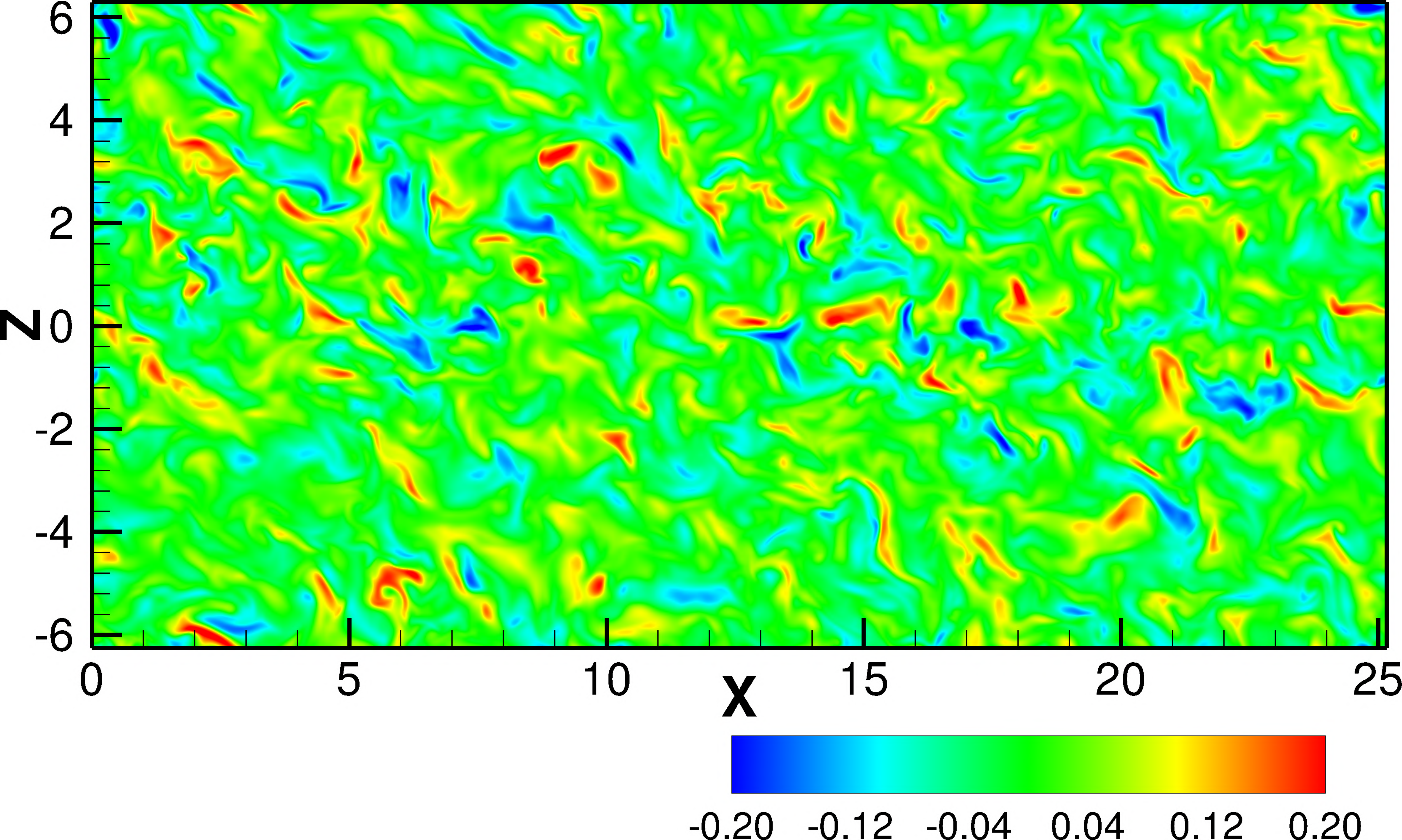}
\end{minipage}
\hspace{0.1cm}
\begin{minipage}[c]{.45\linewidth}
\includegraphics[width = .9\textwidth]{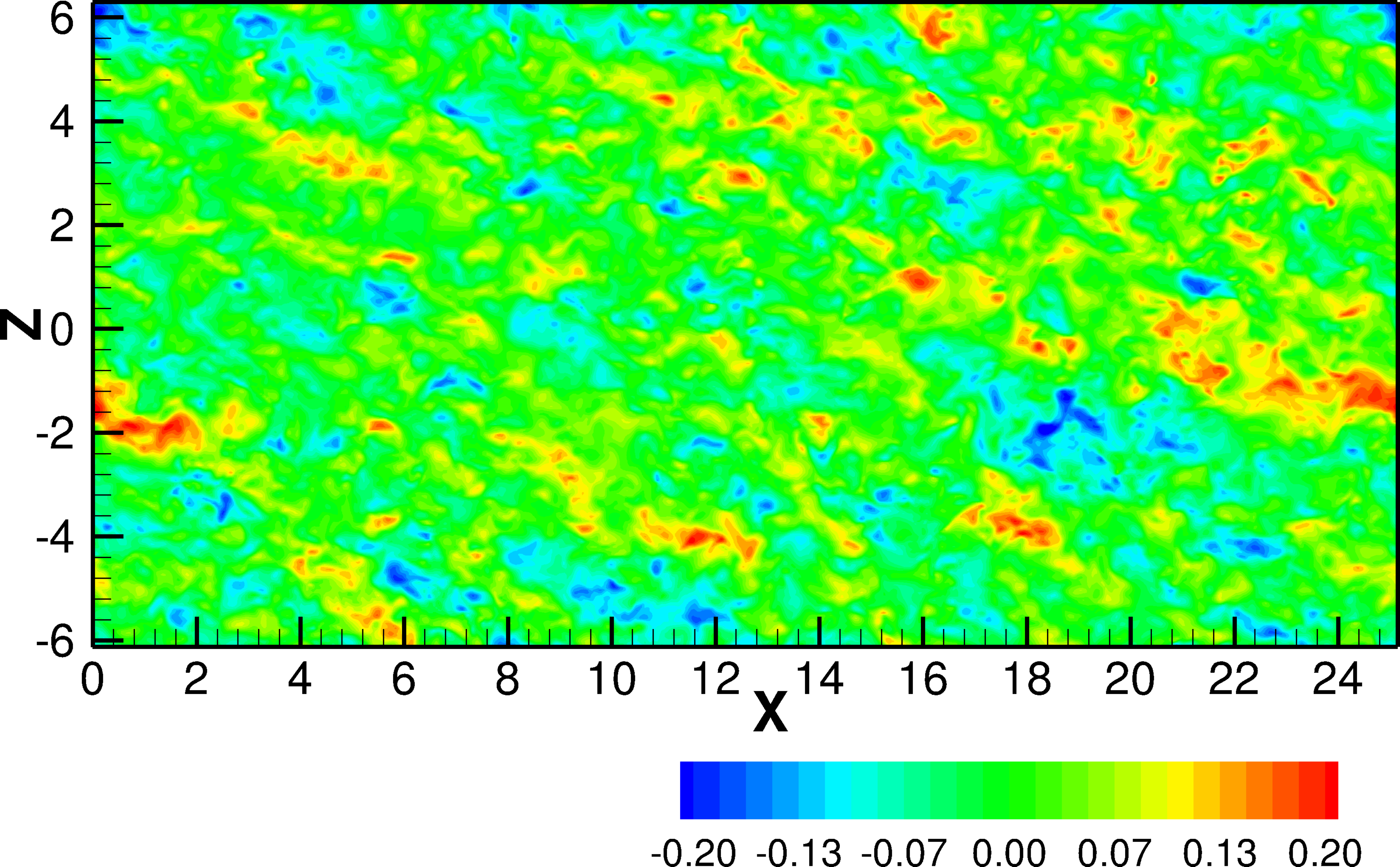}
\end{minipage}
\caption{$XZ$-slices of instantaneous streamwise velocity fluctuation fields for uncontrolled (top row) and controlled (bottom row) turbulent cases at $Re = 2150, 5000$ (left and right columns) in the core of the channel, $y=0$, are presented.}
\label{XZ_slices}
\end{figure}

In Figure \ref{XZ_slices} the $XZ$-slices of the instantaneous streamwise velocity fluctuations are presented in the center of the flow, for the  uncontrolled (top plots) and controlled (bottom plots) turbulent flows at $Re = 2150$ (left column) and $Re = 5000$ (right column). It is well-known that the large-scale eddies form in the core region of the turbulent flow \citep{Tsukahara2006,Pirozzoli2014}.
The completely different picture of streak scales and directions is observed in the case of the controlled flow. The streaks are weakened and have much smaller length scales that corresponds to the calculate length scales. It has to be mentioned that for the controlled uncontrolled flow, a weak spanwise asymmetry is observed. This fact will be discussed in detail in the next section.

\begin{figure}[ht!]
\centering
\begin{minipage}[c]{.48\linewidth}
\includegraphics[width = 1.\textwidth]{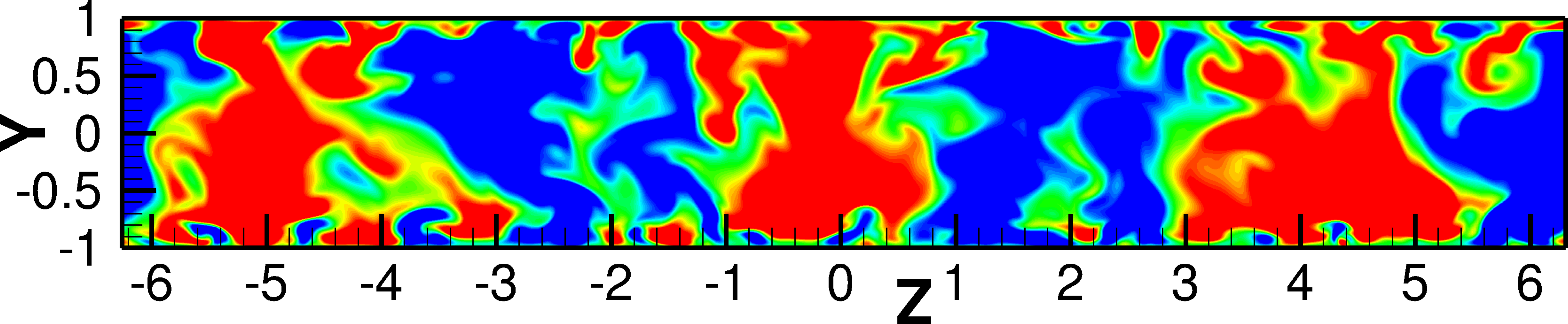}
\end{minipage}
\hfill
\begin{minipage}[c]{.48\linewidth}
\includegraphics[width = 1.\textwidth]{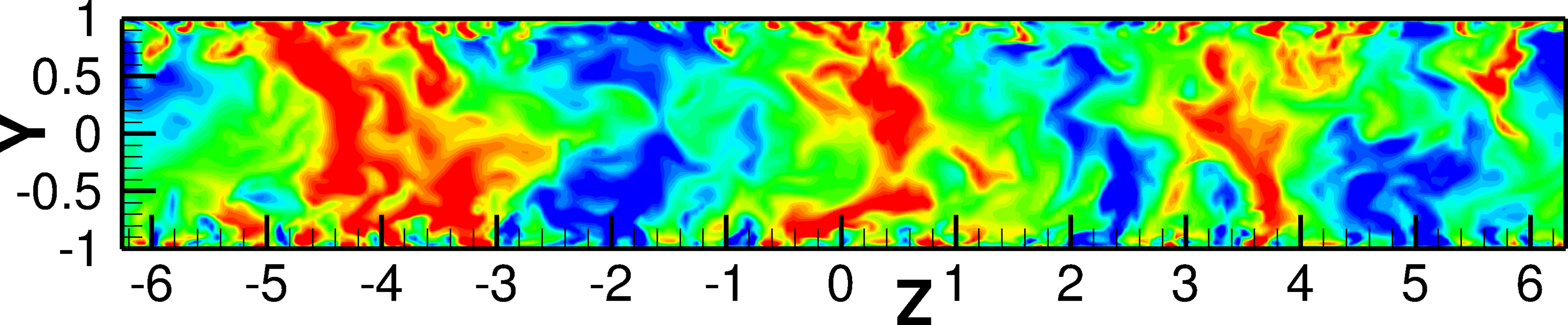}
\end{minipage} \\
\begin{minipage}[c]{.48\linewidth}
\includegraphics[width = 1.\textwidth]{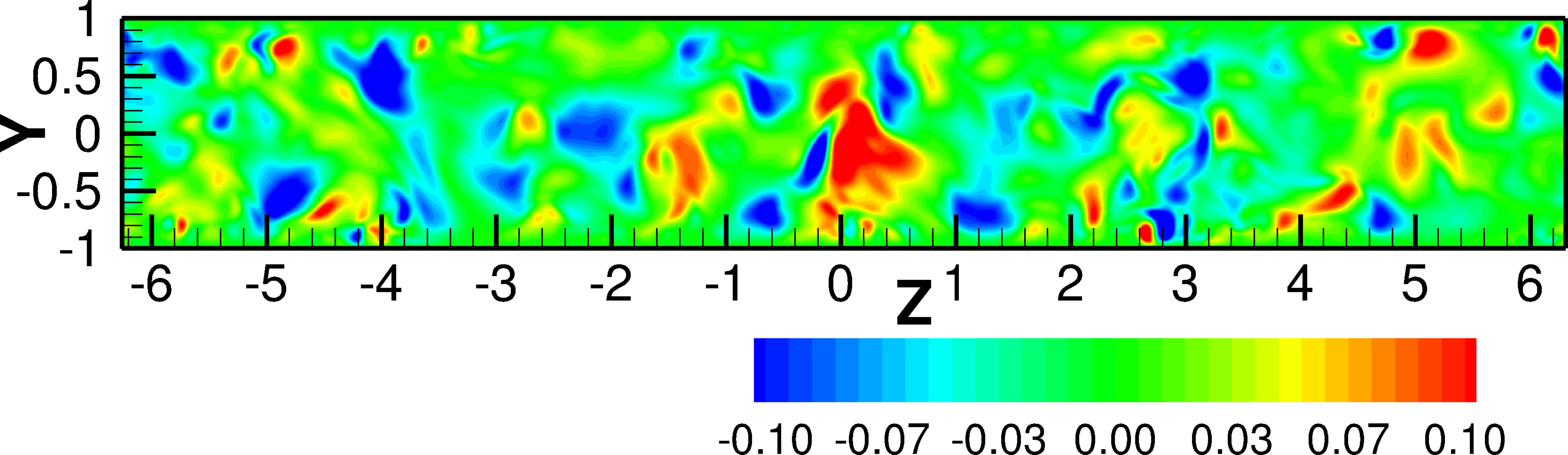}
\end{minipage}
\hfill
\begin{minipage}[c]{.48\linewidth}
\includegraphics[width = 1.\textwidth]{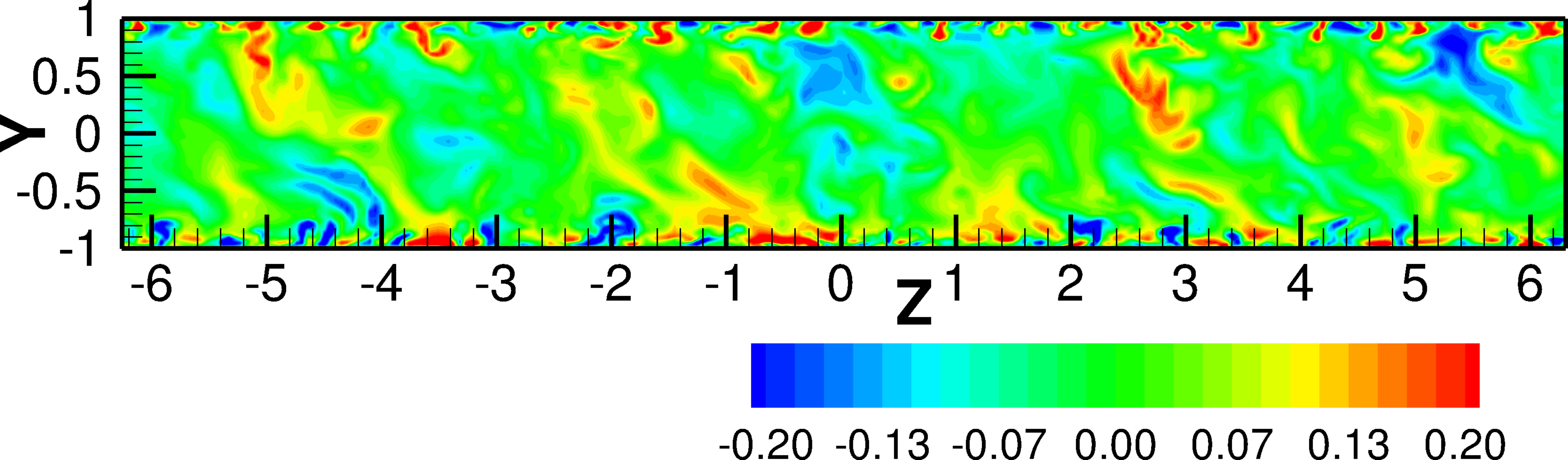}
\end{minipage}
\caption{$ZY$-slices of instantaneous streamwise velocity fluctuation fields for turbulent (top plots) and controlled (bottom plots) cases for $Re = 2150, 5000$ at $x = 12$, are presented.}
\label{ZY_slices}
\end{figure}

In the figure \ref{ZY_slices} $ZY$-slices of instantaneous streamwise velocity fluctuation fields are presented for the uncontrolled and controlled turbulent flows, on the top and bottom plots at $Re = 2150$ and $Re = 5000$ (left and right columns). As it is seen from the figure, in the uncontrolled cases the eddies are spreading from one wall to another. As for the controlled ones, this coherency is destroyed. In the spanwise direction, in the uncontrolled case the strong streamwise rolls (red and blue) are clearly visible, that is not a case in the controlled flow.  

\subsection{Reynolds stresses}
\label{stresses}
\begin{figure*}[ht!]
\begin{minipage}[c]{.24\linewidth}
\includegraphics[width = 1.\textwidth]{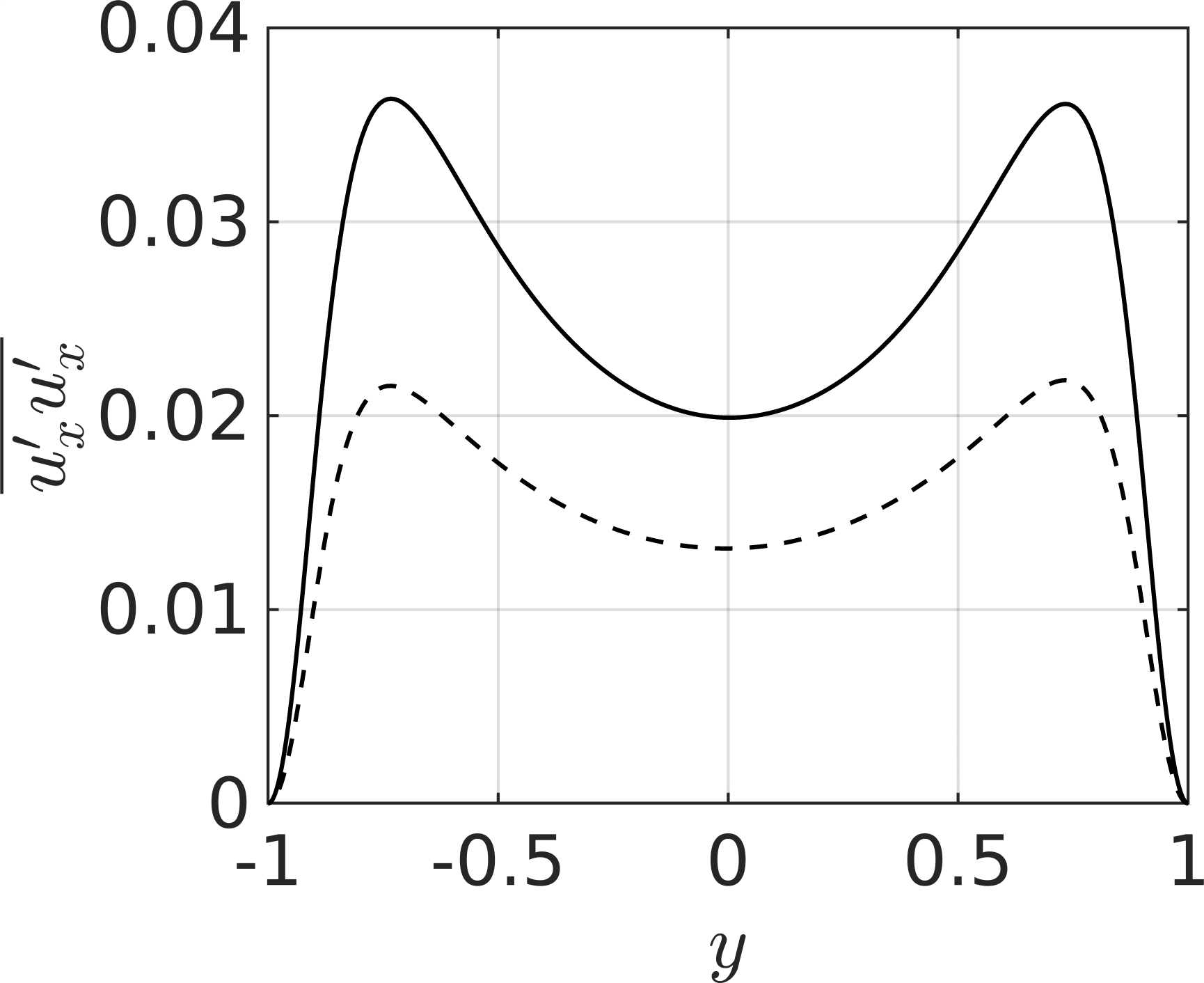}
\end{minipage}
\hfill
\begin{minipage}[c]{.24\linewidth}
\includegraphics[width = 1.\textwidth]{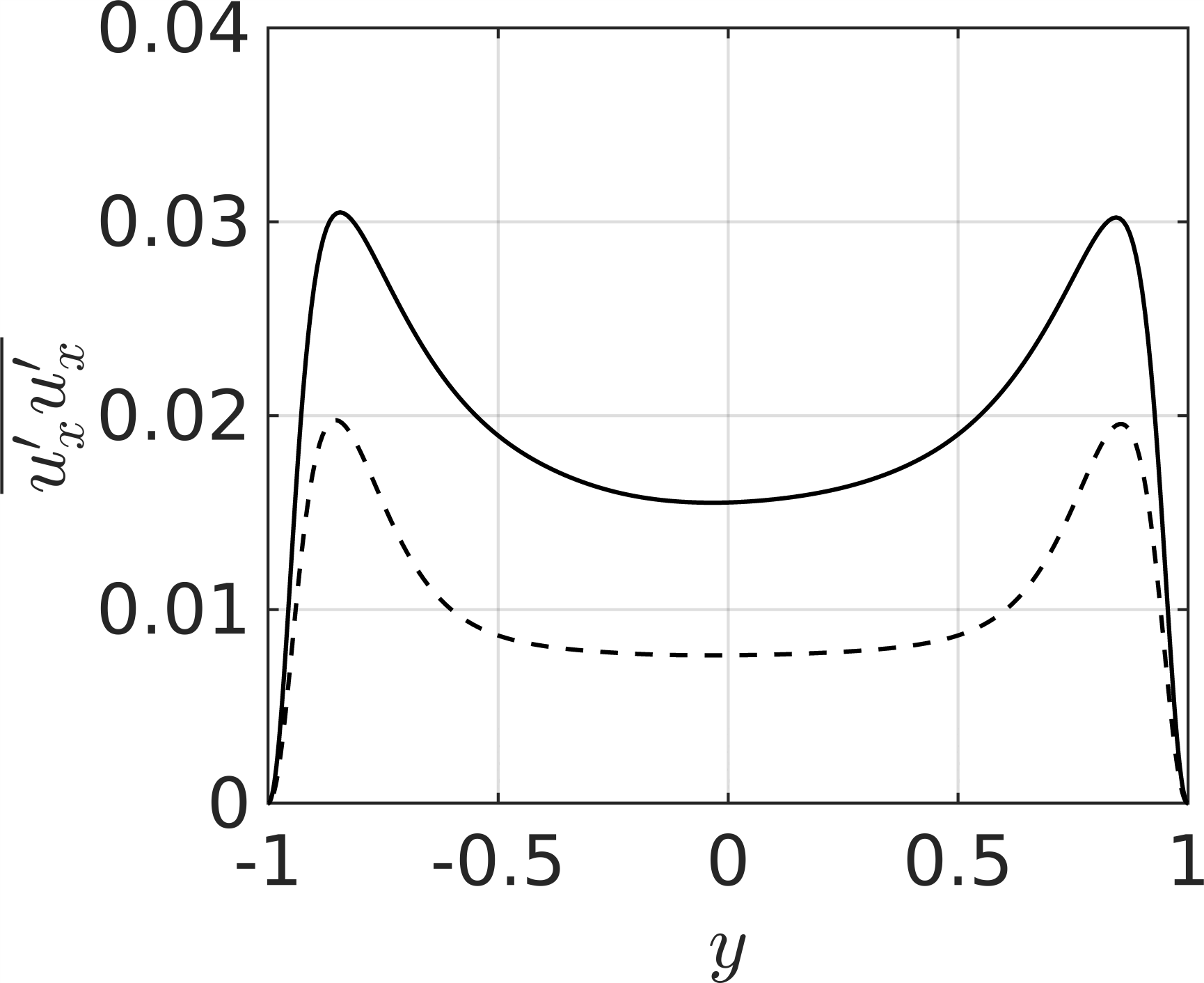}
\end{minipage}
\hfill
\begin{minipage}[c]{.24\linewidth}
\includegraphics[width = 1.\textwidth]{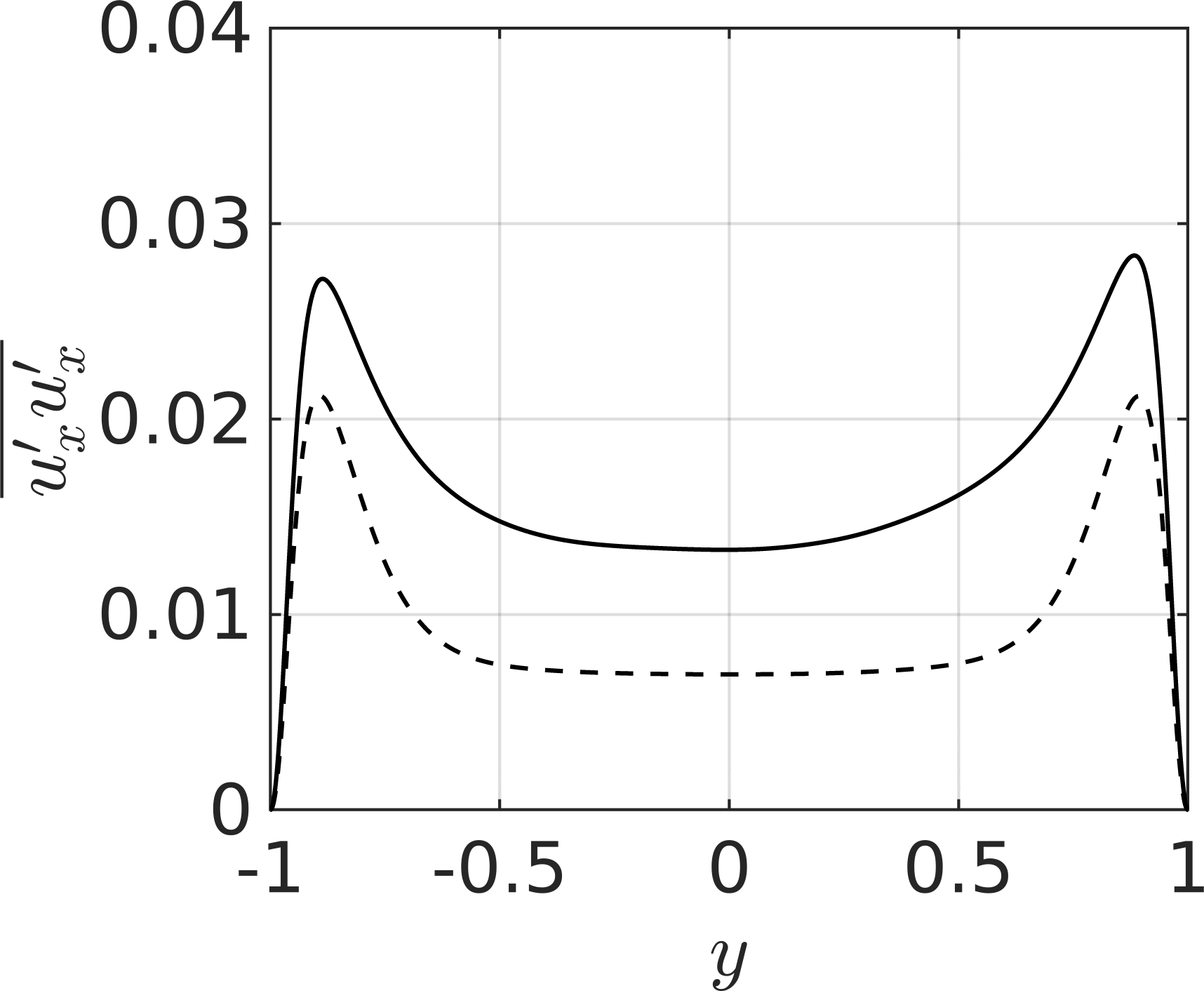}
\end{minipage}
\hfill
\begin{minipage}[c]{.25\linewidth}
\includegraphics[width = 1.\textwidth]{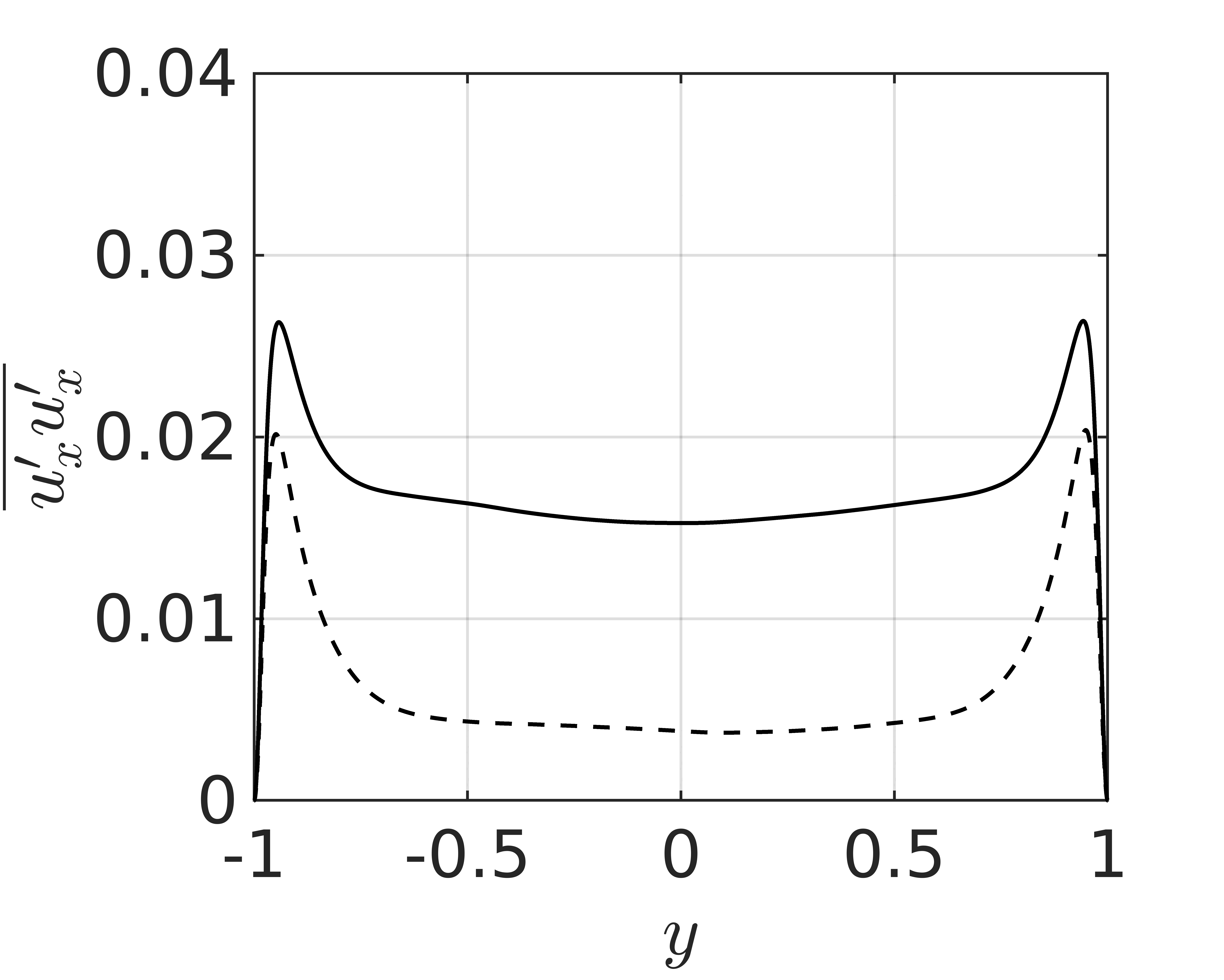}
\end{minipage}\\
\begin{minipage}[c]{.24\linewidth}
\includegraphics[width = 1.\textwidth]{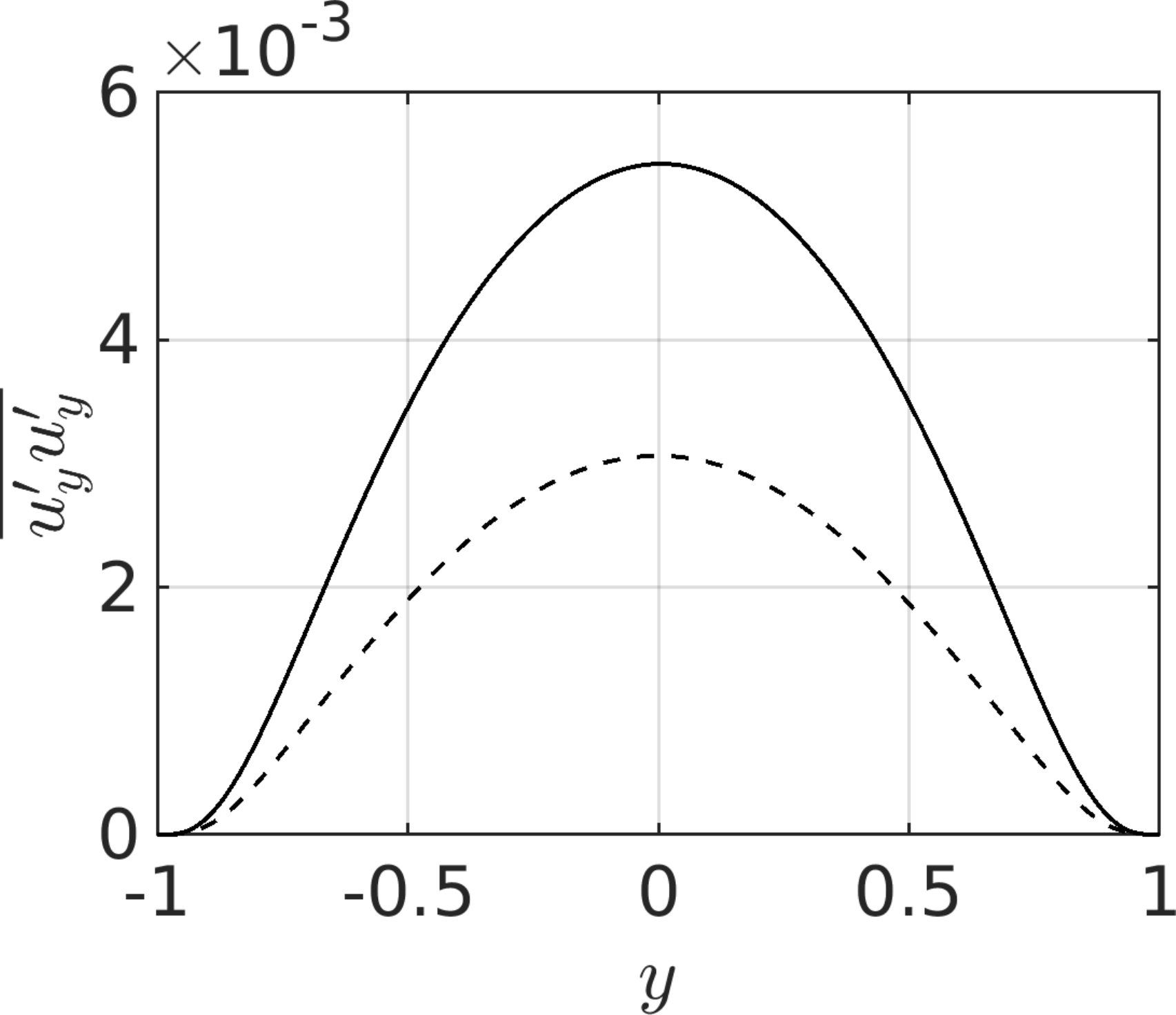}
\end{minipage}
\hfill
\begin{minipage}[c]{.24\linewidth}
\includegraphics[width = 1.\textwidth]{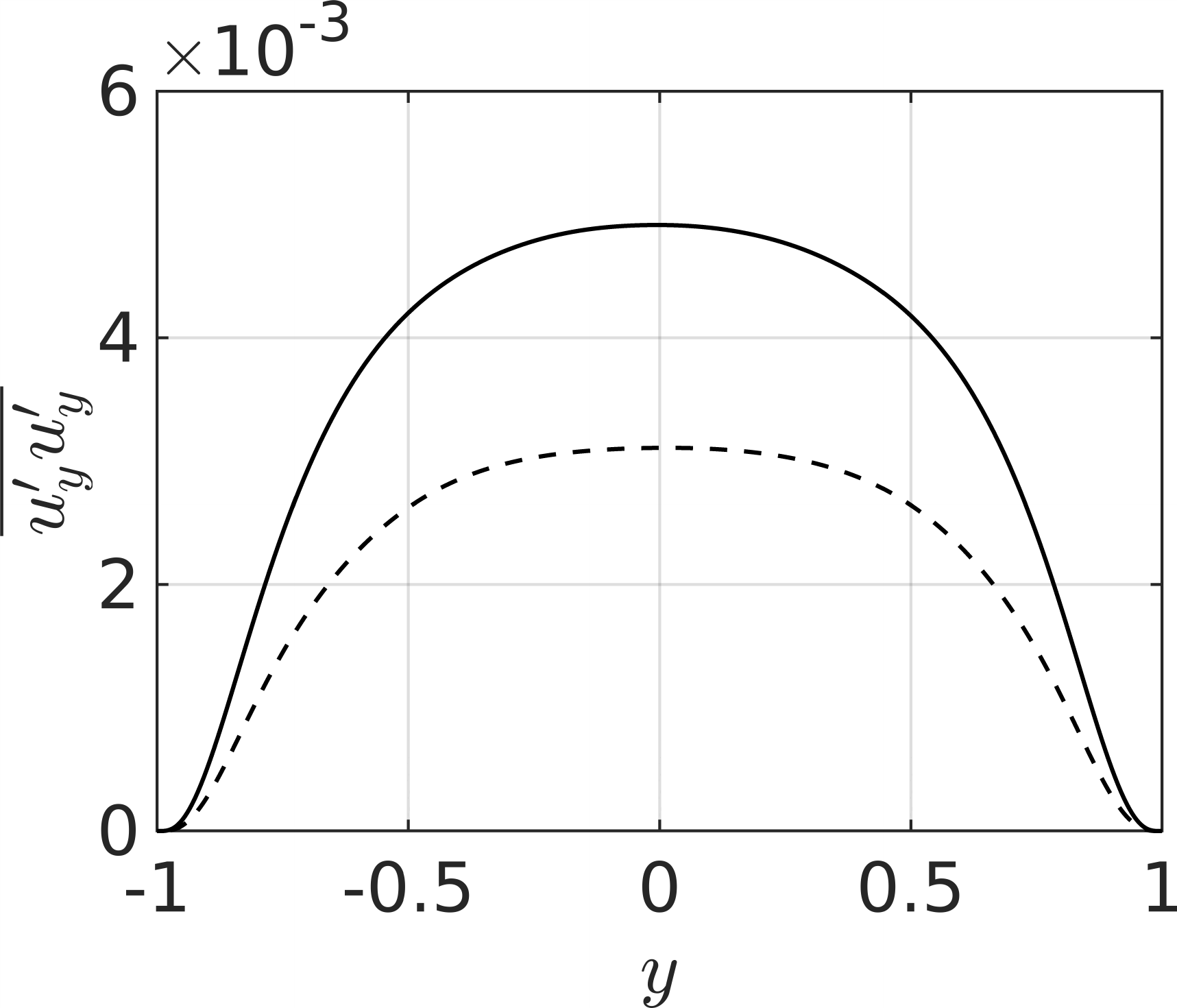}
\end{minipage}
\hfill
\begin{minipage}[c]{.24\linewidth}
\includegraphics[width = 1.\textwidth]{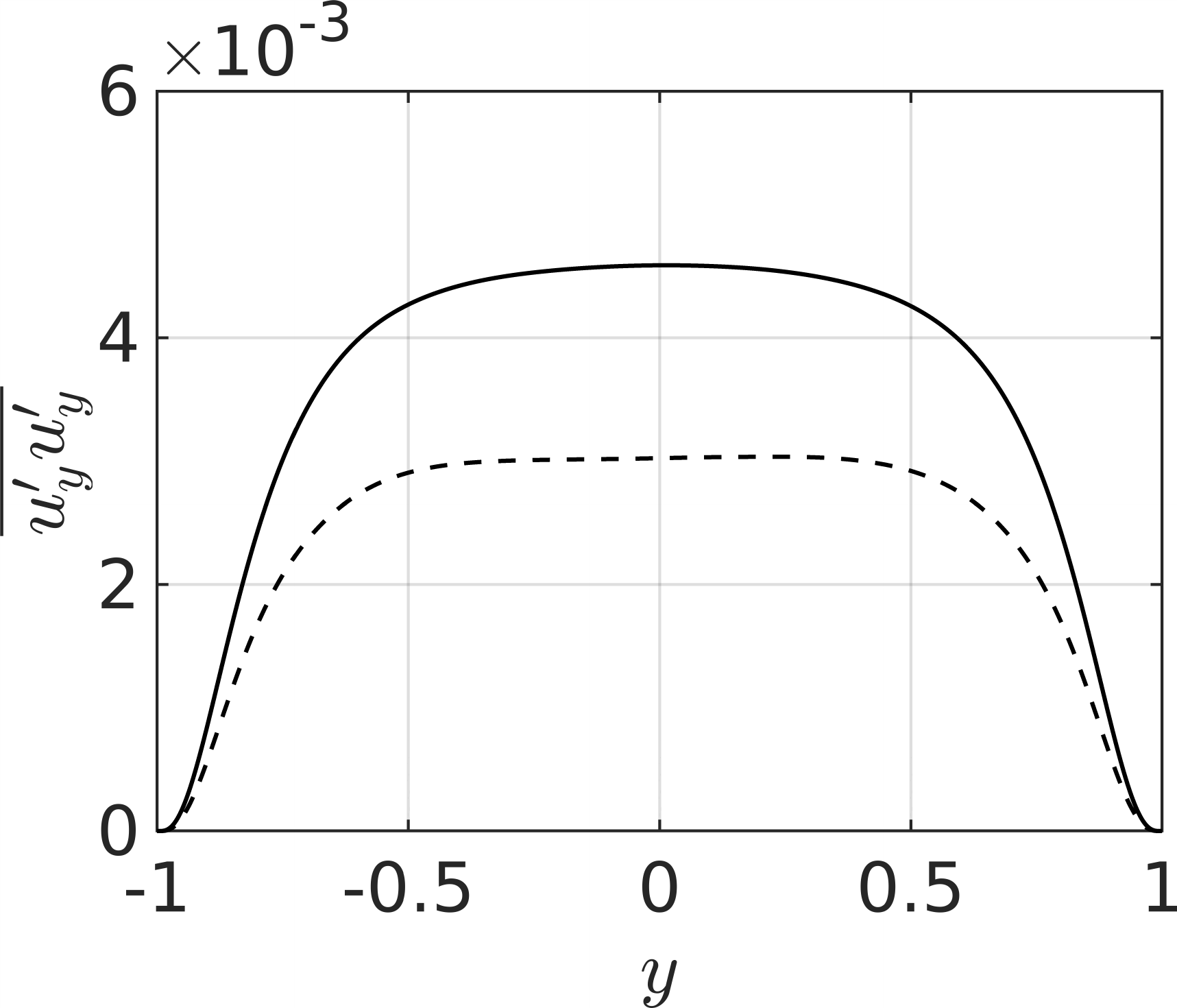}
\end{minipage}
\hfill
\begin{minipage}[c]{.25\linewidth}
\includegraphics[width = 1.\textwidth]{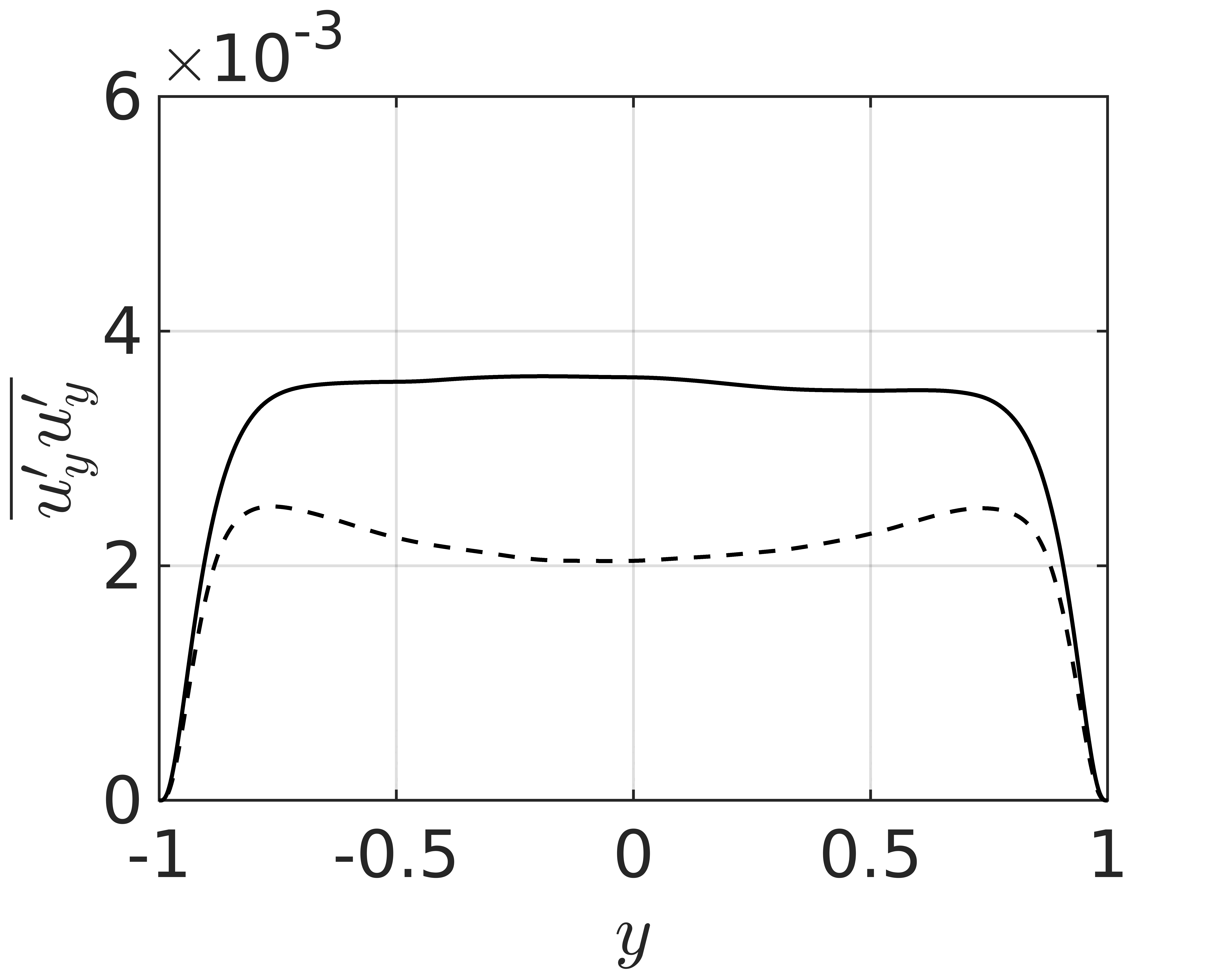}
\end{minipage}\\
\begin{minipage}[c]{.24\linewidth}
\includegraphics[width = 1.\textwidth]{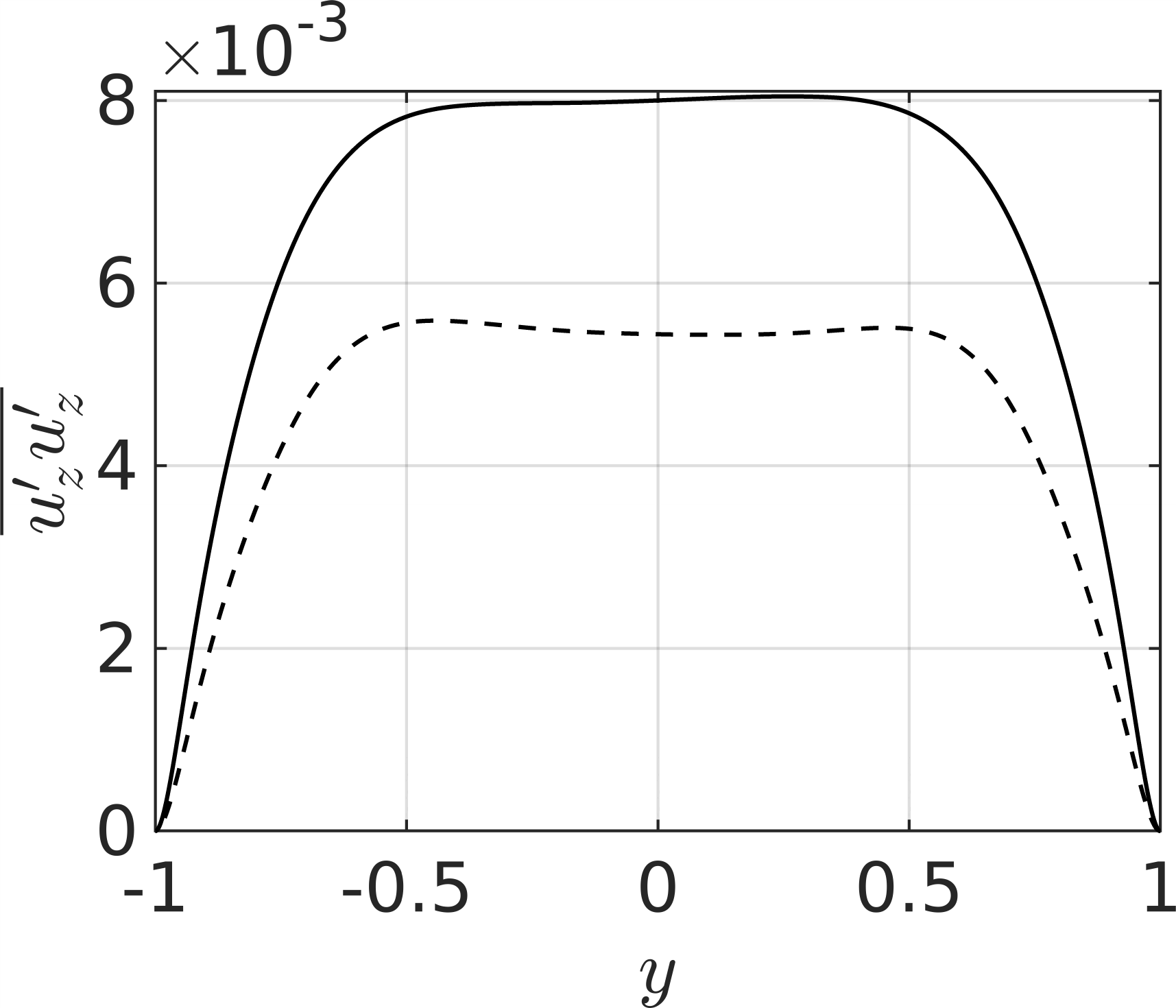}
\end{minipage}
\hfill
\begin{minipage}[c]{.24\linewidth}
\includegraphics[width = 1.\textwidth]{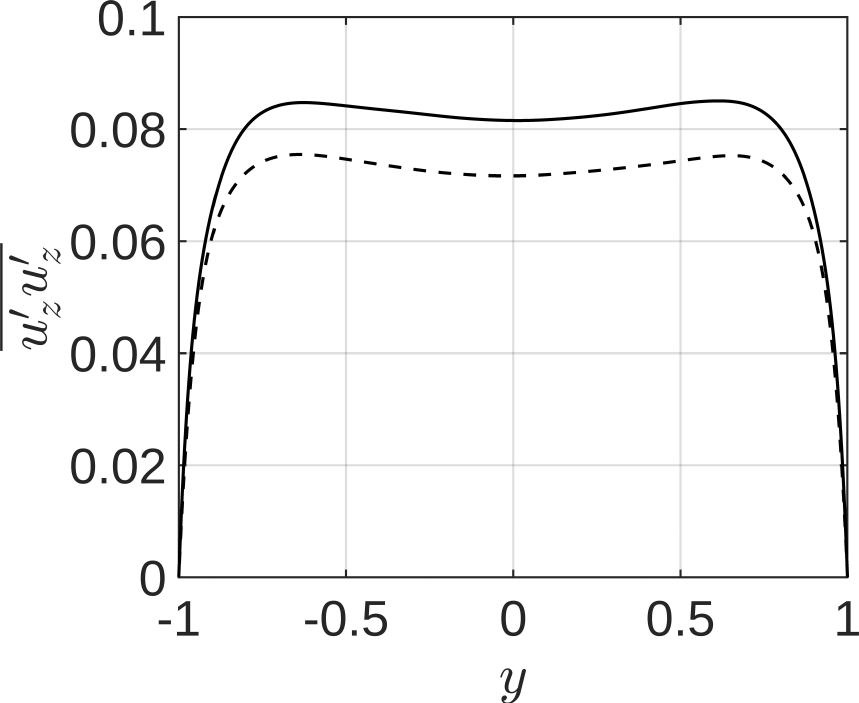}
\end{minipage}
\hfill
\begin{minipage}[c]{.24\linewidth}
\includegraphics[width = 1.\textwidth]{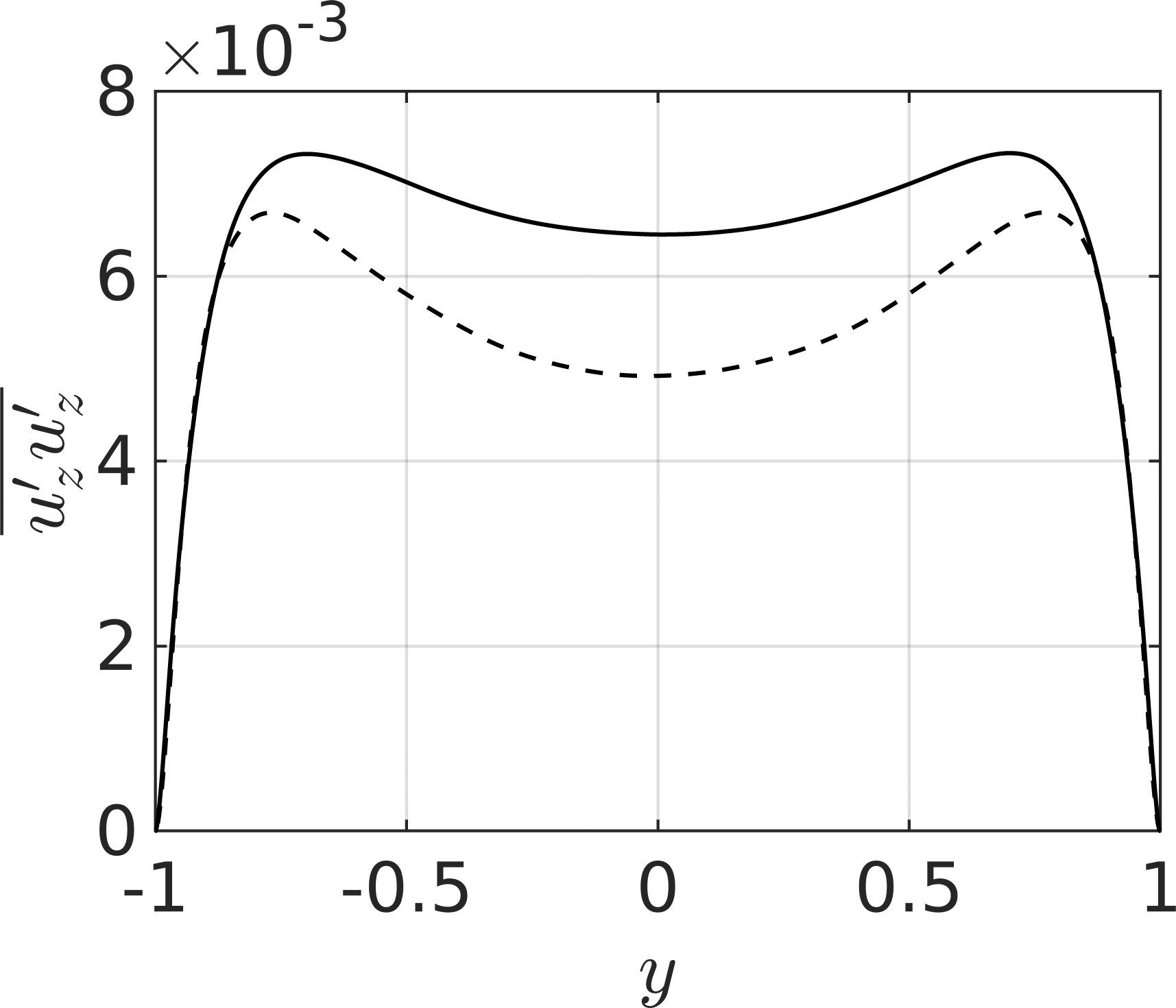}
\end{minipage}
\hfill
\begin{minipage}[c]{.25\linewidth}
\includegraphics[width = 1.\textwidth]{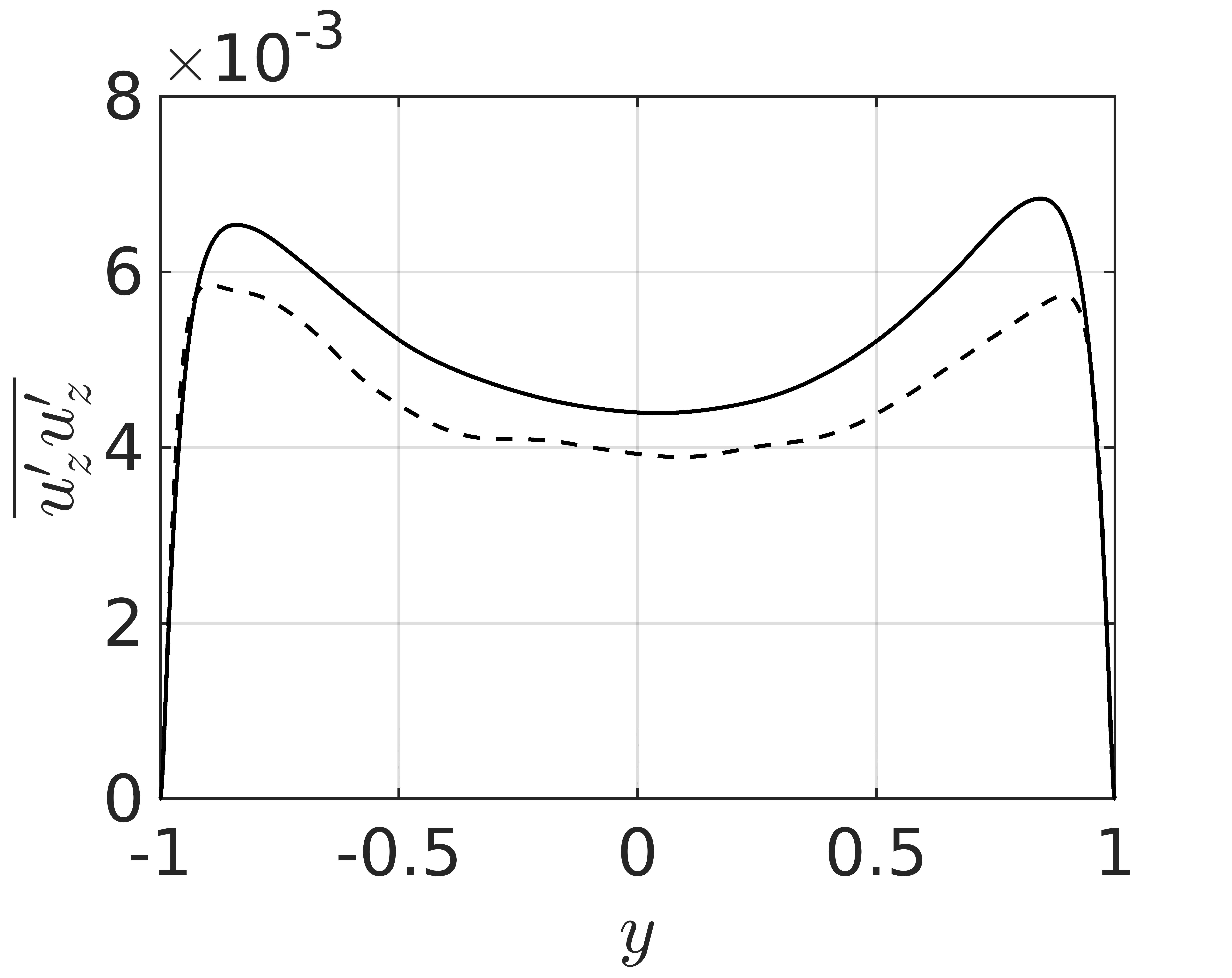}
\end{minipage}
\caption{Reynolds diagonal stresses of  uncontrolled and controlled turbulent flows, solid and dashed lines, correspondingly. $Re= 750, ~1500, ~2150, ~5000$, columns from left to right. One can observe that increase of Reynolds number leads to the decrease of the reduction in $\overline{u_z^\prime u_z^\prime}$, while the other diagonal components show the quantitatively similar behavior.} 
\label{Stat_Stresses}
\end{figure*}

The Reynolds stress diagonal components for the uncontrolled and controlled turbulent flows are presented in Figure \ref{Stat_Stresses} at Reynolds numbers, $Re=750, 1500, ~2150$, left to right columns. The solid lines show the turbulent flow statistics, while the dashed lines correspond to the controlled cases. As one can see, for the controlled turbulent flows, there is a decay of Reynolds stresses. 
\begin{figure*}[ht!]
\begin{minipage}[c]{.24\linewidth}
\includegraphics[width = 1.\textwidth]{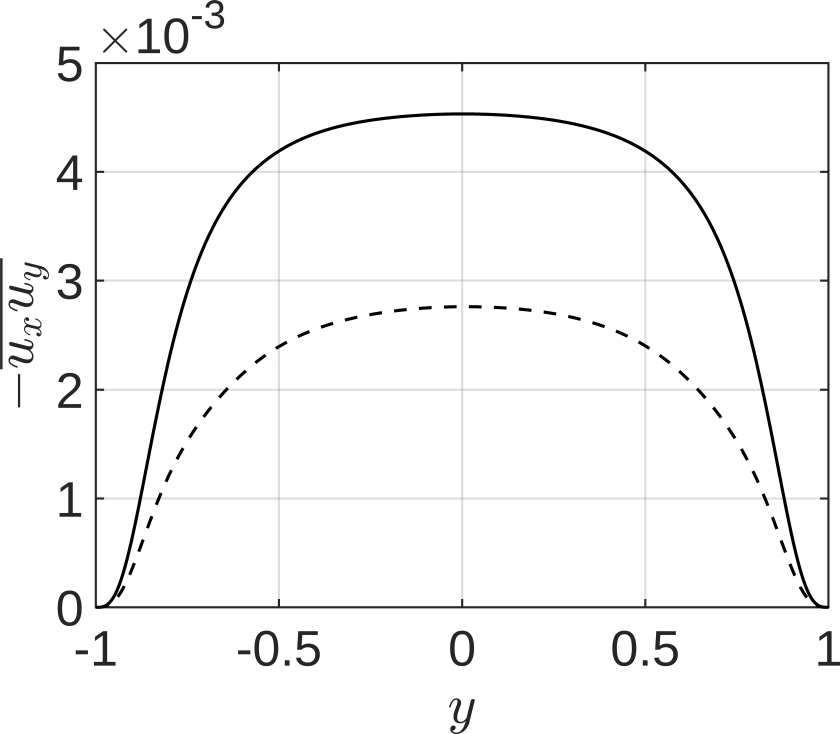}
\end{minipage}
\hfill
\begin{minipage}[c]{.24\linewidth}
\includegraphics[width = 1.\textwidth]{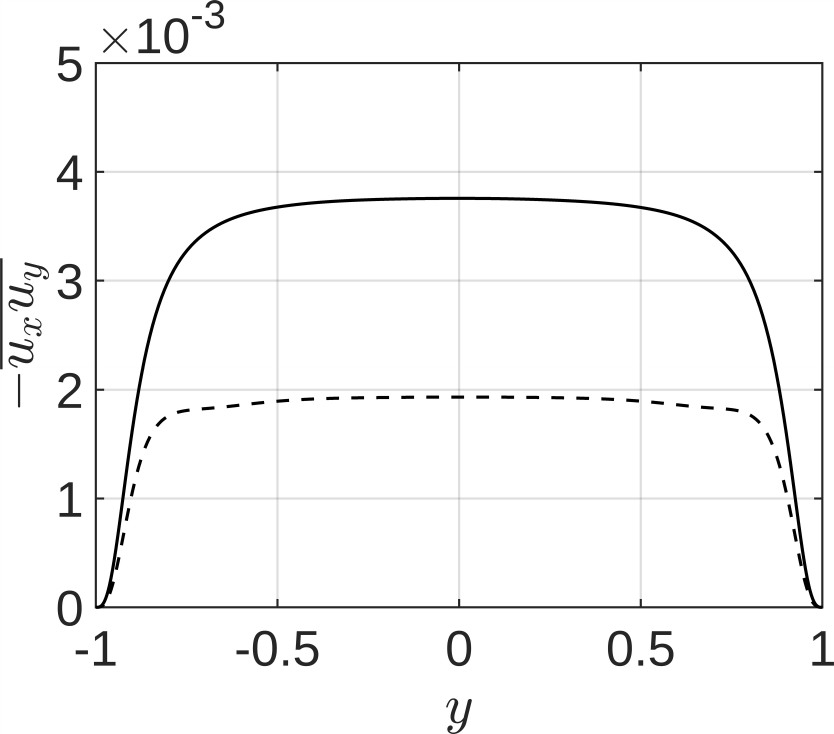}
\end{minipage}
\hfill
\begin{minipage}[c]{.24\linewidth}
\includegraphics[width = 1.\textwidth]{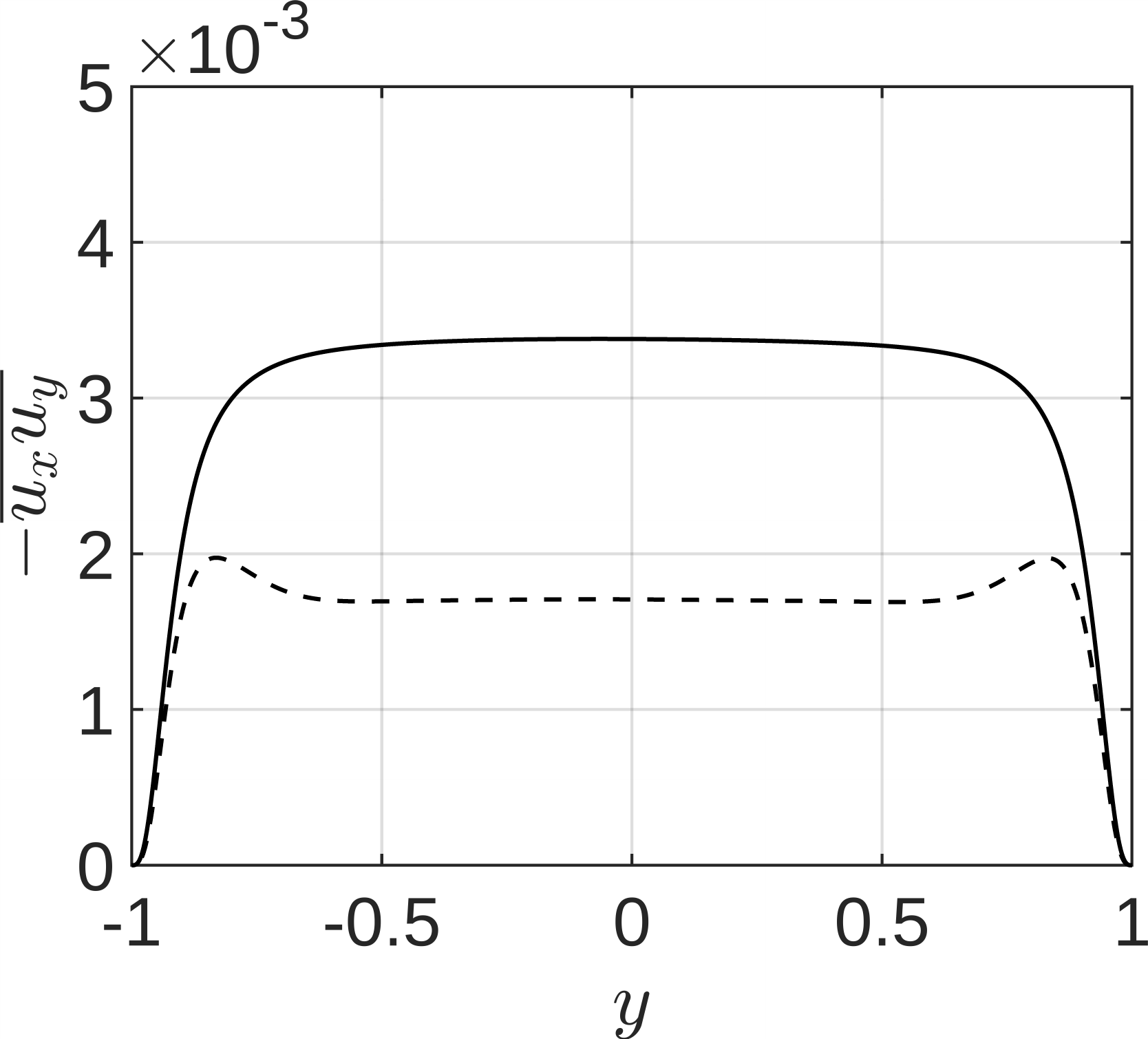}
\end{minipage}
\hfill
\begin{minipage}[c]{.26\linewidth}
\includegraphics[width = 1.\textwidth]{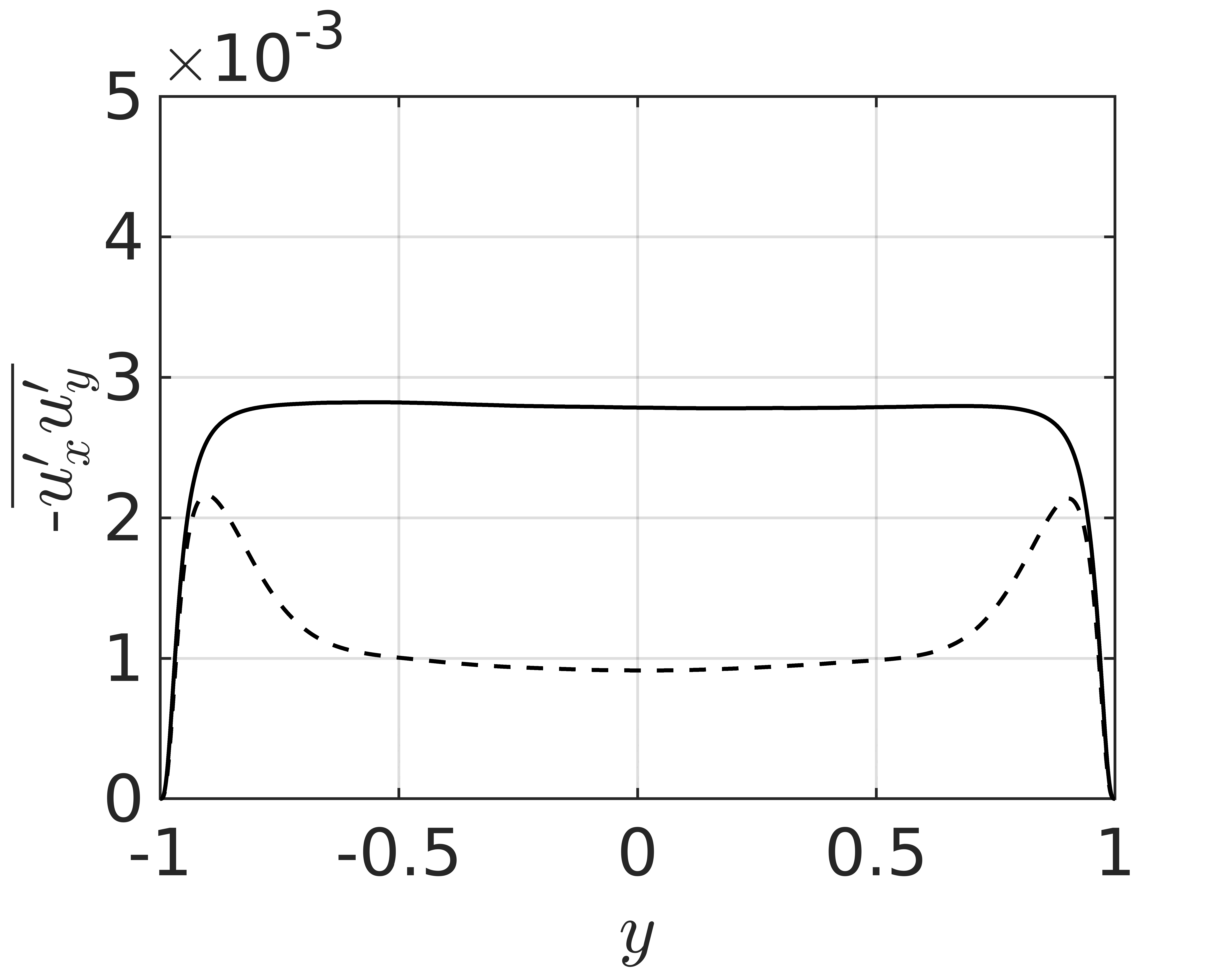}
\end{minipage}\\
\begin{minipage}[c]{.24\linewidth}
\includegraphics[width = 1.\textwidth]{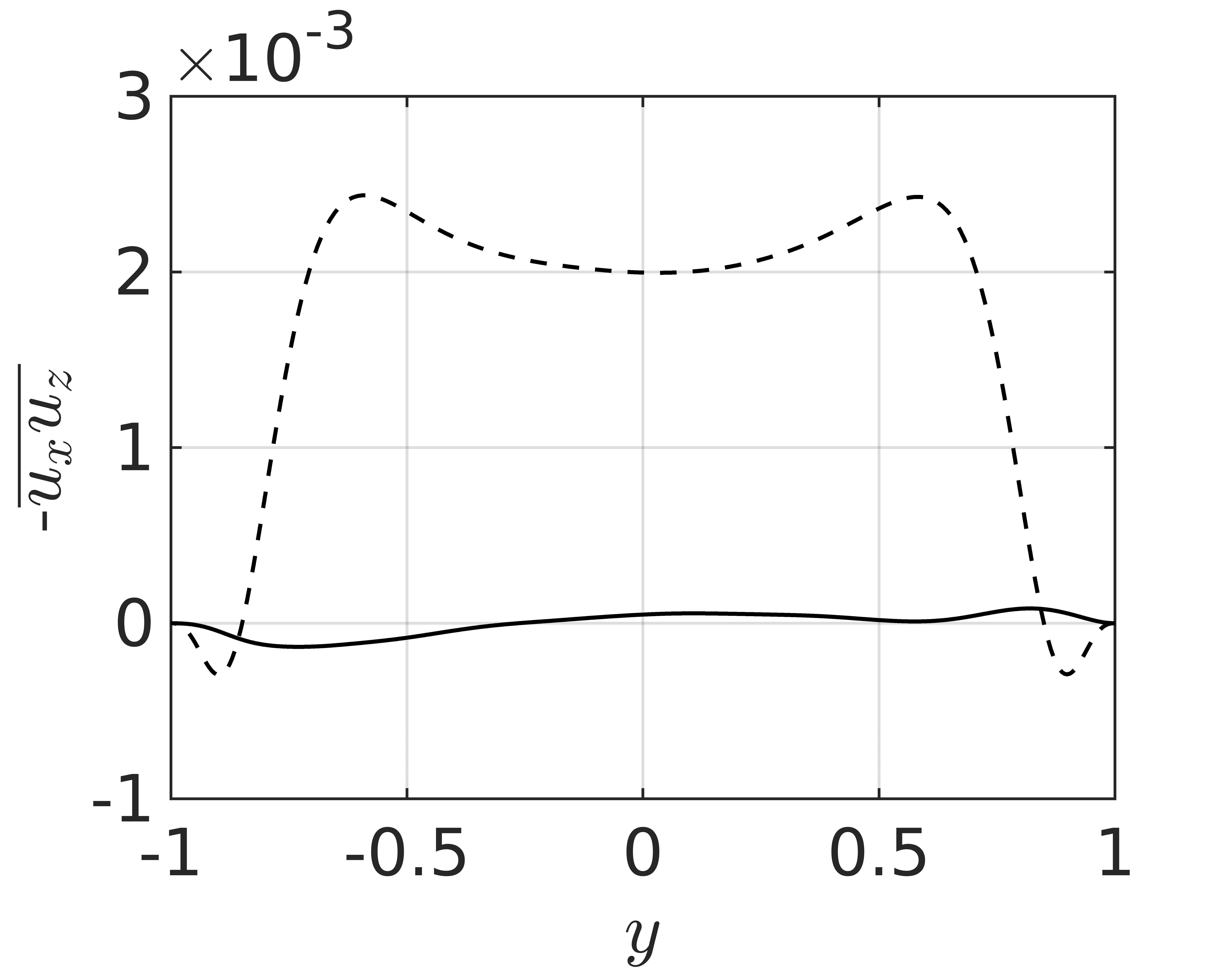}
\end{minipage}
\hfill
\begin{minipage}[c]{.24\linewidth}
\includegraphics[width = 1.\textwidth]{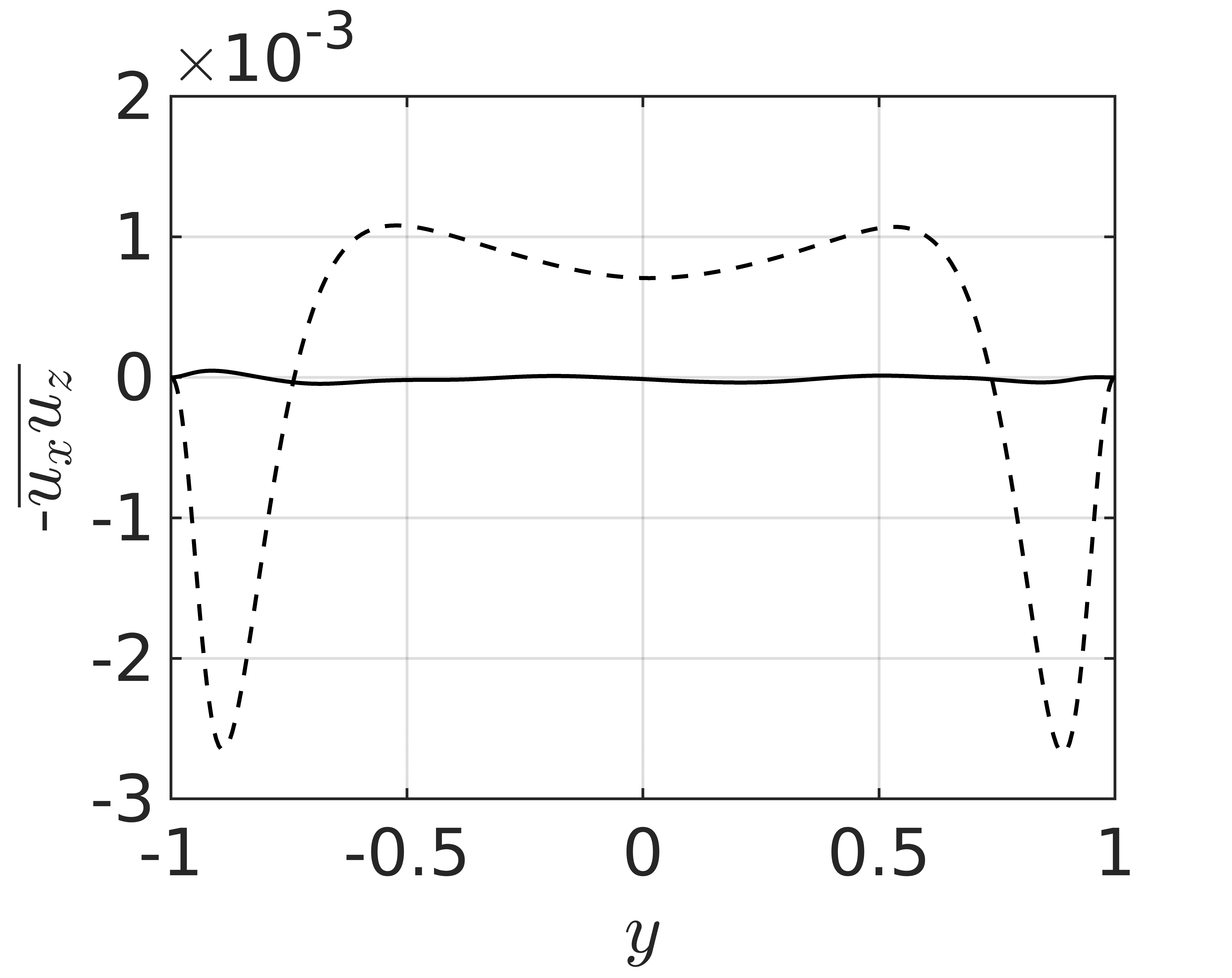}
\end{minipage}
\hfill
\begin{minipage}[c]{.24\linewidth}
\includegraphics[width = 1.\textwidth]{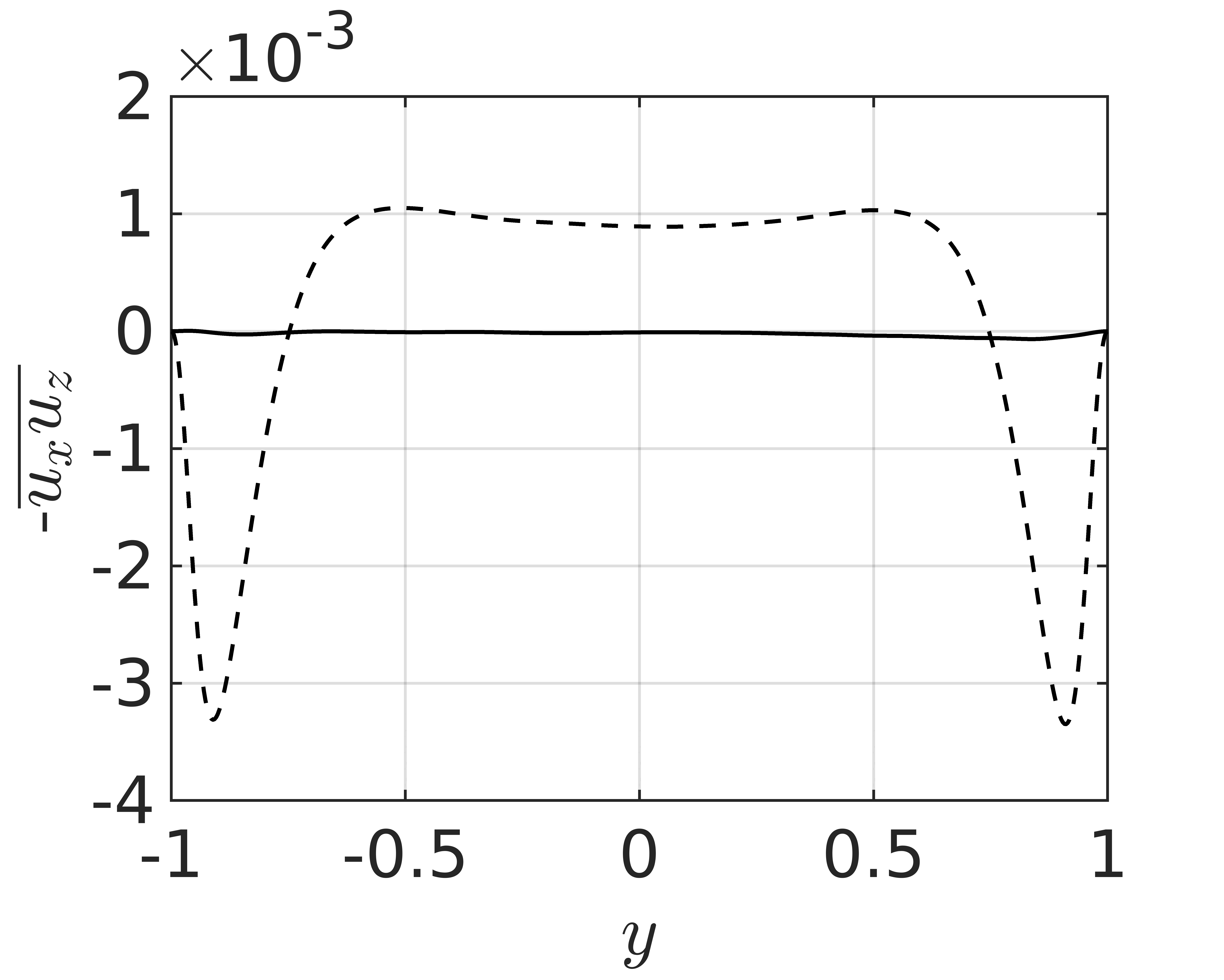}
\end{minipage}
\hfill
\begin{minipage}[c]{.26\linewidth}
\includegraphics[width = 1.\textwidth]{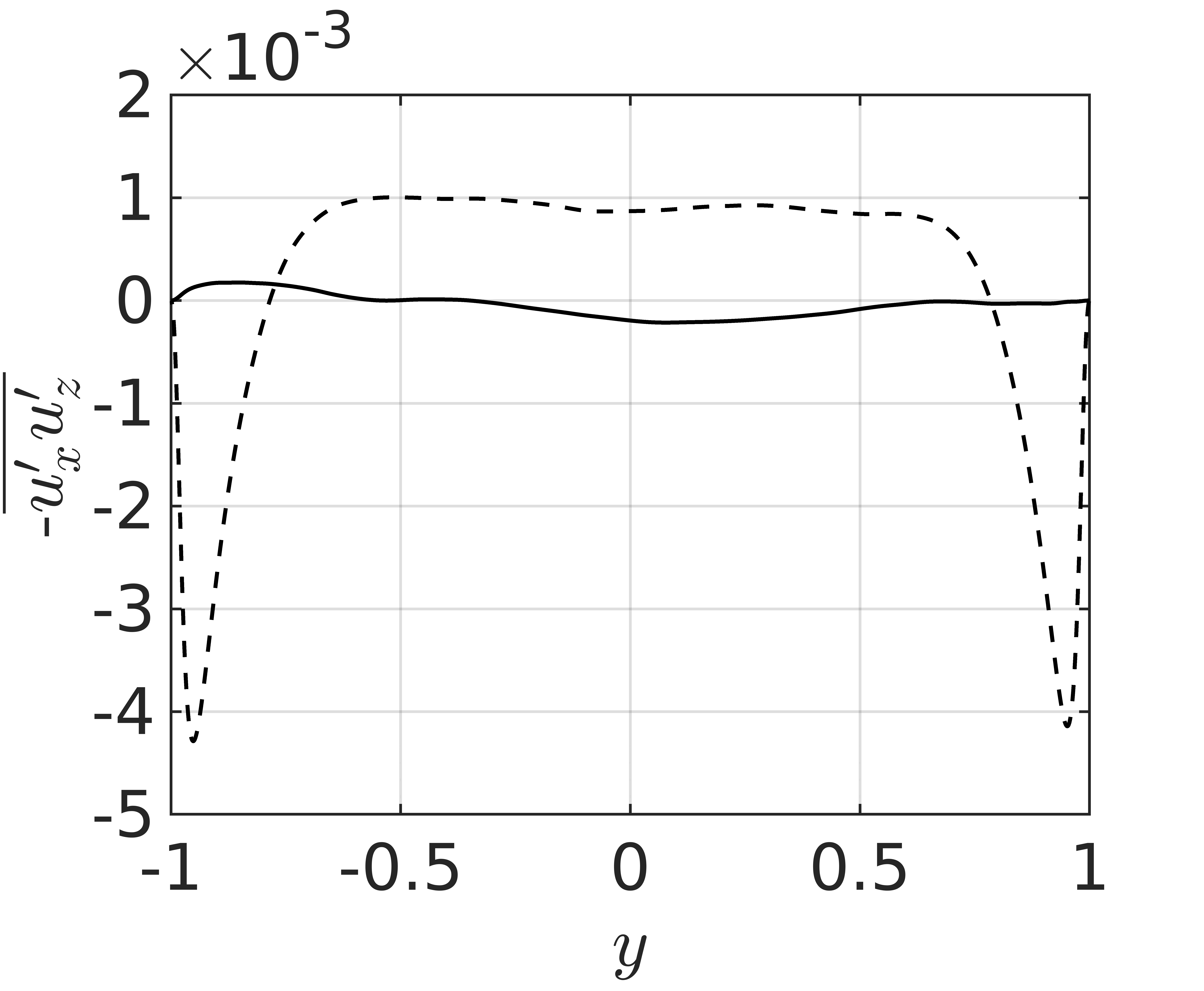}
\end{minipage}\\
\begin{minipage}[c]{.24\linewidth}
\includegraphics[width = 1.\textwidth]{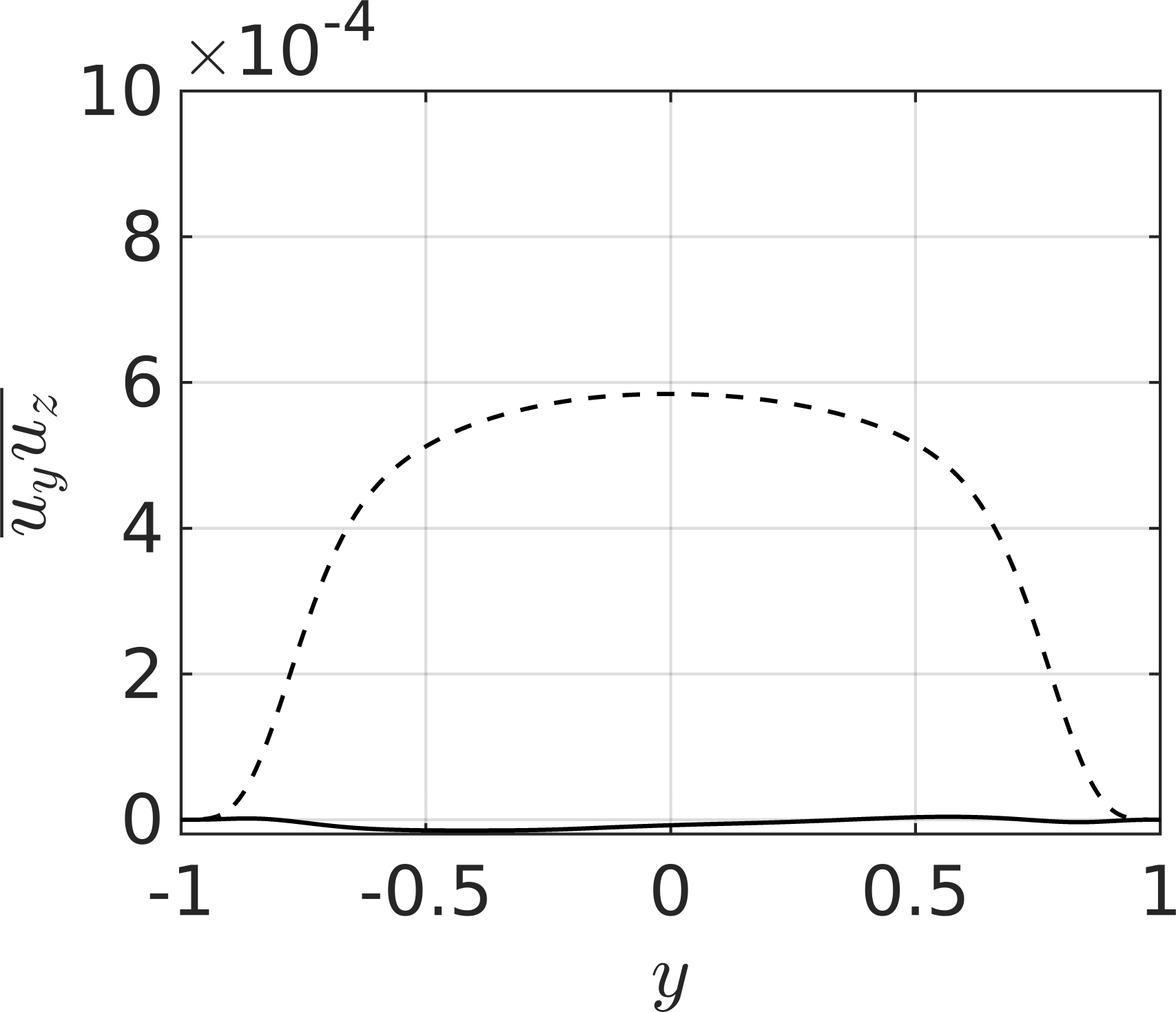}
\end{minipage}
\hfill
\begin{minipage}[c]{.24\linewidth}
\includegraphics[width = 1.\textwidth]{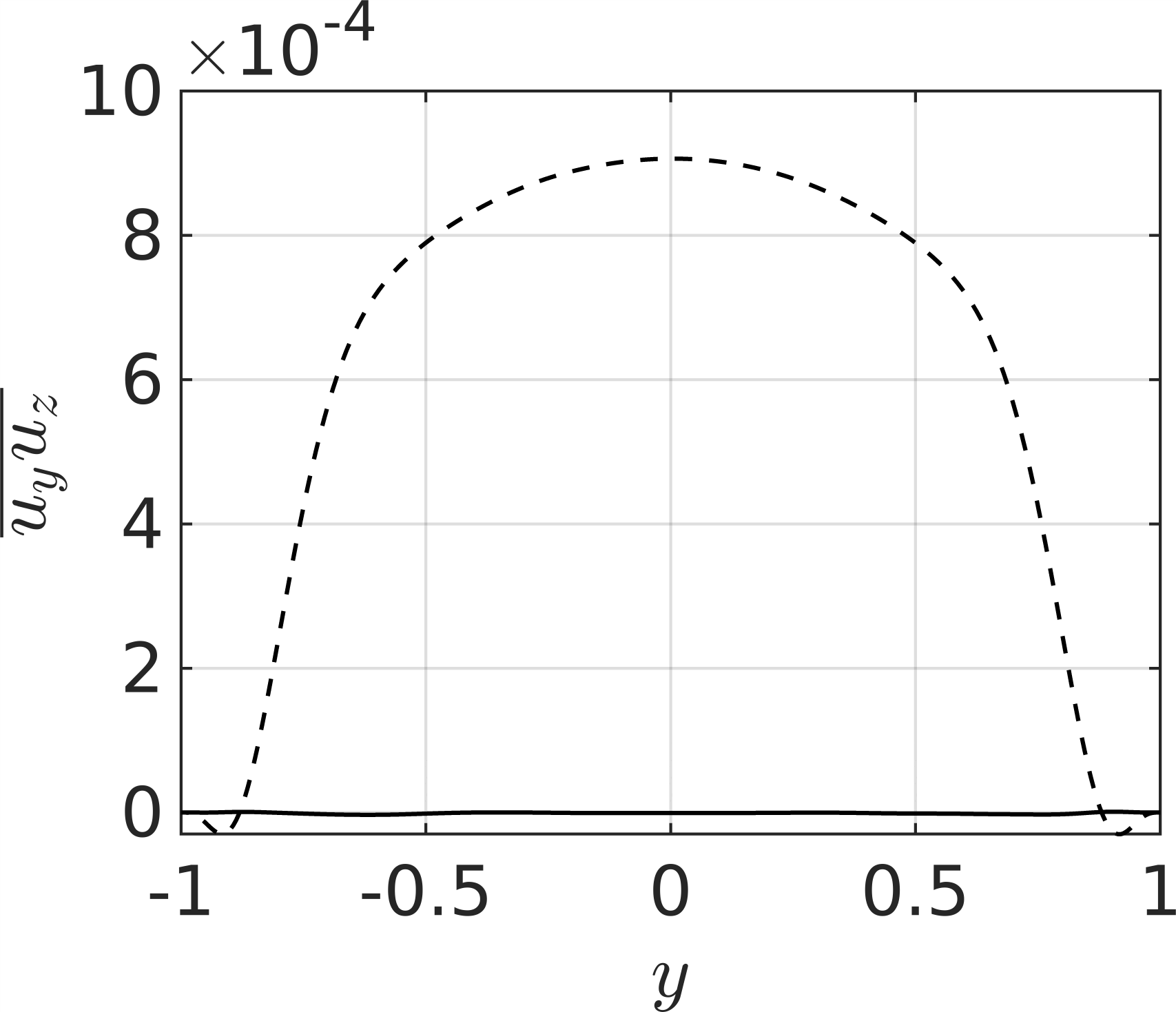}
\end{minipage}
\hfill
\begin{minipage}[c]{.24\linewidth}
\includegraphics[width = 1.\textwidth]{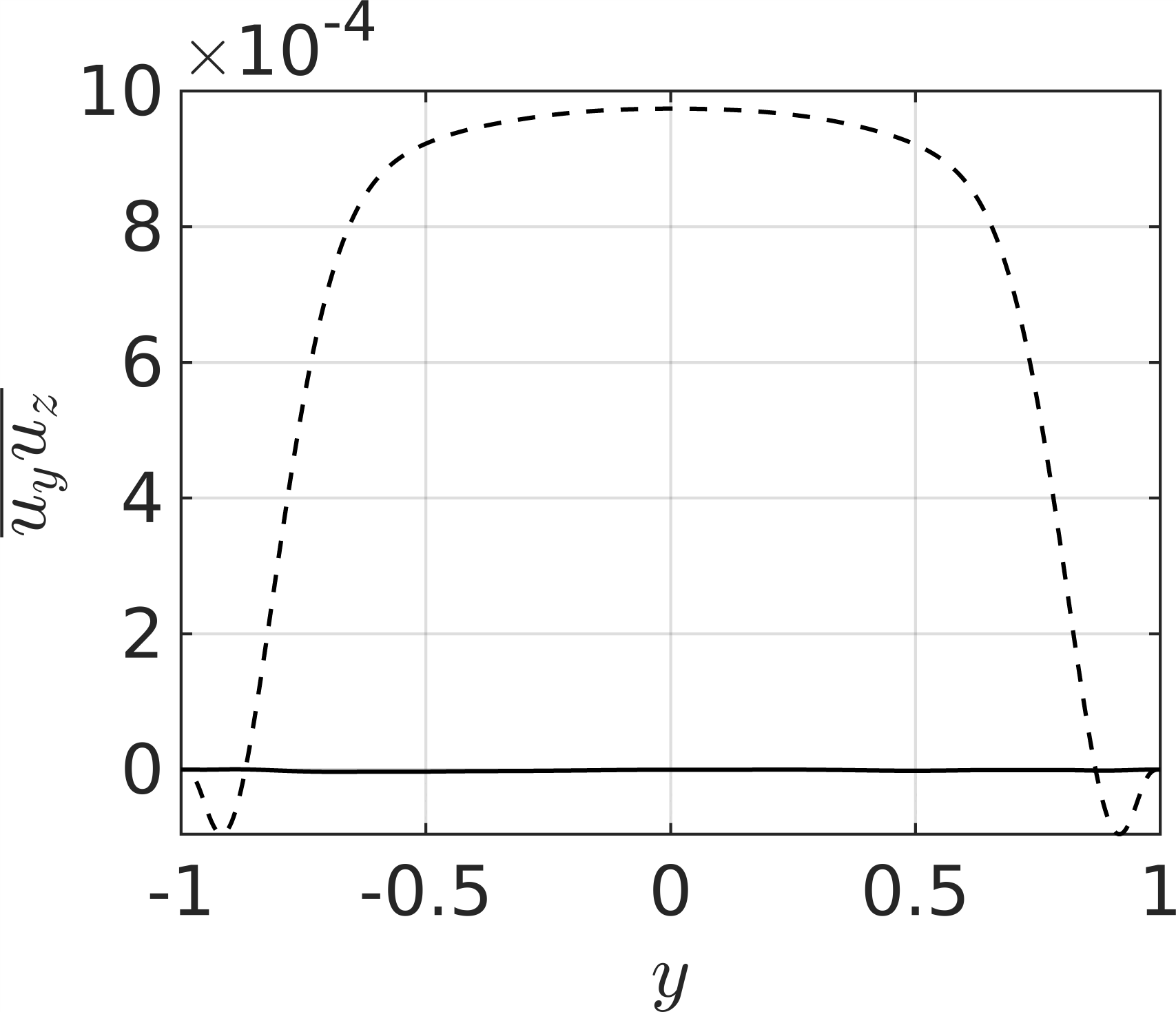}
\end{minipage}
\hfill
\begin{minipage}[c]{.26\linewidth}
\includegraphics[width = 1.\textwidth]{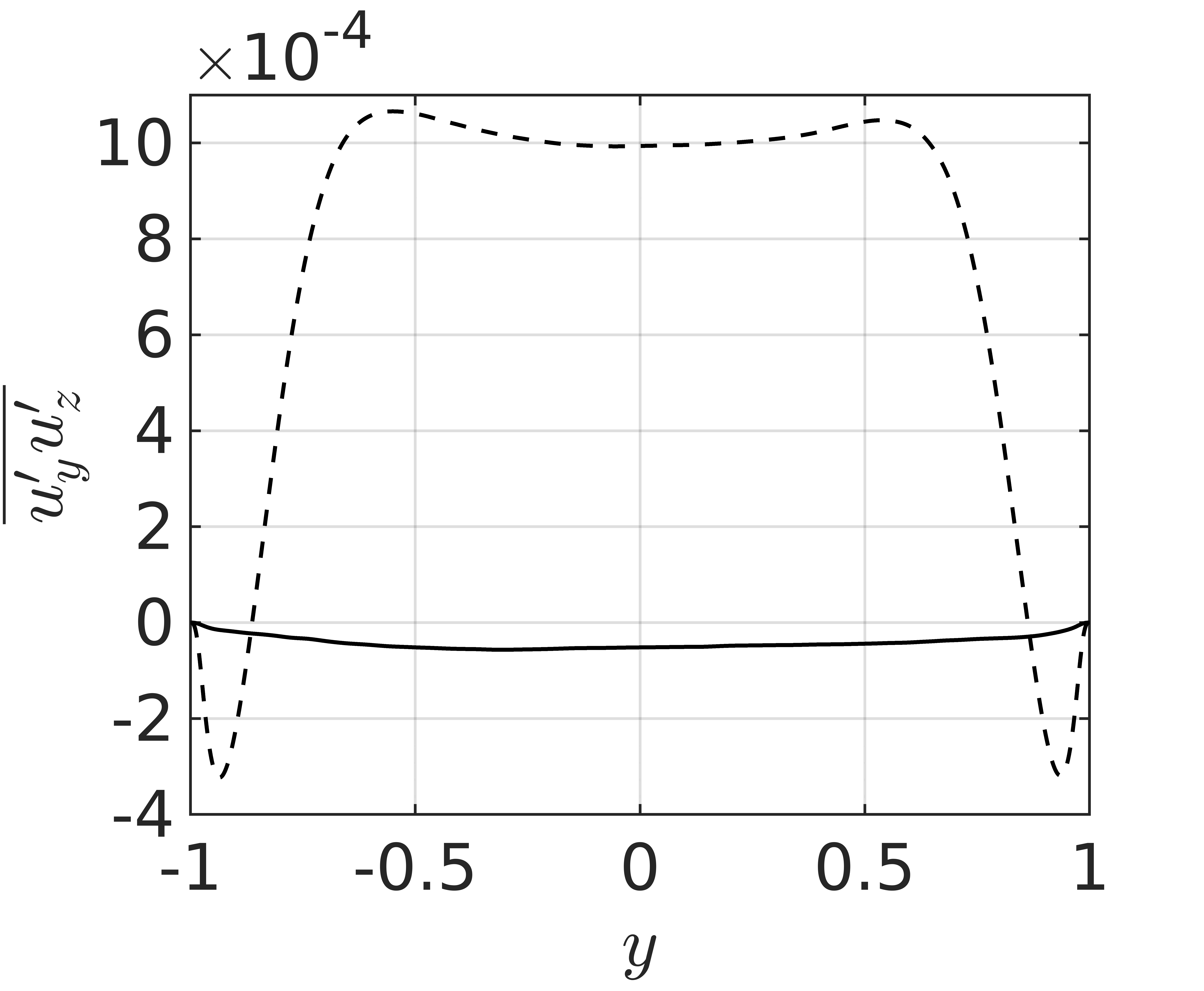}
\end{minipage}
\caption{Reynolds non-diagonal stresses of uncontrolled and controlled turbulent flows, solid and dashed lines, correspondingly.
$Re= 750, ~1500, ~2150, ~5000$, columns from left to right.}
\label{Stat_nondiag_str}
\end{figure*}

The non-diagonal Reynolds stresses are shown in Figure \ref{Stat_nondiag_str}. The first row represents the classical non-diagonal stress $\overline{u_x^\prime u_y^\prime}$ for three different Reynolds numbers. The solid and dashed lines correspond to the uncontrolled and controlled turbulent flows as it was in the previous figure. Strong decay of the stress for the controlled cases are observed for all Reynolds numbers. 

The second and third rows show two other non-diagonal stresses ($\overline{u_x^\prime u_z^\prime}$ and $\overline{u_y^\prime u_z^\prime}$), that usually in the uncontrolled turbulent flows are zero (solid lines on these plots). But in the controlled cases they are non-zero (dashed lines) showing the asymmetry of the flow in spanwise direction. This asymmetry is clearly seen on the instantaneous velocity field (see the $XZ$-slices on the plots \ref{XZ_slices}). The non-zero non-diagonal Reynolds stress $\overline{u_y^\prime u_z^\prime}$ causes the small but non-zero spanwise component of TKE production.

\section{Conclusions}
The efficiency and universality of the control strategy proposed in \citet{Chagelishvili2014} have been demonstrated on the example of turbulent plane Couette flow at various Reynolds numbers ($Re_{\tau} = 52,92,128,270$) and simulation box sizes. 

The essence of the control strategy lies in the imposition of specially designed seed velocity perturbations, which are non-symmetric in the spanwise direction, at the walls of the flow. This special design of the imposed seed velocity field ensures a gain of shear flow energy and breaks the turbulence spanwise reflection symmetry generating nonuniform spanwise mean flow. One has to note that the formed spanwise mean flow is an intrinsic, nonlinear composition of the controlled turbulence and not directly introduced into the system. Overall, the new configuration of the mean flow (the basic streamwise mean flow plus nonuniform spanwise mean flow) changes the self-sustained dynamics of turbulence and results in a considerable reduction of the turbulence level and the production of TKE. For now, to impose the required seed velocity field in the flow, a weak near-wall volume forcing is designed, which is theoretical/hypothetical. However, provides a fairly clear indication of subsequent efforts to implement the discussed control strategy -- replacing the hypothetical forcing with workable purpose-constructed blowing/suction system and, finally, with purpose-designed riblets.

Thus, the main aim of our study was to investigate the efficiency of the control strategy developed when keeping constant the control parameters but change the Reynolds numbers and simulation box sizes -- maintaining constant control parameters -- forcing amplitudes and length scales in parallel and wall-normal directions, localization centers in parallel directions -- constant in all cases, we achieved a significant reduction in TKE production, specifically, within the range of $30-45\%$. A certain decrease in control efficiency at large Reynolds number (to $30\%$) can be avoided by optimization of the control parameters mentioned above (for specifics see $A_i$ and ${l_{i}}$, $i=(x,y,z)$ in Eq. \eqref{VF}). 

In addition to varying Reynolds numbers and simulation boxes, we investigated the efficiency of the control at different locations of the forcing from the wall. It has to be emphasized that all points are located in the viscous sub-layer of the flow. We conducted simulations at two different Reynolds numbers ($Re=1500,~2150$), each for two different points of the forcing localization from the wall. The results presented show that the reduction of TKE production remains unchanged. 

\section*{AUTHOR DECLARATIONS}
The authors have no conflicts to disclose.


\section*{Data Availability Statement}
The data that support the findings of this study are available from the corresponding author upon reasonable request.


\begin{thebibliography}{43}%
\makeatletter
\providecommand \@ifxundefined [1]{%
 \@ifx{#1\undefined}
}%
\providecommand \@ifnum [1]{%
 \ifnum #1\expandafter \@firstoftwo
 \else \expandafter \@secondoftwo
 \fi
}%
\providecommand \@ifx [1]{%
 \ifx #1\expandafter \@firstoftwo
 \else \expandafter \@secondoftwo
 \fi
}%
\providecommand \natexlab [1]{#1}%
\providecommand \enquote  [1]{``#1''}%
\providecommand \bibnamefont  [1]{#1}%
\providecommand \bibfnamefont [1]{#1}%
\providecommand \citenamefont [1]{#1}%
\providecommand \href@noop [0]{\@secondoftwo}%
\providecommand \href [0]{\begingroup \@sanitize@url \@href}%
\providecommand \@href[1]{\@@startlink{#1}\@@href}%
\providecommand \@@href[1]{\endgroup#1\@@endlink}%
\providecommand \@sanitize@url [0]{\catcode `\\12\catcode `\$12\catcode
  `\&12\catcode `\#12\catcode `\^12\catcode `\_12\catcode `\%12\relax}%
\providecommand \@@startlink[1]{}%
\providecommand \@@endlink[0]{}%
\providecommand \url  [0]{\begingroup\@sanitize@url \@url }%
\providecommand \@url [1]{\endgroup\@href {#1}{\urlprefix }}%
\providecommand \urlprefix  [0]{URL }%
\providecommand \Eprint [0]{\href }%
\providecommand \doibase [0]{http://dx.doi.org/}%
\providecommand \selectlanguage [0]{\@gobble}%
\providecommand \bibinfo  [0]{\@secondoftwo}%
\providecommand \bibfield  [0]{\@secondoftwo}%
\providecommand \translation [1]{[#1]}%
\providecommand \BibitemOpen [0]{}%
\providecommand \bibitemStop [0]{}%
\providecommand \bibitemNoStop [0]{.\EOS\space}%
\providecommand \EOS [0]{\spacefactor3000\relax}%
\providecommand \BibitemShut  [1]{\csname bibitem#1\endcsname}%
\let\auto@bib@innerbib\@empty
\bibitem [{\citenamefont {{Gad-el-Hak}}(2000)}]{GadelHak}%
  \BibitemOpen
  \bibfield  {author} {\bibinfo {author} {\bibfnamefont {M.}~\bibnamefont
  {{Gad-el-Hak}}},\ }\href@noop {} {\emph {\bibinfo {title} {Passive, Active
  and Reactive Flow}}}\ (\bibinfo  {publisher} {Cambridge Univ. Press},\
  \bibinfo {address} {Cambridge, UK},\ \bibinfo {year} {2000})\BibitemShut
  {NoStop}%
\bibitem [{\citenamefont {Bewley}(2001)}]{Bewley2001}%
  \BibitemOpen
  \bibfield  {author} {\bibinfo {author} {\bibfnamefont {T.~R.}\ \bibnamefont
  {Bewley}},\ }\bibfield  {title} {\enquote {\bibinfo {title} {Flow control:
  new challenges for a new renaissance},}\ }\href@noop {} {\bibfield  {journal}
  {\bibinfo  {journal} {Progress in Aerospace Sciences}\ }\textbf {\bibinfo
  {volume} {37}},\ \bibinfo {pages} {21--53} (\bibinfo {year}
  {2001})}\BibitemShut {NoStop}%
\bibitem [{\citenamefont {Kim}(2003)}]{Kim2003}%
  \BibitemOpen
  \bibfield  {author} {\bibinfo {author} {\bibfnamefont {J.}~\bibnamefont
  {Kim}},\ }\bibfield  {title} {\enquote {\bibinfo {title} {Control of
  turbulent boundary layers},}\ }\href@noop {} {\bibfield  {journal} {\bibinfo
  {journal} {Physics of Fluids}\ }\textbf {\bibinfo {volume} {15}},\ \bibinfo
  {pages} {1093--1105} (\bibinfo {year} {2003})}\BibitemShut {NoStop}%
\bibitem [{\citenamefont {Dean}\ and\ \citenamefont {Bhushan}(2010)}]{Dean}%
  \BibitemOpen
  \bibfield  {author} {\bibinfo {author} {\bibfnamefont {B.}~\bibnamefont
  {Dean}}\ and\ \bibinfo {author} {\bibfnamefont {B.}~\bibnamefont {Bhushan}},\
  }\bibfield  {title} {\enquote {\bibinfo {title} {Shark-skin surfaces for
  fluid-drag reduction in turbulent flow: A review},}\ }\href@noop {}
  {\bibfield  {journal} {\bibinfo  {journal} {Phil. Trans. R.Soc.A}\ }\textbf
  {\bibinfo {volume} {368}} (\bibinfo {year} {2010})}\BibitemShut {NoStop}%
\bibitem [{\citenamefont {Marusic}\ \emph {et~al.}(2021)\citenamefont
  {Marusic}, \citenamefont {Chandran}, \citenamefont {Rouhi}, \citenamefont
  {Fu}, \citenamefont {Wine}, \citenamefont {Holloway}, \citenamefont {Chung},\
  and\ \citenamefont {Smits}}]{Marusic2021}%
  \BibitemOpen
  \bibfield  {author} {\bibinfo {author} {\bibfnamefont {I.}~\bibnamefont
  {Marusic}}, \bibinfo {author} {\bibfnamefont {D.}~\bibnamefont {Chandran}},
  \bibinfo {author} {\bibfnamefont {A.}~\bibnamefont {Rouhi}}, \bibinfo
  {author} {\bibfnamefont {M.}~\bibnamefont {Fu}}, \bibinfo {author}
  {\bibfnamefont {D.}~\bibnamefont {Wine}}, \bibinfo {author} {\bibfnamefont
  {B.}~\bibnamefont {Holloway}}, \bibinfo {author} {\bibfnamefont
  {D.}~\bibnamefont {Chung}}, \ and\ \bibinfo {author} {\bibfnamefont
  {A.}~\bibnamefont {Smits}},\ }\bibfield  {title} {\enquote {\bibinfo {title}
  {An energy-efficient pathway to turbulent drag reduction},}\ }\href@noop {}
  {\bibfield  {journal} {\bibinfo  {journal} {Nature Communications}\ }\textbf
  {\bibinfo {volume} {12}},\ \bibinfo {pages} {5805} (\bibinfo {year}
  {2021})}\BibitemShut {NoStop}%
\bibitem [{\citenamefont {Ricco}, \citenamefont {Skote},\ and\ \citenamefont
  {Leschziner}(2021)}]{Ricco2021}%
  \BibitemOpen
  \bibfield  {author} {\bibinfo {author} {\bibfnamefont {P.}~\bibnamefont
  {Ricco}}, \bibinfo {author} {\bibfnamefont {M.}~\bibnamefont {Skote}}, \ and\
  \bibinfo {author} {\bibfnamefont {M.}~\bibnamefont {Leschziner}},\ }\bibfield
   {title} {\enquote {\bibinfo {title} {A review of turbulent skin-friction
  drag reduction by near-wall transverse forcing},}\ }\href@noop {} {\bibfield
  {journal} {\bibinfo  {journal} {Progress in Aerospace Sciences}\ }\textbf
  {\bibinfo {volume} {123}},\ \bibinfo {pages} {100713} (\bibinfo {year}
  {2021})}\BibitemShut {NoStop}%
\bibitem [{\citenamefont {Choi}, \citenamefont {Moin},\ and\ \citenamefont
  {Kim}(1993)}]{Choi1993}%
  \BibitemOpen
  \bibfield  {author} {\bibinfo {author} {\bibfnamefont {H.}~\bibnamefont
  {Choi}}, \bibinfo {author} {\bibfnamefont {P.}~\bibnamefont {Moin}}, \ and\
  \bibinfo {author} {\bibfnamefont {J.}~\bibnamefont {Kim}},\ }\bibfield
  {title} {\enquote {\bibinfo {title} {Direct numerical simulation of turbulent
  flow over riblets},}\ }\href@noop {} {\bibfield  {journal} {\bibinfo
  {journal} {J. Fluid Mech.}\ }\textbf {\bibinfo {volume} {255}},\ \bibinfo
  {pages} {503--539} (\bibinfo {year} {1993})}\BibitemShut {NoStop}%
\bibitem [{\citenamefont {Woodcock}, \citenamefont {Sader},\ and\ \citenamefont
  {Marusic}(2012)}]{Marusic2012}%
  \BibitemOpen
  \bibfield  {author} {\bibinfo {author} {\bibfnamefont {J.}~\bibnamefont
  {Woodcock}}, \bibinfo {author} {\bibfnamefont {J.}~\bibnamefont {Sader}}, \
  and\ \bibinfo {author} {\bibfnamefont {I.}~\bibnamefont {Marusic}},\
  }\bibfield  {title} {\enquote {\bibinfo {title} {Induced flow due to blowing
  and suction flow control: and analysis of transpiration},}\ }\href@noop {}
  {\bibfield  {journal} {\bibinfo  {journal} {J. Fluid Mech.}\ }\textbf
  {\bibinfo {volume} {690}},\ \bibinfo {pages} {366--398} (\bibinfo {year}
  {2012})}\BibitemShut {NoStop}%
\bibitem [{\citenamefont {Kametani}\ and\ \citenamefont
  {Fukagata}(2011)}]{Kametani2011}%
  \BibitemOpen
  \bibfield  {author} {\bibinfo {author} {\bibfnamefont {Y.}~\bibnamefont
  {Kametani}}\ and\ \bibinfo {author} {\bibfnamefont {K.}~\bibnamefont
  {Fukagata}},\ }\bibfield  {title} {\enquote {\bibinfo {title} {Direct
  numerical simulation of spatially developing turbulent boundary layers with
  uniform blowing and suction},}\ }\href@noop {} {\bibfield  {journal}
  {\bibinfo  {journal} {J. Fluid Mech.}\ }\textbf {\bibinfo {volume} {681}},\
  \bibinfo {pages} {154--172} (\bibinfo {year} {2011})}\BibitemShut {NoStop}%
\bibitem [{\citenamefont {Baron}\ and\ \citenamefont
  {Quadrio}(1993)}]{Baron1995}%
  \BibitemOpen
  \bibfield  {author} {\bibinfo {author} {\bibfnamefont {O.}~\bibnamefont
  {Baron}}\ and\ \bibinfo {author} {\bibfnamefont {M.}~\bibnamefont
  {Quadrio}},\ }\bibfield  {title} {\enquote {\bibinfo {title} {Turbulent drag
  reduction by spanwise wall oscilattions},}\ }\href@noop {} {\bibfield
  {journal} {\bibinfo  {journal} {Appl. Sci. Res.}\ }\textbf {\bibinfo {volume}
  {55}},\ \bibinfo {pages} {311--326} (\bibinfo {year} {1993})}\BibitemShut
  {NoStop}%
\bibitem [{\citenamefont {Choi}, \citenamefont {DeBisschop},\ and\
  \citenamefont {Clayton}(1998)}]{Choi1998}%
  \BibitemOpen
  \bibfield  {author} {\bibinfo {author} {\bibfnamefont {K.-S.}\ \bibnamefont
  {Choi}}, \bibinfo {author} {\bibfnamefont {J.-R.}\ \bibnamefont
  {DeBisschop}}, \ and\ \bibinfo {author} {\bibfnamefont {B.}~\bibnamefont
  {Clayton}},\ }\bibfield  {title} {\enquote {\bibinfo {title} {Turbulent
  boundary-layer control by means of spanwise-wall oscillation},}\ }\href@noop
  {} {\bibfield  {journal} {\bibinfo  {journal} {AIAA J.}\ }\textbf {\bibinfo
  {volume} {36}} (\bibinfo {year} {1998})}\BibitemShut {NoStop}%
\bibitem [{\citenamefont {Ricco}\ \emph {et~al.}(2012)\citenamefont {Ricco},
  \citenamefont {Ottoneli}, \citenamefont {Hasegawa},\ and\ \citenamefont
  {Quadrio}}]{Quadrio2012}%
  \BibitemOpen
  \bibfield  {author} {\bibinfo {author} {\bibfnamefont {P.}~\bibnamefont
  {Ricco}}, \bibinfo {author} {\bibfnamefont {C.}~\bibnamefont {Ottoneli}},
  \bibinfo {author} {\bibfnamefont {Y.}~\bibnamefont {Hasegawa}}, \ and\
  \bibinfo {author} {\bibfnamefont {M.}~\bibnamefont {Quadrio}},\ }\bibfield
  {title} {\enquote {\bibinfo {title} {Changes in turbulent dissipation in a
  channel flow with oscillating walls},}\ }\href@noop {} {\bibfield  {journal}
  {\bibinfo  {journal} {J. Fluid Mech.}\ }\textbf {\bibinfo {volume} {700}},\
  \bibinfo {pages} {77--104} (\bibinfo {year} {2012})}\BibitemShut {NoStop}%
\bibitem [{\citenamefont {Moarref}\ and\ \citenamefont
  {Jovanovic}(2012)}]{Jovanovic2012}%
  \BibitemOpen
  \bibfield  {author} {\bibinfo {author} {\bibfnamefont {R.}~\bibnamefont
  {Moarref}}\ and\ \bibinfo {author} {\bibfnamefont {M.}~\bibnamefont
  {Jovanovic}},\ }\bibfield  {title} {\enquote {\bibinfo {title} {Model-based
  design of transverse wall oscillations for turbulent drag reduction},}\
  }\href@noop {} {\bibfield  {journal} {\bibinfo  {journal} {Journal of Fluid
  Mechanics}\ }\textbf {\bibinfo {volume} {707}},\ \bibinfo {pages} {205--240}
  (\bibinfo {year} {2012})}\BibitemShut {NoStop}%
\bibitem [{\citenamefont {Touber}\ and\ \citenamefont
  {Leschziner}(2012)}]{Touber2012}%
  \BibitemOpen
  \bibfield  {author} {\bibinfo {author} {\bibfnamefont {E.}~\bibnamefont
  {Touber}}\ and\ \bibinfo {author} {\bibfnamefont {M.}~\bibnamefont
  {Leschziner}},\ }\bibfield  {title} {\enquote {\bibinfo {title} {Near-wall
  streak modification by spanwise osscilatory wall motion and drag-reduction
  mechanisms},}\ }\href@noop {} {\bibfield  {journal} {\bibinfo  {journal} {J.
  Fluid Mech.}\ }\textbf {\bibinfo {volume} {693}},\ \bibinfo {pages}
  {150--200} (\bibinfo {year} {2012})}\BibitemShut {NoStop}%
\bibitem [{\citenamefont {Blesbois}\ \emph {et~al.}(2013)\citenamefont
  {Blesbois}, \citenamefont {S.I.Chernyshenko}, \citenamefont {Touber},\ and\
  \citenamefont {Leschziner}}]{Blesbois2013}%
  \BibitemOpen
  \bibfield  {author} {\bibinfo {author} {\bibfnamefont {O.}~\bibnamefont
  {Blesbois}}, \bibinfo {author} {\bibnamefont {S.I.Chernyshenko}}, \bibinfo
  {author} {\bibfnamefont {E.}~\bibnamefont {Touber}}, \ and\ \bibinfo {author}
  {\bibfnamefont {M.}~\bibnamefont {Leschziner}},\ }\bibfield  {title}
  {\enquote {\bibinfo {title} {Pattern prediction by linear analysis of
  turbulent flow with drag reduction by wall oscillation},}\ }\href@noop {}
  {\bibfield  {journal} {\bibinfo  {journal} {Journal of Fluid Mechanics}\
  }\textbf {\bibinfo {volume} {724}},\ \bibinfo {pages} {607--641} (\bibinfo
  {year} {2013})}\BibitemShut {NoStop}%
\bibitem [{\citenamefont {Agostini}, \citenamefont {Touber},\ and\
  \citenamefont {Leschziner}(2014)}]{Agostini2014}%
  \BibitemOpen
  \bibfield  {author} {\bibinfo {author} {\bibfnamefont {L.}~\bibnamefont
  {Agostini}}, \bibinfo {author} {\bibfnamefont {E.}~\bibnamefont {Touber}}, \
  and\ \bibinfo {author} {\bibfnamefont {M.}~\bibnamefont {Leschziner}},\
  }\bibfield  {title} {\enquote {\bibinfo {title} {Spanwise oscillatory wall
  motion in channel flow: drag-reduction mechanisms inferred from dns-predicted
  phase-wise property variations at ${R}e_{\tau}=1000$.}}\ }\href@noop {}
  {\bibfield  {journal} {\bibinfo  {journal} {Journal of Fluid Mechanics}\
  }\textbf {\bibinfo {volume} {743}},\ \bibinfo {pages} {606--635} (\bibinfo
  {year} {2014})}\BibitemShut {NoStop}%
\bibitem [{\citenamefont {Yudhistira}\ and\ \citenamefont
  {Skote}(2014)}]{Skote2011}%
  \BibitemOpen
  \bibfield  {author} {\bibinfo {author} {\bibfnamefont {I.}~\bibnamefont
  {Yudhistira}}\ and\ \bibinfo {author} {\bibfnamefont {M.}~\bibnamefont
  {Skote}},\ }\bibfield  {title} {\enquote {\bibinfo {title} {Direct numerical
  simulation of a turbulent boundary layer over an oscillating wall},}\
  }\href@noop {} {\bibfield  {journal} {\bibinfo  {journal} {Journal of
  turbulence}\ }\textbf {\bibinfo {volume} {12}},\ \bibinfo {pages} {1--17}
  (\bibinfo {year} {2014})}\BibitemShut {NoStop}%
\bibitem [{\citenamefont {Skote}(2013)}]{Skote2013}%
  \BibitemOpen
  \bibfield  {author} {\bibinfo {author} {\bibfnamefont {M.}~\bibnamefont
  {Skote}},\ }\bibfield  {title} {\enquote {\bibinfo {title} {Comparison
  between spatial and temporal wall oscillations in turbulent boundary layer
  flows},}\ }\href@noop {} {\bibfield  {journal} {\bibinfo  {journal} {Journal
  of Fluid Mechanics}\ }\textbf {\bibinfo {volume} {730}},\ \bibinfo {pages}
  {273--294} (\bibinfo {year} {2013})}\BibitemShut {NoStop}%
\bibitem [{\citenamefont {Skote}(2014)}]{Skote2014}%
  \BibitemOpen
  \bibfield  {author} {\bibinfo {author} {\bibfnamefont {M.}~\bibnamefont
  {Skote}},\ }\bibfield  {title} {\enquote {\bibinfo {title} {Scaling of the
  velocity profile in strongly drag reduced turbulent flows over an oscillating
  wall},}\ }\href@noop {} {\bibfield  {journal} {\bibinfo  {journal}
  {International Journal of Heat and Fluid Flow}\ }\textbf {\bibinfo {volume}
  {50}},\ \bibinfo {pages} {352–358} (\bibinfo {year} {2014})}\BibitemShut
  {NoStop}%
\bibitem [{\citenamefont {Hack}\ and\ \citenamefont {Zaki}(2014)}]{Zaki2014}%
  \BibitemOpen
  \bibfield  {author} {\bibinfo {author} {\bibfnamefont {M.}~\bibnamefont
  {Hack}}\ and\ \bibinfo {author} {\bibfnamefont {T.~A.}\ \bibnamefont
  {Zaki}},\ }\bibfield  {title} {\enquote {\bibinfo {title} {The influence of
  harmonic wall motion on transitional boundary layers},}\ }\href@noop {}
  {\bibfield  {journal} {\bibinfo  {journal} {Journal of Fluid Mechanics}\
  }\textbf {\bibinfo {volume} {760}},\ \bibinfo {pages} {63--94} (\bibinfo
  {year} {2014})}\BibitemShut {NoStop}%
\bibitem [{\citenamefont {Karniadakis}\ and\ \citenamefont
  {Choi}(2003)}]{Karniadakis2003}%
  \BibitemOpen
  \bibfield  {author} {\bibinfo {author} {\bibfnamefont {G.~E.}\ \bibnamefont
  {Karniadakis}}\ and\ \bibinfo {author} {\bibfnamefont {K.-S.}\ \bibnamefont
  {Choi}},\ }\bibfield  {title} {\enquote {\bibinfo {title} {Mechanisms on
  transverse motions in turbulent wall flows},}\ }\href@noop {} {\bibfield
  {journal} {\bibinfo  {journal} {Annu Rev. Fluid Mech.}\ }\textbf {\bibinfo
  {volume} {35}},\ \bibinfo {pages} {45--62} (\bibinfo {year}
  {2003})}\BibitemShut {NoStop}%
\bibitem [{\citenamefont {Quadrio}, \citenamefont {Ricco},\ and\ \citenamefont
  {Viotti}(2009)}]{Quadrio2009}%
  \BibitemOpen
  \bibfield  {author} {\bibinfo {author} {\bibfnamefont {M.}~\bibnamefont
  {Quadrio}}, \bibinfo {author} {\bibfnamefont {P.}~\bibnamefont {Ricco}}, \
  and\ \bibinfo {author} {\bibfnamefont {C.}~\bibnamefont {Viotti}},\
  }\bibfield  {title} {\enquote {\bibinfo {title} {Streamwise-travelling waves
  of spanwise wall velocity for turbulent drag reduction},}\ }\href@noop {}
  {\bibfield  {journal} {\bibinfo  {journal} {J. Fluid Mech.}\ }\textbf
  {\bibinfo {volume} {627}},\ \bibinfo {pages} {161--178} (\bibinfo {year}
  {2009})}\BibitemShut {NoStop}%
\bibitem [{\citenamefont {Duque-Daza}\ \emph {et~al.}(2012)\citenamefont
  {Duque-Daza}, \citenamefont {Baig}, \citenamefont {Lockerby}, \citenamefont
  {Chernyshenko},\ and\ \citenamefont {Davies}}]{Duque2012}%
  \BibitemOpen
  \bibfield  {author} {\bibinfo {author} {\bibfnamefont {C.}~\bibnamefont
  {Duque-Daza}}, \bibinfo {author} {\bibfnamefont {M.}~\bibnamefont {Baig}},
  \bibinfo {author} {\bibfnamefont {D.}~\bibnamefont {Lockerby}}, \bibinfo
  {author} {\bibfnamefont {S.}~\bibnamefont {Chernyshenko}}, \ and\ \bibinfo
  {author} {\bibfnamefont {C.}~\bibnamefont {Davies}},\ }\bibfield  {title}
  {\enquote {\bibinfo {title} {Modelling turbulent skin-friction control using
  linearized navier-stokes equations},}\ }\href@noop {} {\bibfield  {journal}
  {\bibinfo  {journal} {Journal of Fluid Mechanics}\ }\textbf {\bibinfo
  {volume} {702}},\ \bibinfo {pages} {403--414} (\bibinfo {year}
  {2012})}\BibitemShut {NoStop}%
\bibitem [{\citenamefont {Gallorini}, \citenamefont {Quadrio},\ and\
  \citenamefont {Gatti}(2022)}]{Gallorini2022}%
  \BibitemOpen
  \bibfield  {author} {\bibinfo {author} {\bibfnamefont {E.}~\bibnamefont
  {Gallorini}}, \bibinfo {author} {\bibfnamefont {M.}~\bibnamefont {Quadrio}},
  \ and\ \bibinfo {author} {\bibfnamefont {D.}~\bibnamefont {Gatti}},\
  }\bibfield  {title} {\enquote {\bibinfo {title} {Coherent near-wall
  structures and drag reduction by spanwise forcing},}\ }\href@noop {}
  {\bibfield  {journal} {\bibinfo  {journal} {Physical Review Fluids}\ }\textbf
  {\bibinfo {volume} {7}},\ \bibinfo {pages} {114602} (\bibinfo {year}
  {2022})}\BibitemShut {NoStop}%
\bibitem [{\citenamefont {Fukagata}, \citenamefont {Iwamoto},\ and\
  \citenamefont {Hasegawa}(2024)}]{Fukagata2024}%
  \BibitemOpen
  \bibfield  {author} {\bibinfo {author} {\bibfnamefont {K.}~\bibnamefont
  {Fukagata}}, \bibinfo {author} {\bibfnamefont {K.}~\bibnamefont {Iwamoto}}, \
  and\ \bibinfo {author} {\bibfnamefont {Y.}~\bibnamefont {Hasegawa}},\
  }\bibfield  {title} {\enquote {\bibinfo {title} {Turbulent drag reduction by
  streamwise traveling waves of wall-normal forcing},}\ }\href@noop {}
  {\bibfield  {journal} {\bibinfo  {journal} {Annual Review of Fluid
  Mechanics}\ }\textbf {\bibinfo {volume} {56}},\ \bibinfo {pages} {45--66}
  (\bibinfo {year} {2024})}\BibitemShut {NoStop}%
\bibitem [{\citenamefont {Choi}, \citenamefont {Moin},\ and\ \citenamefont
  {Kim}(1994)}]{Choi1994}%
  \BibitemOpen
  \bibfield  {author} {\bibinfo {author} {\bibfnamefont {H.}~\bibnamefont
  {Choi}}, \bibinfo {author} {\bibfnamefont {P.}~\bibnamefont {Moin}}, \ and\
  \bibinfo {author} {\bibfnamefont {J.}~\bibnamefont {Kim}},\ }\bibfield
  {title} {\enquote {\bibinfo {title} {Active turbulence control for drag
  reduction in wall-bounded flows},}\ }\href@noop {} {\bibfield  {journal}
  {\bibinfo  {journal} {J. Fluid Mech.}\ }\textbf {\bibinfo {volume} {262}},\
  \bibinfo {pages} {75--110} (\bibinfo {year} {1994})}\BibitemShut {NoStop}%
\bibitem [{\citenamefont {Stroh}\ \emph {et~al.}(2015)\citenamefont {Stroh},
  \citenamefont {Frohnapfel}, \citenamefont {Schlatter},\ and\ \citenamefont
  {Hasegawa}}]{Stroh2015}%
  \BibitemOpen
  \bibfield  {author} {\bibinfo {author} {\bibfnamefont {A.}~\bibnamefont
  {Stroh}}, \bibinfo {author} {\bibfnamefont {B.}~\bibnamefont {Frohnapfel}},
  \bibinfo {author} {\bibfnamefont {P.}~\bibnamefont {Schlatter}}, \ and\
  \bibinfo {author} {\bibfnamefont {Y.}~\bibnamefont {Hasegawa}},\ }\bibfield
  {title} {\enquote {\bibinfo {title} {A comparison of opposition control in
  turbulent boundary layer and turbulent channel flow},}\ }\href@noop {}
  {\bibfield  {journal} {\bibinfo  {journal} {Physics of Fluids}\ }\textbf
  {\bibinfo {volume} {27}},\ \bibinfo {pages} {075101} (\bibinfo {year}
  {2015})}\BibitemShut {NoStop}%
\bibitem [{\citenamefont {Xia}, \citenamefont {Xuang},\ and\ \citenamefont
  {Xu}(2015)}]{Xia2015}%
  \BibitemOpen
  \bibfield  {author} {\bibinfo {author} {\bibfnamefont {Q.}~\bibnamefont
  {Xia}}, \bibinfo {author} {\bibfnamefont {W.}~\bibnamefont {Xuang}}, \ and\
  \bibinfo {author} {\bibfnamefont {C.}~\bibnamefont {Xu}},\ }\bibfield
  {title} {\enquote {\bibinfo {title} {Direct numerical simulation of spatially
  developing turbulent boundary layers with opposition control},}\ }\href@noop
  {} {\bibfield  {journal} {\bibinfo  {journal} {Fluid Dynamic Research}\
  }\textbf {\bibinfo {volume} {47}},\ \bibinfo {pages} {025503} (\bibinfo
  {year} {2015})}\BibitemShut {NoStop}%
\bibitem [{\citenamefont {Cheng}\ \emph {et~al.}(2021)\citenamefont {Cheng},
  \citenamefont {Qiao}, \citenamefont {Zhang}, \citenamefont {Quadrio},\ and\
  \citenamefont {Zhou}}]{Cheng2021}%
  \BibitemOpen
  \bibfield  {author} {\bibinfo {author} {\bibfnamefont {X.}~\bibnamefont
  {Cheng}}, \bibinfo {author} {\bibfnamefont {Z.}~\bibnamefont {Qiao}},
  \bibinfo {author} {\bibfnamefont {X.}~\bibnamefont {Zhang}}, \bibinfo
  {author} {\bibfnamefont {M.}~\bibnamefont {Quadrio}}, \ and\ \bibinfo
  {author} {\bibfnamefont {Y.}~\bibnamefont {Zhou}},\ }\bibfield  {title}
  {\enquote {\bibinfo {title} {Skin-friction reduction using periodic blowing
  through streamwise slits},}\ }\href@noop {} {\bibfield  {journal} {\bibinfo
  {journal} {Journal of Fluid Mechanics}\ }\textbf {\bibinfo {volume} {920}},\
  \bibinfo {pages} {A50} (\bibinfo {year} {2021})}\BibitemShut {NoStop}%
\bibitem [{\citenamefont {Rouhi}\ \emph {et~al.}(2023)\citenamefont {Rouhi},
  \citenamefont {Fu}, \citenamefont {Chandran}, \citenamefont {Zampiron},
  \citenamefont {Smits},\ and\ \citenamefont {Marusic}}]{Marusic2023a}%
  \BibitemOpen
  \bibfield  {author} {\bibinfo {author} {\bibfnamefont {A.}~\bibnamefont
  {Rouhi}}, \bibinfo {author} {\bibfnamefont {M.}~\bibnamefont {Fu}}, \bibinfo
  {author} {\bibfnamefont {D.}~\bibnamefont {Chandran}}, \bibinfo {author}
  {\bibfnamefont {A.}~\bibnamefont {Zampiron}}, \bibinfo {author}
  {\bibfnamefont {A.}~\bibnamefont {Smits}}, \ and\ \bibinfo {author}
  {\bibfnamefont {I.}~\bibnamefont {Marusic}},\ }\bibfield  {title} {\enquote
  {\bibinfo {title} {Turbulent drag reduction by spanwise wall forcing. part 1.
  large-eddy simulations},}\ }\href@noop {} {\bibfield  {journal} {\bibinfo
  {journal} {Journal of Fluid Mechanics}\ }\textbf {\bibinfo {volume} {968}},\
  \bibinfo {pages} {A6} (\bibinfo {year} {2023})}\BibitemShut {NoStop}%
\bibitem [{\citenamefont {Chandran}\ \emph {et~al.}(2023)\citenamefont
  {Chandran}, \citenamefont {Zampiron}, \citenamefont {Rouhi}, \citenamefont
  {Fu}, \citenamefont {Wine}, \citenamefont {Holloway}, \citenamefont {Smits},\
  and\ \citenamefont {Marusic}}]{Marusic2023b}%
  \BibitemOpen
  \bibfield  {author} {\bibinfo {author} {\bibfnamefont {D.}~\bibnamefont
  {Chandran}}, \bibinfo {author} {\bibfnamefont {A.}~\bibnamefont {Zampiron}},
  \bibinfo {author} {\bibfnamefont {A.}~\bibnamefont {Rouhi}}, \bibinfo
  {author} {\bibfnamefont {M.}~\bibnamefont {Fu}}, \bibinfo {author}
  {\bibfnamefont {D.}~\bibnamefont {Wine}}, \bibinfo {author} {\bibfnamefont
  {B.}~\bibnamefont {Holloway}}, \bibinfo {author} {\bibfnamefont
  {A.}~\bibnamefont {Smits}}, \ and\ \bibinfo {author} {\bibfnamefont
  {I.}~\bibnamefont {Marusic}},\ }\bibfield  {title} {\enquote {\bibinfo
  {title} {Turbulent drag reduction by spanwise wall forcing. part 2.
  high-reynolds-number experiments},}\ }\href@noop {} {\bibfield  {journal}
  {\bibinfo  {journal} {Journal of Fluid Mechanics}\ }\textbf {\bibinfo
  {volume} {968}},\ \bibinfo {pages} {A7} (\bibinfo {year} {2023})}\BibitemShut
  {NoStop}%
\bibitem [{\citenamefont {Kametani}\ \emph {et~al.}(2015)\citenamefont
  {Kametani}, \citenamefont {Fukagata}, \citenamefont {\"Orl\"u},\ and\
  \citenamefont {Schlatter}}]{Kametani2015}%
  \BibitemOpen
  \bibfield  {author} {\bibinfo {author} {\bibfnamefont {Y.}~\bibnamefont
  {Kametani}}, \bibinfo {author} {\bibfnamefont {K.}~\bibnamefont {Fukagata}},
  \bibinfo {author} {\bibfnamefont {R.}~\bibnamefont {\"Orl\"u}}, \ and\
  \bibinfo {author} {\bibfnamefont {P.}~\bibnamefont {Schlatter}},\ }\bibfield
  {title} {\enquote {\bibinfo {title} {Effect of uniform blowing/suction in a
  turbulent boundary layer at moderate reynolds number},}\ }\href@noop {}
  {\bibfield  {journal} {\bibinfo  {journal} {Intl J. Heat Fluid Flow}\
  }\textbf {\bibinfo {volume} {55}},\ \bibinfo {pages} {132–142} (\bibinfo
  {year} {2015})}\BibitemShut {NoStop}%
\bibitem [{\citenamefont {Kametani}\ \emph {et~al.}(2016)\citenamefont
  {Kametani}, \citenamefont {Fukagata}, \citenamefont {\"Orl\"u},\ and\
  \citenamefont {Schlatter}}]{Kametani2016}%
  \BibitemOpen
  \bibfield  {author} {\bibinfo {author} {\bibfnamefont {Y.}~\bibnamefont
  {Kametani}}, \bibinfo {author} {\bibfnamefont {K.}~\bibnamefont {Fukagata}},
  \bibinfo {author} {\bibfnamefont {R.}~\bibnamefont {\"Orl\"u}}, \ and\
  \bibinfo {author} {\bibfnamefont {P.}~\bibnamefont {Schlatter}},\ }\bibfield
  {title} {\enquote {\bibinfo {title} {Drag reduction in spatially developing
  turbulent boundary layers by spatially intermittent blowing at constant
  mass-flux},}\ }\href@noop {} {\bibfield  {journal} {\bibinfo  {journal}
  {Journal of Turbulence}\ }\textbf {\bibinfo {volume} {17}},\ \bibinfo {pages}
  {913--929} (\bibinfo {year} {2016})}\BibitemShut {NoStop}%
\bibitem [{\citenamefont {Fukagata}, \citenamefont {Iwamoto},\ and\
  \citenamefont {Kasagi}(2002)}]{FIK}%
  \BibitemOpen
  \bibfield  {author} {\bibinfo {author} {\bibfnamefont {K.}~\bibnamefont
  {Fukagata}}, \bibinfo {author} {\bibfnamefont {K.}~\bibnamefont {Iwamoto}}, \
  and\ \bibinfo {author} {\bibfnamefont {N.}~\bibnamefont {Kasagi}},\
  }\bibfield  {title} {\enquote {\bibinfo {title} {Contribution of reynolds
  stress distribution to the skin friction wall-bounded flows},}\ }\href@noop
  {} {\bibfield  {journal} {\bibinfo  {journal} {J. Fluid Mech.}\ }\textbf
  {\bibinfo {volume} {14}},\ \bibinfo {pages} {L73--L76} (\bibinfo {year}
  {2002})}\BibitemShut {NoStop}%
\bibitem [{\citenamefont {Stroh}\ \emph {et~al.}(2016)\citenamefont {Stroh},
  \citenamefont {Hasegawa}, \citenamefont {Schlatter},\ and\ \citenamefont
  {Frohnapfel}}]{Stroh2016}%
  \BibitemOpen
  \bibfield  {author} {\bibinfo {author} {\bibfnamefont {A.}~\bibnamefont
  {Stroh}}, \bibinfo {author} {\bibfnamefont {Y.}~\bibnamefont {Hasegawa}},
  \bibinfo {author} {\bibfnamefont {P.}~\bibnamefont {Schlatter}}, \ and\
  \bibinfo {author} {\bibfnamefont {B.}~\bibnamefont {Frohnapfel}},\ }\bibfield
   {title} {\enquote {\bibinfo {title} {Global effect of local skin friction
  drag reduction in spatially developing turbulent boundary layer},}\
  }\href@noop {} {\bibfield  {journal} {\bibinfo  {journal} {Journal of Fluid
  Mechanics}\ }\textbf {\bibinfo {volume} {805}},\ \bibinfo {pages} {303--321}
  (\bibinfo {year} {2016})}\BibitemShut {NoStop}%
\bibitem [{\citenamefont {Chagelishvili}\ \emph {et~al.}(2014)\citenamefont
  {Chagelishvili}, \citenamefont {Khujadze}, \citenamefont {Foysi},\ and\
  \citenamefont {Oberlack}}]{Chagelishvili2014}%
  \BibitemOpen
  \bibfield  {author} {\bibinfo {author} {\bibfnamefont {G.}~\bibnamefont
  {Chagelishvili}}, \bibinfo {author} {\bibfnamefont {G.}~\bibnamefont
  {Khujadze}}, \bibinfo {author} {\bibfnamefont {H.}~\bibnamefont {Foysi}}, \
  and\ \bibinfo {author} {\bibfnamefont {M.}~\bibnamefont {Oberlack}},\
  }\bibfield  {title} {\enquote {\bibinfo {title} {Spanwise reflection symmetry
  breaking and turbulence control: Plane couette flow},}\ }\href@noop {}
  {\bibfield  {journal} {\bibinfo  {journal} {Journal of Fluid Mechanics}\
  }\textbf {\bibinfo {volume} {745}},\ \bibinfo {pages} {300--320} (\bibinfo
  {year} {2014})}\BibitemShut {NoStop}%
\bibitem [{\citenamefont {Farrell}\ and\ \citenamefont
  {Ioannou}(1993)}]{Farr931}%
  \BibitemOpen
  \bibfield  {author} {\bibinfo {author} {\bibfnamefont {B.~F.}\ \bibnamefont
  {Farrell}}\ and\ \bibinfo {author} {\bibfnamefont {P.~J.}\ \bibnamefont
  {Ioannou}},\ }\bibfield  {title} {\enquote {\bibinfo {title} {Optimal
  excitation of three-dimensional perturbations in viscous constant shear
  flow},}\ }\href@noop {} {\bibfield  {journal} {\bibinfo  {journal} {Phys.
  Fluids A}\ }\textbf {\bibinfo {volume} {5}},\ \bibinfo {pages} {1390--1400}
  (\bibinfo {year} {1993})}\BibitemShut {NoStop}%
\bibitem [{\citenamefont {Farrell}\ and\ \citenamefont
  {Ioannou}(2000)}]{Farrell2000}%
  \BibitemOpen
  \bibfield  {author} {\bibinfo {author} {\bibfnamefont {B.~F.}\ \bibnamefont
  {Farrell}}\ and\ \bibinfo {author} {\bibfnamefont {P.~J.}\ \bibnamefont
  {Ioannou}},\ }\bibfield  {title} {\enquote {\bibinfo {title} {{Transient and
  asymptotic growth of two-dimensional perturbations in viscous compressible
  shear flow}},}\ }\href@noop {} {\bibfield  {journal} {\bibinfo  {journal}
  {{Phys. Fluids}}\ }\textbf {\bibinfo {volume} {12}},\ \bibinfo {pages} {3021
  -- 3028} (\bibinfo {year} {2000})}\BibitemShut {NoStop}%
\bibitem [{\citenamefont {Chevalier}\ \emph {et~al.}(2007)\citenamefont
  {Chevalier}, \citenamefont {Schlatter}, \citenamefont {Lundbladh},\ and\
  \citenamefont {Henningson}}]{Chevalier2007}%
  \BibitemOpen
  \bibfield  {author} {\bibinfo {author} {\bibfnamefont {M.}~\bibnamefont
  {Chevalier}}, \bibinfo {author} {\bibfnamefont {P.}~\bibnamefont
  {Schlatter}}, \bibinfo {author} {\bibfnamefont {A.}~\bibnamefont
  {Lundbladh}}, \ and\ \bibinfo {author} {\bibfnamefont {D.~S.}\ \bibnamefont
  {Henningson}},\ }\href@noop {} {\enquote {\bibinfo {title} {{SIMSON -- A
  pseudo-spectral solver for incompressible boundary layer flows}},}\ }\bibinfo
  {type} {Tech. Rep.}\ (\bibinfo  {institution} {KTH Stockholm},\ \bibinfo
  {year} {2007})\BibitemShut {NoStop}%
\bibitem [{\citenamefont {Tsukahara}, \citenamefont {Kawamura},\ and\
  \citenamefont {Shingai}(2006)}]{Tsukahara2006}%
  \BibitemOpen
  \bibfield  {author} {\bibinfo {author} {\bibfnamefont {T.}~\bibnamefont
  {Tsukahara}}, \bibinfo {author} {\bibfnamefont {H.}~\bibnamefont {Kawamura}},
  \ and\ \bibinfo {author} {\bibfnamefont {K.}~\bibnamefont {Shingai}},\
  }\bibfield  {title} {\enquote {\bibinfo {title} {{DNS of turbulent Couette
  flow with emphasis on the large-scale structure in the core region}},}\
  }\href@noop {} {\bibfield  {journal} {\bibinfo  {journal} {JTurb}\ }\textbf
  {\bibinfo {volume} {7}},\ \bibinfo {pages} {1--16} (\bibinfo {year}
  {2006})}\BibitemShut {NoStop}%
\bibitem [{\citenamefont {Pirozzoli}, \citenamefont {Bernardini},\ and\
  \citenamefont {Orlandi}(2014)}]{Pirozzoli2014}%
  \BibitemOpen
  \bibfield  {author} {\bibinfo {author} {\bibfnamefont {S.}~\bibnamefont
  {Pirozzoli}}, \bibinfo {author} {\bibfnamefont {M.}~\bibnamefont
  {Bernardini}}, \ and\ \bibinfo {author} {\bibfnamefont {P.}~\bibnamefont
  {Orlandi}},\ }\bibfield  {title} {\enquote {\bibinfo {title} {Turbulence
  statistics in couette flow at high reynolds number},}\ }\href@noop {}
  {\bibfield  {journal} {\bibinfo  {journal} {journal of Fluid Mechanics}\
  }\textbf {\bibinfo {volume} {758}},\ \bibinfo {pages} {327--343} (\bibinfo
  {year} {2014})}\BibitemShut {NoStop}%
\bibitem [{\citenamefont {Lee}\ and\ \citenamefont
  {Moser}(2018)}]{LeeMoser2018}%
  \BibitemOpen
  \bibfield  {author} {\bibinfo {author} {\bibfnamefont {M.}~\bibnamefont
  {Lee}}\ and\ \bibinfo {author} {\bibfnamefont {R.}~\bibnamefont {Moser}},\
  }\bibfield  {title} {\enquote {\bibinfo {title} {Extreme-scale motions in
  turbulent plane couette flows},}\ }\href@noop {} {\bibfield  {journal}
  {\bibinfo  {journal} {Journal of Fluid Mechanics}\ }\textbf {\bibinfo
  {volume} {842}},\ \bibinfo {pages} {128--145} (\bibinfo {year}
  {2018})}\BibitemShut {NoStop}%
\bibitem [{\citenamefont {Craik}\ and\ \citenamefont
  {Criminale}(1986)}]{Craik1986}%
  \BibitemOpen
  \bibfield  {author} {\bibinfo {author} {\bibfnamefont {A.~D.~D.}\
  \bibnamefont {Craik}}\ and\ \bibinfo {author} {\bibfnamefont {W.~O.}\
  \bibnamefont {Criminale}},\ }\bibfield  {title} {\enquote {\bibinfo {title}
  {{Evolution of Wavelike Disturbances in Shear Flows: A Class of Exact
  Solutions of the Navier-Stokes Equations}},}\ }\href@noop {} {\bibfield
  {journal} {\bibinfo  {journal} {Proc. R. Soc. Lond. A}\ }\textbf {\bibinfo
  {volume} {406}},\ \bibinfo {pages} {13--26} (\bibinfo {year}
  {1986})}\BibitemShut {NoStop}%
\end{thebibliography}
\end{document}